# Non-Adiabatic Field on Quantum Phase Space: A Century after Ehrenfest

*Baihua Wu, Xin He, and Jian Liu\**

Beijing National Laboratory for Molecular Sciences, Institute of Theoretical and Computational Chemistry, College of Chemistry and Molecular Engineering,

Peking University, Beijing 100871, China

AUTHOR INFORMATION

**Corresponding Author**

\* Electronic mail: jianliupku@pku.edu.cn

ABSTRACT.

Non-adiabatic transition dynamics lies at the core of many electron/hole transfer, photo-activated, and vacuum field-coupled processes. About a century after Ehrenfest proposed "*Phasenraum*" and the Ehrenfest theorem, we report a conceptually novel trajectory-based non-adiabatic dynamics approach, non-adiabatic field (NAF) based on a generalized exact coordinate-momentum phase space formulation of quantum mechanics. It does not employ the conventional Born-Oppenheimer or Ehrenfest trajectory in the non-adiabatic coupling region. Instead, in NAF the equations of motion of the independent trajectory involve a non-adiabatic nuclear force term in addition to an adiabatic nuclear force term of a single electronic state. A few benchmark tests for gas phase and condensed phase systems indicate that NAF offers a practical tool to capture the correct correlation of electronic and nuclear dynamics for processes where the states keep coupled all the time as well as for the asymptotic region where the coupling of electronic states vanishes.

**TOC GRAPHICS**

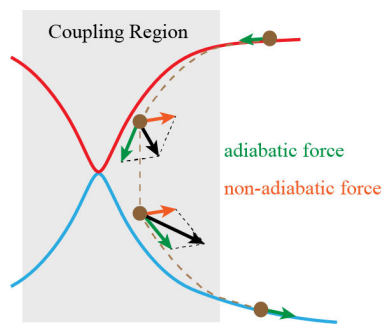

Since Paul Ehrenfest and Tatyana Ehrenfest first used "*Phasenraum*" in 1911, the expression "phase space" has been widely employed[1, 2]. In classical mechanics, phase space with coordinate-momentum variables presents a fundamental concept and tool[3, 4]. Phase space formulations with coordinate-momentum variables also offer an exact interpretation of quantum mechanics for continuous-variable systems[5-9], for which pioneering works were demonstrated by Weyl and Wigner[5, 6] and a unified scheme was later presented by Cohen[10]. More recently, we have proposed a generalized exact coordinate-momentum phase space formulation of quantum mechanics for composite systems, where both continuous and discrete variables are involved[11-18]. It maps discrete degrees of freedom (DOFs) onto *constraint* coordinate-momentum *phase space* (CPS) and continuous DOFs onto infinite coordinate-momentum phase space. Trajectory-based dynamics is derived from the symplectic structure of mapping phase space where the generated mapping Hamiltonian is conserved. The generalized exact coordinate-momentum phase space formulation then offers a framework to develop approximate but practically useful dynamics approaches for studying composite systems[16, 17] that include coupled multi-electronic-state systems that appear in such as photochemistry, charge (electron/hole) and electronic energy transfer, cavity modified phenomena, etc[19-29]. Such important processes involve the quantum mechanical behavior of both electrons and nuclei in the context of non-adiabatic dynamics, where electronic DOFs are often described by the discrete (adiabatic or diabatic) state representation and nuclear DOFs are depicted in continuous space.

Paul Ehrenfest also proposed the Ehrenfest theorem[30] in 1927. It has led to a simple and convenient mean field approach, also called Ehrenfest dynamics. Early examples of application of the Ehrenfest theorem to chemical dynamics include refs [31-34]. Although it is not clear to us when the Ehrenfest theorem was first used to study electronically non-adiabatic transition processes, Ehrenfest dynamics has been a conventional practical approach in the field of non-adiabatic dynamics. (Please see more discussion in Section S8 of the Supporting Information.) It involves the independent trajectory, for which the evolution of nuclei is governed by the mean field force that is independent of the representation of the electronic basis sets. Ehrenfest dynamics is useful for describing the dynamical behavior in the coupling region but fails to capture the bifurcation characteristic for the nuclear motion in the asymptotic region where the coupling of electronic states vanishes. Another prevailing category of trajectory-based non-adiabatic methods employs the independent Born-Oppenheimer trajectory, of which the nuclear force is produced from a single adiabatic potential energy surface (PES). For example, surface hopping (SH) methods[35-41] (pioneered by Tully[42, 43]) have been developed by using various hopping mechanisms to connect two independent Born-Oppenheimer trajectories on two different adiabatic PESes in the coupling region. Such methods are capable of depicting the correct physical picture for the nuclear motion in the asymptotic region but are often less efficient and/or less accurate in the coupling region. It is often challenging for Born-Oppenheimer trajectory-based methods to describe dynamical processes where the states remain coupled all the time, especially in the low temperature regime.

As described in ref [17], three key elements for a trajectory-based quantum dynamics method are the equations of motion (EOMs) of the trajectory, the initial condition of the trajectory, and the integral expression for evaluation of the time-dependent physical property. The generalized exact coordinate-momentum phase space formulation of quantum mechanics provides a new stage to investigate such key elements. In the Perspective, we focus on the investigation of the EOMs in the electronically adiabatic representation based on our previous work in refs [15, 17]. We propose a conceptually novel trajectory-based approach—nonadiabatic field (NAF) by maintaining the nonadiabatic term of the Ehrenfest-like force in addition to a single-state adiabatic nuclear force in the EOMs for nuclear mapping variables. The purpose of the Perspective is to show that NAF is capable of capturing important features in both coupling and asymptotic regions for non-adiabatic molecular dynamics.

**Generalized Coordinate-Momentum Phase Space Formulation of Quantum Mechanics.** Consider the full Hamiltonian of nuclei and electrons of the molecular system,

$$\hat{H} = \frac{1}{2}\hat{\mathbf{P}}^T \mathbf{M}^{-1}\hat{\mathbf{P}} + \hat{H}_{el}(\hat{\mathbf{R}}) \quad , \tag{1}$$

where $\hat{H}_{el}(\hat{\mathbf{R}})$ is the electronic Hamiltonian, $\mathbf{M} = \text{diag}\{m_j\}$ is the diagonal nuclear mass matrix, and $\{\mathbf{R}, \mathbf{P}\}$ are the coordinates and momenta of nuclear degrees of freedom (DOFs). Assume that $\{|n\rangle\}, n \in \{1,\cdots,F\}$ is the "complete" set of diabatic electronic states and $\{|\phi_k(\mathbf{R})\rangle\}, n \in \{1,\cdots,F\}$ is the "complete" set of adiabatic electronic states. ($F$ is in general infinite when the set of electronic states is rigorously complete.) The representation of $\hat{H}_{el}(\hat{\mathbf{R}})$ in the diabatic basis reads

$$\hat{H}_{el}(\mathbf{R}) = \sum_{n,m} V_{nm}(\mathbf{R})|n\rangle\langle m| \equiv \mathbf{V}(\mathbf{R}) \tag{2}$$

and that in the adiabatic basis is

$$\hat{H}_{el}(\mathbf{R}) = \sum_k E_k(\mathbf{R})|\phi_k(\mathbf{R})\rangle\langle\phi_k(\mathbf{R})| \quad , \tag{3}$$

where $E_k(\mathbf{R})$ denotes the adiabatic potential energy surface of the *k*-th adiabatic electronic state. The expression of eq (3) for the electronic Hamiltonian $\hat{H}_{el}(\mathbf{R})$ in eq (1) first appeared for phase space mapping methods for non-adiabatic dynamics in refs [17, 44]. The generalized coordinate-momentum phase space formulation maps the coupled $F$-electronic-state Hamiltonian operator (e.g., in the diabatic basis for eqs (1)-(2)), $\hat{H} = \sum_{n,m=1}^{F} \left[ \frac{1}{2} \hat{\mathbf{P}}^T \mathbf{M}^{-1} \hat{\mathbf{P}} \delta_{nm} + V_{nm}(\hat{\mathbf{R}}) \right] |n\rangle\langle m|$, to

$$H(\mathbf{R},\mathbf{P};\mathbf{x},\mathbf{p};\boldsymbol{\Gamma}) = \frac{1}{2}\mathbf{P}^T\mathbf{M}^{-1}\mathbf{P} + \sum_{n,m=1}^{F}\left[\frac{1}{2}\left(x^{(n)}x^{(m)} + p^{(n)}p^{(m)}\right) - \Gamma_{nm}\right]V_{mn}(\mathbf{R}) \; , \qquad (4)$$

where $\{\mathbf{x},\mathbf{p}\} = \{x^{(1)},\cdots,x^{(F)},p^{(1)},\cdots,p^{(F)}\}$ are the coordinate and momentum variables for representing the discrete electronic states, and $\boldsymbol{\Gamma}$ is the Hermitian commutator matrix whose elements are variables. For convenience, the potential energy matrix $V_{nm}(\hat{\mathbf{R}})$ is set to be a real symmetric matrix. Commutator matrix $\boldsymbol{\Gamma}$ was heuristically proposed from the general commutation relation, $\left[\hat{x}^{(m)},\hat{p}^{(n)}\right] + \left[\hat{x}^{(n)},\hat{p}^{(m)}\right] = 2i(\Gamma_{nm} + \Gamma_{mn})$ for the underlying mapping electronic DOFs which are neither bosons nor fermions[11, 15] (with the convention $\hbar = 1$ for electronic DOFs throughout the Perspective). Mapping Hamiltonian eq (4) is rigorously produced from the one-to-one correspondence relation between any quantum operator $\hat{A}$ and its corresponding phase space function

$$A_C(\mathbf{R},\mathbf{P};\mathbf{x},\mathbf{p};\boldsymbol{\Gamma}) = \text{Tr}_{n,e}\left[\hat{A}\hat{K}_{nuc}(\mathbf{R},\mathbf{P}) \otimes \hat{K}_{ele}(\mathbf{x},\mathbf{p};\boldsymbol{\Gamma})\right]. \qquad (5)$$

Here, $\text{Tr}_n[\ ]$ and $\text{Tr}_e[\ ]$ represent the trace over nuclear and electronic DOFs, respectively, and the kernel of the Wigner function[6] is used as the mapping kernel for continuous (nuclear) DOFs,

$$\hat{K}_{nuc}(\mathbf{R},\mathbf{P}) = \left(\frac{\hbar}{2\pi}\right)^N \int d\boldsymbol{\zeta}d\boldsymbol{\eta}\, e^{i\boldsymbol{\zeta}\cdot(\hat{\mathbf{R}}-\mathbf{R}) + i\boldsymbol{\eta}\cdot(\hat{\mathbf{P}}-\mathbf{P})}, \qquad (6)$$

where $N$ represents the number of dimensions of nuclear DOFs. The mapping relation, eq (5), implies a broad class of mapping kernels on CPS for discrete (electronic) DOFs, e.g.,

$$\hat{K}_{ele}(\mathbf{x},\mathbf{p},\mathbf{\Gamma}) = \sum_{n,m=1}^{F} \left[ \frac{1}{2}\left(x^{(n)}+ip^{(n)}\right)\left(x^{(m)}-ip^{(m)}\right) - \Gamma_{nm} \right] |n\rangle\langle m| \; . \tag{7}$$

It is straightforward to show that the mathematical structure of the mapping CPS for eq (7) is related to the quotient space $U(F)/U(F-r)$, namely the complex Stiefel manifold[45-47]. Here, $1 \leq r < F$.

Consider the simplest case, $r=1$. Commutator matrix $\mathbf{\Gamma}$ becomes a constant matrix, i.e., the mapping CPS is reduced to the quotient space $U(F)/U(F-1)$,

$$\mathcal{S}(\mathbf{x},\mathbf{p};\gamma) = \delta\left( \sum_{n=1}^{F} \frac{\left(x^{(n)}\right)^2 + \left(p^{(n)}\right)^2}{2} - (1+F\gamma) \right) \; , \tag{8}$$

where parameter $\gamma \in (-1/F, \infty)$. The mapping kernel on CPS for discrete (electronic) DOFs, eq (7), is then

$$\hat{K}_{ele}(\mathbf{x},\mathbf{p};\gamma) = \sum_{n,m=1}^{F} \left[ \frac{1}{2}\left(x^{(n)}+ip^{(n)}\right)\left(x^{(m)}-ip^{(m)}\right) - \gamma \right] |n\rangle\langle m| \; , \tag{9}$$

and mapping Hamiltonian eq (4) then leads to a model reminiscent of the celebrated Meyer-Miller mapping Hamiltonian, which in 1979 Meyer and Miller originally proposed by applying the Langer correction[48] and in 1997 Stock and Thoss derived by using the Schwinger oscillator model of angular momentum[49]. The generalized coordinate-momentum phase space formulation suggests that the physical meaning of parameter $\gamma$ is beyond the conventional zero-point-energy parameter indicated in refs [48, 49], which takes even negative values and is a special case of the commutator matrix related to the complex Stiefel manifold[11, 13-17, 47]. The trace of a product of two quantum operators is expressed as an integral of two functions on mapping coordinate-momentum phase space, that is,

$$\mathrm{Tr}_{n,e}\left[\hat{A}\hat{B}\right]=\int d\boldsymbol{\mu}_{\mathrm{nuc}}\left(\mathbf{R},\mathbf{P}\right)\int_{\mathcal{S}(\mathbf{x},\mathbf{p};\gamma)}d\boldsymbol{\mu}_{\mathrm{ele}}\left(\mathbf{x},\mathbf{p};\gamma\right)A_{C}\left(\mathbf{R},\mathbf{P};\mathbf{x},\mathbf{p};\gamma\right)\tilde{B}_{C}\left(\mathbf{R},\mathbf{P};\mathbf{x},\mathbf{p};\gamma\right) \quad (10)$$

with

$$\tilde{B}_{C}\left(\mathbf{R},\mathbf{P};\mathbf{x},\mathbf{p};\gamma\right)=\mathrm{Tr}_{n,e}\left[\hat{K}_{\mathrm{nuc}}^{-1}\left(\mathbf{R},\mathbf{P}\right)\otimes\hat{K}_{\mathrm{ele}}^{-1}\left(\mathbf{x},\mathbf{p};\gamma\right)\hat{B}\right], \quad (11)$$

$d\boldsymbol{\mu}_{\mathrm{nuc}}\left(\mathbf{R},\mathbf{P}\right)=(2\pi\hbar)^{-N}d\mathbf{R}d\mathbf{P}$ and $d\boldsymbol{\mu}_{\mathrm{ele}}\left(\mathbf{x},\mathbf{p};\gamma\right)=Fd\mathbf{x}d\mathbf{p}$ as the integration measure on infinite phase space for continuous nuclear DOFs and that on CPS for discrete electronic DOFs, respectively. The integral over the mapping CPS variables for electronic DOFs in eq (10) is performed as

$$\int_{\mathcal{S}(\mathbf{x},\mathbf{p})}Fd\mathbf{x}d\mathbf{p}\,g(\mathbf{x},\mathbf{p})=\int Fd\mathbf{x}d\mathbf{p}\frac{1}{\Omega}\mathcal{S}(\mathbf{x},\mathbf{p})g(\mathbf{x},\mathbf{p}), \quad (12)$$

where $\Omega=\int d\mathbf{x}d\mathbf{p}\mathcal{S}(\mathbf{x},\mathbf{p})$ is the area of constraint phase space $\mathcal{S}(\mathbf{x},\mathbf{p})$ as the normalization constant. In eq (11) the inverse mapping kernel for continuous (nuclear) DOFs is

$$\hat{K}_{nuc}^{-1}\left(\mathbf{R},\mathbf{P}\right)=\left(\frac{\hbar}{2\pi}\right)^{N}\int d\boldsymbol{\zeta}d\boldsymbol{\eta}e^{i\boldsymbol{\zeta}\cdot(\hat{\mathbf{R}}-\mathbf{R})+i\boldsymbol{\eta}\cdot(\hat{\mathbf{P}}-\mathbf{P})}. \quad (13)$$

The corresponding inverse mapping kernel for finite discrete electronic DOFs is, however, not unique[47], even when CPS is the quotient space $U(F)/U(F-1)$. For simplicity, in the Perspective, we use the inverse mapping kernel of refs [13-15],

$$\hat{K}_{\mathrm{ele}}^{-1}\left(\mathbf{x},\mathbf{p};\gamma\right)=\sum_{n,m=1}^{F}\left[\frac{1+F}{2(1+F\gamma)^{2}}\left(x^{(n)}+ip^{(n)}\right)\left(x^{(m)}-ip^{(m)}\right)-\frac{1-\gamma}{1+F\gamma}\delta_{nm}\right]|n\rangle\langle m| \quad (14)$$

for demonstration. More inverse mapping kernels will be discussed in our forthcoming paper.

**Equations of Motion on the Generalized Coordinate-Momentum Phase Space.** When operator $\hat{B}$ is replaced by its Heisenberg operator $\hat{B}(t)$, the evaluation of eq (10) is often approximated by trajectory-based dynamics on mapping phase space, e.g.,

$$\mathrm{Tr}_{n,e}\left[\hat{A}(0)\hat{B}(t)\right] = \int d\boldsymbol{\mu}_{\mathrm{nuc}}(\mathbf{R},\mathbf{P})\int_{\mathcal{S}(\mathbf{x},\mathbf{p};\gamma)} d\boldsymbol{\mu}_{\mathrm{ele}}(\mathbf{x},\mathbf{p};\gamma) A_C(\mathbf{R},\mathbf{P};\mathbf{x},\mathbf{p};\gamma) \tilde{B}_C(\mathbf{R},\mathbf{P};\mathbf{x},\mathbf{p};\gamma;t)$$
$$\approx \int d\boldsymbol{\mu}_{\mathrm{nuc}}(\mathbf{R},\mathbf{P})\int_{\mathcal{S}(\mathbf{x},\mathbf{p};\gamma)} d\boldsymbol{\mu}_{\mathrm{ele}}(\mathbf{x},\mathbf{p};\gamma) A_C(\mathbf{R},\mathbf{P};\mathbf{x},\mathbf{p};\gamma) \tilde{B}_C(\mathbf{R}_t,\mathbf{P}_t;\mathbf{x}_t,\mathbf{p}_t;\gamma)$$
(15)

In addition to the integral form of eq (15) for evaluating the time-dependent physical property, the EOMs for the trajectory *compatible* with the phase space integral expression are another key element.

It is straightforward to follow refs [15, 17] to show that the symplectic structure of mapping Hamiltonian eq (4) leads to the EOMs in the diabatic representation,

$$\begin{aligned}
\dot{\mathbf{x}} + i\dot{\mathbf{p}} &= -i\mathbf{V}(\mathbf{R})(\mathbf{x}+i\mathbf{p}) \\
\dot{\mathbf{R}} &= \mathbf{M}^{-1}\mathbf{P} \\
\dot{\mathbf{P}} &= -\sum_{n,m=1}^{F}(\nabla_{\mathbf{R}}V_{mn}(\mathbf{R}))\left[\frac{1}{2}(x^{(n)}x^{(m)}+p^{(n)}p^{(m)})-\Gamma_{nm}\right] \\
\dot{\boldsymbol{\Gamma}} &= i[\boldsymbol{\Gamma}\mathbf{V}(\mathbf{R})-\mathbf{V}(\mathbf{R})\boldsymbol{\Gamma}]
\end{aligned}$$
(16)

which conserve the value of the mapping Hamiltonian. Applying the covariance relation under the diabatic-to-adiabatic transformation (as explicitly discussed in Section 4.1 of ref [17]), one obtains the corresponding EOMs in the adiabatic representation

$$\dot{\tilde{\mathbf{x}}}(\mathbf{R}) + i\dot{\tilde{\mathbf{p}}}(\mathbf{R}) = -i\mathbf{V}^{(\mathrm{eff})}(\mathbf{R},\mathbf{P})(\tilde{\mathbf{x}}(\mathbf{R})+i\tilde{\mathbf{p}}(\mathbf{R}))$$
(17)

$$\dot{\mathbf{R}} = \mathbf{M}^{-1}\mathbf{P}$$
(18)

$$\dot{\mathbf{P}} = -\sum_{k}\nabla_{\mathbf{R}}E_k(\mathbf{R})\left[\frac{1}{2}\left(\left(\tilde{x}^{(k)}\right)^2+\left(\tilde{p}^{(k)}\right)^2\right)-\tilde{\Gamma}_{kk}\right]$$
$$-\sum_{k\neq l}\left[(E_k(\mathbf{R})-E_l(\mathbf{R}))\mathbf{d}_{lk}(\mathbf{R})\right]\left[\frac{1}{2}(\tilde{x}^{(k)}\tilde{x}^{(l)}+\tilde{p}^{(k)}\tilde{p}^{(l)})-\tilde{\Gamma}_{kl}\right]$$
(19)

$$\dot{\tilde{\boldsymbol{\Gamma}}} = i\left[\tilde{\boldsymbol{\Gamma}}\mathbf{V}^{(\mathrm{eff})}(\mathbf{R},\mathbf{P})-\mathbf{V}^{(\mathrm{eff})}(\mathbf{R},\mathbf{P})\tilde{\boldsymbol{\Gamma}}\right]$$
(20)

In eq (17), $\{\tilde{\mathbf{x}}, \tilde{\mathbf{p}}, \tilde{\boldsymbol{\Gamma}}\} \equiv \{\tilde{\mathbf{x}}(\mathbf{R}), \tilde{\mathbf{p}}(\mathbf{R}), \tilde{\boldsymbol{\Gamma}}(\mathbf{R})\}$ are the mapping phase space variables and commutator matrix for discrete electronic DOFs in the adiabatic representation, and the element of the effective potential matrix, $\mathbf{V}^{(\text{eff})}$, is a function of the nuclear phase space variables,

$$V_{nk}^{(\text{eff})}(\mathbf{R}, \mathbf{P}) = E_n(\mathbf{R})\delta_{nk} - i\dot{\mathbf{R}} \cdot \mathbf{d}_{nk}(\mathbf{R}) = E_n(\mathbf{R})\delta_{nk} - i\mathbf{M}^{-1}\mathbf{P} \cdot \mathbf{d}_{nk}(\mathbf{R}) \quad . \tag{21}$$

Eq (19) is produced because the non-adiabatic coupling vector in the adiabatic representation,

$$\mathbf{d}_{mn}(\mathbf{R}) = \left\langle \phi_m(\mathbf{R}) \middle| \frac{\partial \phi_n(\mathbf{R})}{\partial \mathbf{R}} \right\rangle \quad , \tag{22}$$

satisfies $\mathbf{d}_{mn}(\mathbf{R}) = -\mathbf{d}_{nm}(\mathbf{R})$ when the electronic wavefunction of the basis set is often real for molecular systems. (It is trivial to extend the framework to cases where the electronic wavefunction of the basis set is complex, e.g., the spin adiabatic state of molecular systems that include spin-orbit coupling terms.) The $J$-th component of the left-hand side (LHS) of eq (22) is $d_{mn}^{(J)}(\mathbf{R})$. It is straightforward to see that $-i\mathbf{d}^{(J)}(\mathbf{R})$ is a Hermitian matrix of the electronic state DOFs. Vector $-i\mathbf{d}(\mathbf{R})$ indicates a non-abelian gauge field[50], which is a generalization of the vector potential of the electromagnetic field[51]. The EOMs in the adiabatic representation, eqs (17) -(20), also conserve the mapping Hamiltonian (obtained in the diabatic representation), eq (4), which in the adiabatic representation becomes

$$\begin{aligned} H_C(\tilde{\mathbf{R}}, \tilde{\mathbf{P}}, \tilde{\mathbf{x}}, \tilde{\mathbf{p}}) &\equiv H_C(\mathbf{R}, \mathbf{P}, \tilde{\mathbf{x}}, \tilde{\mathbf{p}}) \\ &= \frac{1}{2}\mathbf{P}(\tilde{\mathbf{P}}, \tilde{\mathbf{x}}, \tilde{\mathbf{p}}, \tilde{\mathbf{R}})^T \mathbf{M}^{-1} \mathbf{P}(\tilde{\mathbf{P}}, \tilde{\mathbf{x}}, \tilde{\mathbf{p}}, \tilde{\mathbf{R}}) \\ &+ \sum_{n=1}^{F} E_n(\tilde{\mathbf{R}})\left(\frac{1}{2}\left(\left(\tilde{x}^{(n)}(\tilde{\mathbf{R}})\right)^2 + \left(\tilde{p}^{(n)}(\tilde{\mathbf{R}})\right)^2\right) - \tilde{\Gamma}_{nn}\right) \end{aligned} \tag{23}$$

of which the canonical variables on mapping phase space for nuclear DOFs in the adiabatic representation, $\{\tilde{\mathbf{R}}, \tilde{\mathbf{P}}\}$, are

$$\tilde{\mathbf{R}} = \mathbf{R}$$
$$\tilde{\mathbf{P}} = \mathbf{P} + i\sum_{m,n}\left[\frac{1}{2}\left(\tilde{x}^{(n)}+i\tilde{p}^{(n)}\right)\left(\tilde{x}^{(m)}-i\tilde{p}^{(m)}\right)-\tilde{\Gamma}_{nm}\right]\mathbf{d}_{mn}(\mathbf{R}) \quad . \tag{24}$$

The mapping diabatic momentum, $\mathbf{P}$, is intrinsically the kinematic momentum of the adiabatic representation. Owing to the strategy suggested by Cotton *et al.* for the Meyer-Miller mapping Hamiltonian model[52], it is more convenient to employ the EOMs for $\{\mathbf{R},\mathbf{P},\tilde{\mathbf{x}},\tilde{\mathbf{p}},\tilde{\boldsymbol{\Gamma}}\}$ to avoid the derivative of non-adiabatic coupling terms. The generalized phase space formulation makes it clear that the kinematic nuclear momentum of the adiabatic representation is intrinsically the mapping nuclear momentum, $\mathbf{P}$, in the diabatic representation[17]. It is then not surprising that the EOMs eqs. (17)-(20) can *not* be generated by the mapping Hamiltonian eq (23), although its value (the mapping energy) is conserved.

As also discussed in Appendix 2 of the supporting information of ref [17], the EOMs in the adiabatic representation, eqs (17)-(20), are valid only when the gauge field tensor,

$$\frac{\partial(-i\mathbf{d}^{(J)})}{\partial \tilde{R}_I} - \frac{\partial(-i\mathbf{d}^{(I)})}{\partial \tilde{R}_J} + i[-i\mathbf{d}^{(I)},-i\mathbf{d}^{(J)}]_{\text{ele}} \quad , \tag{25}$$

is always zero. This holds only when infinite adiabatic states are considered, or equivalently, when the diabatic representation exists[17]. Since a physical process involves only finite (active) energy, when $F$ electronic states are effectively complete to describe the process, the gauge field tensor, eq (25), is nearly zero and may be neglected with caution.

In addition, as we have first explicitly indicated in the fewest-switches surface hopping (FSSH) applications of ref [17], the kinematic momentum of the adiabatic representation (i.e., the mapping nuclear momentum in the diabatic representation) was also inherently used in SH methods. E. g., the "energy" of the hopping trajectory on the PES of the occupied adiabatic state (the $j_{occ}$-th state), $E_{SH} \equiv \frac{1}{2}\mathbf{P}^T\mathbf{M}^{-1}\mathbf{P} + E_{j_{occ}}(\mathbf{R})$, in most SH methods[36, 37, 41, 43] intrinsically involves $\mathbf{P}$, the kinematic momentum of the adiabatic representation, or equivalently the mapping nuclear momentum in the diabatic representation. This is what we have implemented for FSSH in the adiabatic representation in comparison to CPS mapping approaches in refs [16-18].

Ehrenfest-like dynamics governed by eq (16) in the diabatic representation or by eqs (17)-(20) in the adiabatic representation (inherently generated from the symplectic structure of the diabatic mapping Hamiltonian, eq (4)) can be employed for the trajectory with the initial condition eq (27) in the phase space integral expression eq (15) (in either the diabatic or adiabatic representation) for time-dependent properties. It leads to the classical mapping model with commutator variables (CMMcv), a CPS non-adiabatic dynamics approach that works well for typical condensed phase benchmark models where the states remain coupled all the time but less well for Tully's gas phase scattering models where asymptotic regions exist[15]. As discussed in ref [17], the EOMs on mapping phase space (of our CPS approaches) are covariant under the transformation of electronic basis sets, e.g., independent of whether the adiabatic or diabatic representation is employed. It is a merit of Ehrenfest-like dynamics. Cotton *et al.* have already drawn a similar conclusion for the Meyer-Miller mapping Hamiltonian model[52].

In the Perspective we will focus on the EOMs on quantum phase space in the adiabatic representation. This is due to two major reasons:

1) The previous investigation of nuclear quantum statistic mechanical properties by the imaginary time path integral approach for non-adiabatic systems indicates that, compared to exact thermodynamic properties, the adiabatic representation leads to more favorable estimations when the number of path integral beads is one, i.e., in the nuclear classical limit[53]. The adiabatic representation is then also preferable for further theoretical investigation of the EOMs of the real-time trajectory for nuclear DOFs in the generalized coordinate-momentum phase space formulation.

2) It is relatively easy to apply electronic structure methods to obtain the electronically adiabatic basis sets as well as the adiabatic nuclear force on the Born-Oppenheimer PES for most real molecular systems, where the electronically diabatic basis sets are often not rigorously defined.

**Initial Condition for Electronic Phase Space Variables in the Adiabatic Representation.** As described in ref [15] and its Supporting Information, provided that the $j_{occ}$-th diabatic electronic state is occupied at the beginning, the initial condition $\Gamma(0)$ for the commutator matrix is

$$\Gamma_{nm}(0) = \left(\frac{1}{2}\left(\left(x^{(n)}(0)\right)^2 + \left(p^{(n)}(0)\right)^2\right) - \delta_{n,j_{occ}}\right)\delta_{nm} \quad . \tag{26}$$

Electronic phase space variables $\{\tilde{\mathbf{x}}(0), \tilde{\mathbf{p}}(0), \tilde{\Gamma}(0)\}$ in the adiabatic presentation can then be obtained from $\{\mathbf{x}(0), \mathbf{p}(0), \Gamma(0)\}$ by the diabatic-to-adiabatic transformation, before the EOMs in the adiabatic representation are used for propagation of the trajectory.

Alternatively, when $j_{occ}$-th adiabatic electronic state is occupied, the initial condition $\tilde{\Gamma}(0)$ for the commutator matrix is

$$\tilde{\Gamma}_{nm}(0) = \left(\frac{1}{2}\left(\left(\tilde{x}^{(n)}(0)\right)^2 + \left(\tilde{p}^{(n)}(0)\right)^2\right) - \delta_{n,j_{occ}}\right)\delta_{nm} \quad . \tag{27}$$

In either case, the initial values of electronic phase space variables, $\left(x^{(n)}(0), p^{(n)}(0)\right)$ in the diabatic representation or $\left(\tilde{x}^{(n)}(0), \tilde{p}^{(n)}(0)\right)$ in the adiabatic representation, are uniformly sampled from CPS defined by eq (8).

The symplectic integrator of the evolution of electronic phase space variables in the adiabatic representation (eqs (17) and (20)) at each nuclear phase point $(\mathbf{R}(t), \mathbf{P}(t))$ reads

$$\tilde{\mathbf{g}}(t+\Delta t) = \tilde{\mathbf{U}}(\mathbf{R},\mathbf{P};\Delta t)\tilde{\mathbf{g}}(t)$$
$$\tilde{\mathbf{\Gamma}}(t+\Delta t) = \tilde{\mathbf{U}}(\mathbf{R},\mathbf{P};\Delta t)\tilde{\mathbf{\Gamma}}(t)\tilde{\mathbf{U}}^\dagger(\mathbf{R},\mathbf{P};\Delta t)$$
(28)

Here, $\tilde{\mathbf{g}}(t) = \tilde{\mathbf{x}}(\mathbf{R}(t);t) + i\tilde{\mathbf{p}}(\mathbf{R}(t);t)$ and $\tilde{\mathbf{U}}(\mathbf{R}(t),\mathbf{P}(t);\Delta t) = \exp\left[-i\Delta t \mathbf{V}^{(\text{eff})}(\mathbf{R}(t),\mathbf{P}(t))\right]$ are used. We then examine the EOMs for nuclear phase space variables, especially the nuclear force.

**Adiabatic and Non-Adiabatic Nuclear Force Terms.** For simplicity we define

$$\tilde{\rho}_{kl}(\tilde{\mathbf{x}},\tilde{\mathbf{p}},\tilde{\mathbf{\Gamma}}) = \frac{1}{2}\left(\tilde{x}^{(k)}\tilde{x}^{(l)} + \tilde{p}^{(k)}\tilde{p}^{(l)}\right) - \tilde{\Gamma}_{kl} \quad . \tag{29}$$

The EOM of eq (19) becomes

$$\dot{\mathbf{P}} = -\sum_k \nabla_{\mathbf{R}} E_k(\mathbf{R})\tilde{\rho}_{kk} - \sum_{k\neq l}\left[(E_k(\mathbf{R}) - E_l(\mathbf{R}))\mathbf{d}_{lk}(\mathbf{R})\right]\tilde{\rho}_{kl} \quad . \tag{30}$$

It is important to understand the two terms of the Ehrenfest-like force (for updating $\mathbf{P}$) in the right-hand side (RHS) of eq (30) in the adiabatic representation. The first term is the weighted adiabatic force, i.e., the sum of weighted gradients of all adiabatic PESes. The second term of the RHS of eq (30) is from the contribution related to non-adiabatic couplings between different (adiabatic) electronic states. Such a term is the *intrinsic* non-adiabatic nuclear force that accounts for non-adiabatic transition processes, which should *never* be ignored in the EOMs for nuclear variables in the coupling region. The non-adiabatic nuclear force, the second term of the RHS of eq (30),

$$\mathbf{f}_{non-adia} = -\sum_{k\neq l}\left[(E_k(\mathbf{R}) - E_l(\mathbf{R}))\mathbf{d}_{lk}(\mathbf{R})\right]\tilde{\rho}_{kl} \quad , \tag{31}$$

naturally vanishes in the asymptotic region where the non-adiabatic coupling disappears.

On the other hand, the weighted adiabatic force, the first term of the RHS of eq (30), accounts for the unphysical picture of nuclear dynamics even in the asymptotic region. Instead of the average adiabatic force, the adiabatic force term should be from the gradient of the adiabatic PES of a single electronic state. A few non-adiabatic field approaches can be proposed to address the problem of the adiabatic force term, while maintaining eq (31), the non-adiabatic nuclear force term, in eq (19), i.e., in the EOM for updating **P**, the kinematic nuclear momentum of the adiabatic representation. That is, gauge field tensor eq (25) in the non-adiabatic field is effectively zero, and the non-adiabatic field involves the original non-adiabatic force term (eq (31)) in addition to a single-state adiabatic nuclear force.

**Non-Adiabatic Field.** There exist two simple but reasonable approaches. The first approach is to stochastically choose an ingredient in the sum of the first term in the RHS of eq (30) based on its "weight". I. e., it is more physical to keep a single-state adiabatic force (the gradient of the adiabatic PES), rather than the weighted/average adiabatic force $-\sum_k \nabla_\mathbf{R} E_k(\mathbf{R})\tilde{\rho}_{kk}$, for the first term of the RHS of eq (19) of the EOMs. The second approach is to select the dominant ingredient in the sum of the first term in the RHS of eq (30), i.e., the single-state adiabatic nuclear force that takes the largest weight. This ignores the contribution of adiabatic force ingredients with smaller weights. Meanwhile, in either approach, the non-adiabatic nuclear force term eq (31), the second term of the RHS of eq (19), stays the same. As shown in Section S4 of the Supporting Information of the Perspective, the two approaches lead to almost the same results for several benchmark condensed phase model systems.

Because the single-state adiabatic nuclear force component is switched more frequently in the former approach than the latter one, we choose the latter approach for simplicity. Eq (19) then becomes

$$\dot{\mathbf{P}} = -\sum_k \nabla_\mathbf{R} E_k(\mathbf{R})\left(\prod_{j\neq k} h\left(\tilde{\rho}_{kk} - \tilde{\rho}_{jj}\right)\right) - \sum_{k\neq l}\left[\left(E_k(\mathbf{R}) - E_l(\mathbf{R})\right)\mathbf{d}_{lk}(\mathbf{R})\right]\tilde{\rho}_{kl} \quad . \tag{32}$$

Here, $h(y) := \begin{cases} 1 & (y \geq 0) \\ 0 & (y < 0) \end{cases}$ is the Heaviside function. The new EOMs now include eqs (17)-(18), (20) and eq (32). In addition, the initial value of mapping Hamiltonian eq (23) or mapping energy should be also conserved during the evolution. Because we choose the dominant single-state adiabatic nuclear force in the EOMs for each time step, it suggests that in mapping Hamiltonian eq (23), we should keep the corresponding occupied single-state adiabatic potential. That is, when the $j_{occ}$-th adiabatic state PES contributes the adiabatic nuclear force, mapping Hamiltonian becomes

$$H_C(\mathbf{R}_0, \mathbf{P}_0, \tilde{\mathbf{x}}_0, \tilde{\mathbf{p}}_0) = H_{NAF}(\mathbf{R}, \mathbf{P}, \tilde{\mathbf{x}}(\mathbf{R}), \tilde{\mathbf{p}}(\mathbf{R})) \equiv \frac{1}{2}\mathbf{P}^T \mathbf{M}^{-1} \mathbf{P} + E_{j_{occ}}(\mathbf{R}) \quad . \tag{33}$$

Similar to Ehrenfest-like dynamics on mapping phase space, although the mapping Hamiltonian, eq (33), should be conserved during the evaluation, it should *not* be used to yield the EOMs for mapping phase space variables in the adiabatic representation. The kinematic nuclear momentum of the adiabatic representation, $\mathbf{P}$, is rescaled along the momentum vector (after the update of $\mathbf{P}$ in eq (32)) every time step to satisfy the mapping energy conservation.

When the adiabatic electronic state that has the largest "weight" is switched, the kinematic nuclear momentum, $\mathbf{P}$, should also be rescaled along its original direction such that the mapping energy is conserved. If the dominate weight is switched from the $j_{old}$-th state component to the $j_{new}$-th state component, but it is impossible to rescale the momentum to conserve the mapping energy, i.e., $H_C(\mathbf{R}_0, \mathbf{P}_0, \tilde{\mathbf{x}}_0, \tilde{\mathbf{p}}_0) < E_{j_{new}}(\mathbf{R})$, then the switching of the adiabatic nuclear force component in eq (32) is unphysical. In such a case, the switching is frustrated/prohibited and the single-state adiabatic nuclear force (the first term of the RHS of eq (32)) for the evolution of nuclear phase space variables $\{\mathbf{R}, \mathbf{P}\}$ is still from $-\nabla_{\mathbf{R}} E_{j_{old}}(\mathbf{R})$, the gradient of the previously occupied adiabatic PES even though its "weight" is now not the largest. The integrator of the EOMs of NAF for each time step is described in Section S3 of the Supporting Information.

The EOMs of NAF (e.g., eqs (17)-(18), (20) and eq (32)) on quantum phase space are *not* covariant under the transformation of electronic basis sets. NAF should be used in the adiabatic representation. Because the EOMs of NAF are obtained in the generalized quantum coordinate-momentum phase space formulation, we also use eq (15) for the phase space integral expression of time-dependent properties, and eq (27) for the initial condition for electronic mapping variables on CPS. They are the three key elements of NAF, the trajectory-based quantum dynamics method that we propose in the Perspective.

NAF is exact in the frozen-nuclei limit. When non-adiabatic couplings are constant and nonzero, and only dynamics of the electronic DOFs are involved, NAF recovers the time-dependent Schrodinger equation. NAF also satisfies the Born-Oppenheimer limit (for the classical nuclear force), where non-adiabatic couplings vanish and a single adiabatic state is occupied. Finally, NAF is exact in the Landau-Zener limit[54-57], where the electronic DOFs evolve under the time-dependent Hamiltonian defined by $\mathbf{R}(t)$, the coordinate of the nuclear trajectory. We further show the numerical performance of NAF for a few benchmark model systems. Parameter $\gamma$ of eq (15), the phase space integral expression of time-dependent properties, is chosen in region $\left[\left(\sqrt{F+1}-1\right)/F, 1/2\right]$ as suggested in ref [15].

For comparison, we also test the EOMs of NAF with the initial condition and expression for evaluation of time-dependent (electronic) properties in the conventional mean field approach (NAF-Ehrenfest). This is to demonstrate the importance of the initial condition and that of integral expression. The conventional Ehrenfest dynamics and FSSH approaches are also tested for comparison. We employ a variant of Tully's FSSH method[43] as described in ref [37]. This FSSH algorithm is reviewed in Section S7 of the Supporting Information of this Perspective. For fair comparison among trajectory-based non-adiabatic dynamics methods, the initial condition for nuclear DOFs is sampled from the Wigner distribution on nuclear coordinate-momentum phase space. We consider a few typical benchmark condensed phase and gas phase non-adiabatic models. Numerically exact results of most of these models are available only in the diabatic representation. We perform simulations for all trajectory-based dynamics methods in the adiabatic representation and the results are then transformed to those in the diabatic representation for the purpose of comparison.

**Spin-Boson Models.** The spin-boson model depicts a two-state system bilinearly coupled with harmonic bath DOFs of a dissipative environment in the condensed phase[58]. It also serves as a prototype model for electron/energy transfer/transport in chemical and biological reactions. Numerically exact results of the spin-boson model can be achieved by quasi-adiabatic propagator path integral (QuAPI)[59-61] and more efficient small matrix PI (SMatPI)[62, 63], hierarchy equations of motion (HEOM)[64-68], (multi-layer) multi-configuration time-dependent Hartree [(ML-)MCTDH][69-71], and time-dependent density matrix renormalization group (TD-DMRG)[72]. The spin-boson model then offers benchmark tests for investigating trajectory-based non-adiabatic dynamics methods. We consider the initial condition that the bath modes are at the thermal equilibrium (i.e., the quantum Boltzmann distribution for the pure bath Hamiltonian operator) and the system is in the excited state (the higher energy level). The discretization scheme of refs [73, 74] is used for obtaining discrete bath modes for the (Ohmic or Debye) spectral density. More numerical details are presented in Section S1-A of the Supporting Information of this Perspective.

Except Ehrenfest dynamics and NAF-Ehrenfest which use the conventional mean field framework, most methods (including FSSH and NAF) perform well for spin-boson models in the high temperature region, as shown in Figure S5 of the Supporting Information. Here we focus on much more challenging cases in the low temperature regime. Figure 1 shows the comparison of exact data to numerical results produced by Ehrenfest dynamics, NAF-Ehrenfest, FSSH, and NAF. Neither Ehrenfest dynamics nor FSSH is capable of capturing the correct asymptotic behaviors for population terms (i.e., diagonal elements of the electronic reduced matrix) as well as for "coherence" terms (i.e., off-diagonal elements of the electronic reduced matrix). NAF-Ehrenfest only slightly improves over Ehrenfest dynamics. In contrast, the NAF results show overall good agreement with numerically exact data in the challenging model tests. Although NAF-Ehrenfest and NAF share the same EOMs for real-time dynamics, the two methods employ different formulations of the time-dependent property, as well as different phase space initial conditions. The significant difference between the numerical performance of NAF-Ehrenfest and that of NAF suggests that the formulation of the time-dependent property and that of the phase space initial condition of the trajectory are also important for real-time dynamics methods. NAF with either $\gamma = \left(\sqrt{F+1} - 1\right)/F$ or $\gamma = 1/2$ in the integral formulation of time-dependent properties eq (15) leads to nearly the same results. It suggests that NAF is robust when parameter $\gamma$ lies in the region, $\left[\left(\sqrt{F+1}-1\right)/F, 1/2\right]$.

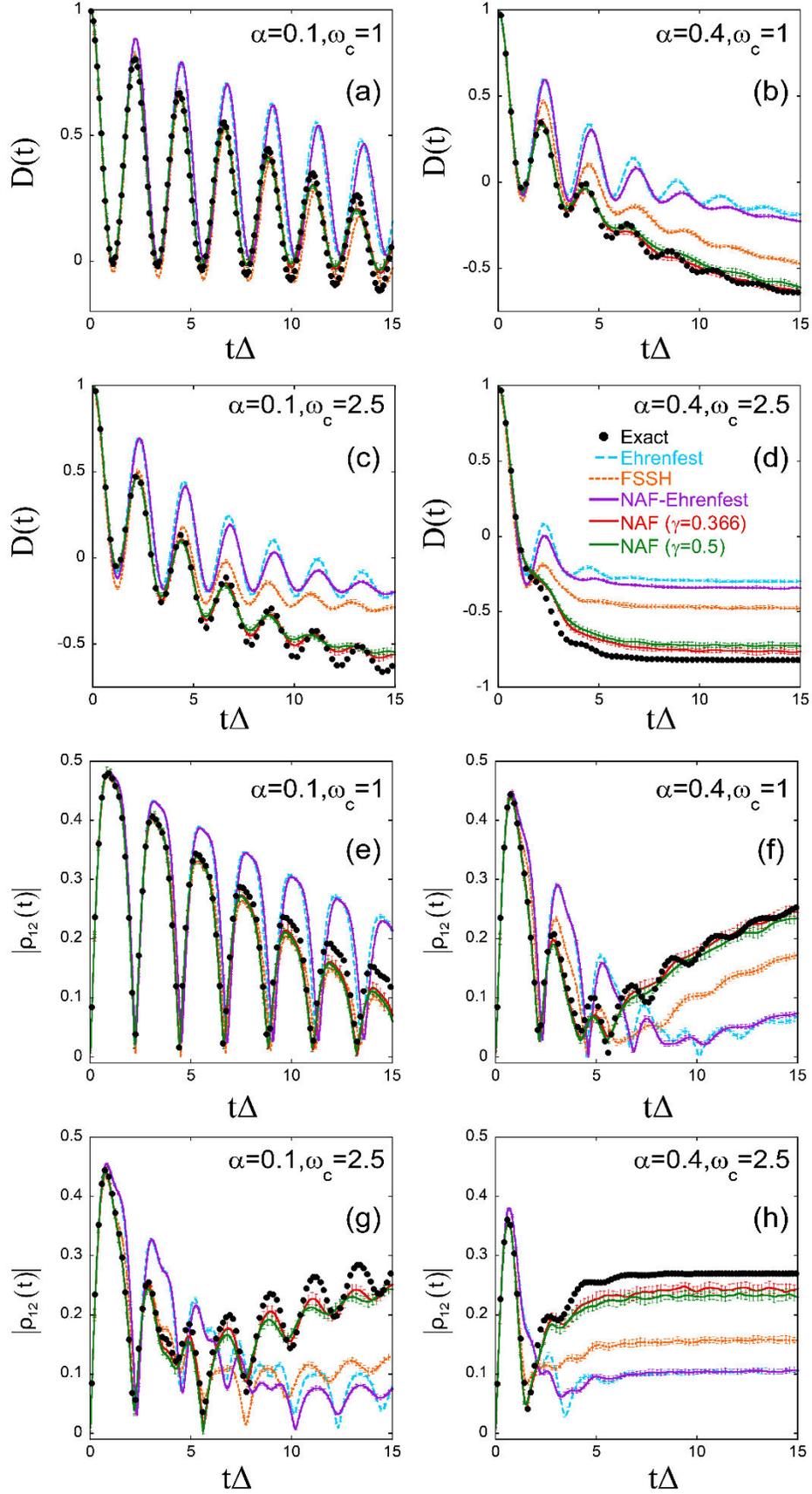

**Figure 1.** Panels (a)-(d): The population difference between state 1 and state 2, $D(t)$ as a function of time for spin-boson models with Ohmic spectral density at $\beta = 5$. Panels (a)-(d) represent the results of the models with parameters $\{\alpha = 0.1, \omega_c = 1\}$, $\{\alpha = 0.4, \omega_c = 1\}$, $\{\alpha = 0.1, \omega_c = 2.5\}$ and $\{\alpha = 0.4, \omega_c = 2.5\}$, respectively. Black points: Exact results produced by eHEOM. Cyan long-dashed lines: Ehrenfest dynamics. Orange short-dashed lines: FSSH. Purple, red and green solid lines: NAF-Ehrenfest, NAF ($\gamma = 0.366$) and NAF ($\gamma = 0.5$), respectively. Panels (e)-(h) are the same as Panels (a)-(d) but for the module of the off-diagonal term $|\rho_{12}(t)|$. Three hundred discrete bath modes are used to obtain converged results. Please see Section S1-A of the Supporting Information.

**Seven-Site Model for the FMO Monomer.** The Fenna–Matthews–Olson (FMO) complex of green sulfur bacteria is a prototype system to study photosynthetic organisms[23, 75-81]. We consider the site-exciton model of the FMO monomer of ref [78], which describes a seven-site system coupled with a harmonic bath defined by the Debye spectral density. It offers a benchmark model for testing various non-adiabatic dynamics methods. Initially, the bath modes are at the thermal equilibrium and the first site (pigment) is occupied. Numerical details are listed in Section S1-B of the Supporting Information.

We study the evolution of both population and "coherence" terms for the electronic DOFs and plot the results in Figure 2. It is evident that Ehrenfest dynamics performs poorly even since relatively short time and is not capable of even qualitatively capturing the steady-state behavior in the long time limit. NAF-Ehrenfest changes little in comparison to Ehrenfest dynamics. Interestingly, the numerical performance of the FSSH results (produced by the algorithm in ref [37]) is also similar to that of the data produced by Ehrenfest dynamics for this test case. In comparison, NAF is competent in quantitatively describing both population and "coherence" terms, even for the asymptotic behavior in the long time limit. The NAF results are relatively insensitive to parameter $\gamma$ in region $\left[\left(\sqrt{F+1}-1\right)/F, 1/2\right]$.

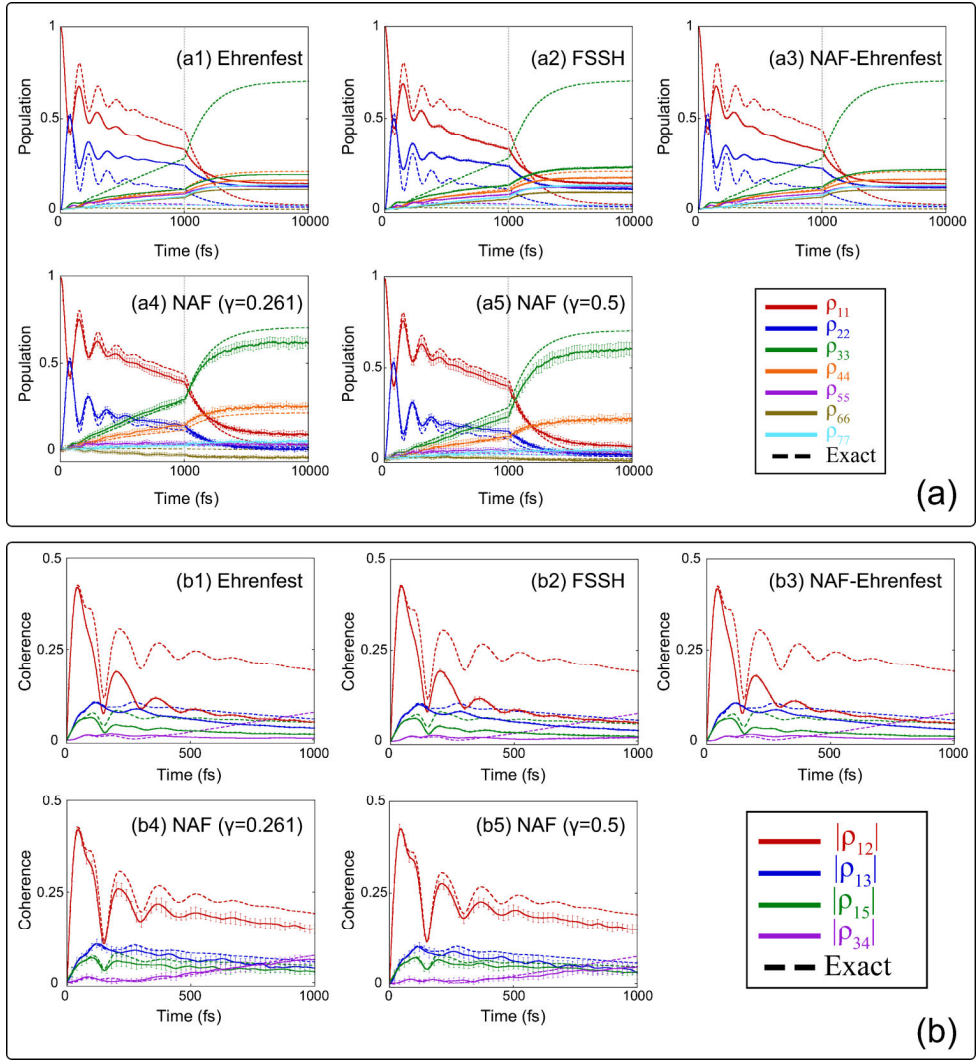

**Figure 2.** Panel (a): The population dynamics of the FMO model at $T=77$ K. The initially occupied site is the first site. Panels (a1)-(a5) denote the results of Ehrenfest dynamics, FSSH, NAF-Ehrenfest, NAF ($\gamma = 0.261$) and NAF ($\gamma = 0.5$), respectively, where the red, blue, green, orange, purple, brown and cyan solid lines represent the population of state 1-7, respectively, and the exact results produced by HEOM are also presented by dashed lines with corresponding colors. Panel (b) is the same as Panel (a) but for the coherence terms. In Panels (b1)-(b5), the red, blue, green and purple solid lines represent the module of the off-diagonal terms $\rho_{12}$,

$\rho_{l3}$, $\rho_{l5}$ and $\rho_{34}$, respectively, and the exact results produced by HEOM are also presented by dashed lines with corresponding colors. Fifty discrete bath modes for each site are employed to obtain converged results. Please see Section S1-B of the Supporting Information.

**Atom-in-Cavity Models of Cavity Quantum Electrodynamic.** It has been reported that a few important phenomena appear in cavity quantum electrodynamics (cQED) especially when a matter is strongly coupled to the vacuum field of a confined optical cavity[28, 82-84]. We test two benchmark cQED models that depict a multi-level atom imprisoned in a one-dimensional lossless cavity[85-88]. The atomic system is coupled to multi-cavity-modes. The first model involves three atomic levels, and the second one is a reduced two-level model where only the two lowest levels are considered. At the beginning, the highest atomic level is activated and all cavity modes are in the vacuum state. The spontaneous emission first happens and releases a photon, which propagates in the cavity and is reflected to meet the atom. The re-absorption and re-emission processes then follow. Numerical details of the two models are described in Section S1-C of the Supporting Information of the Perspective.

As shown in Figure 3, while both Ehrenfest dynamics and FSSH results demonstrate significant deviation even since quite short time, NAF-Ehrenfest improves little over Ehrenfest dynamics. In contrast, NAF yields much more accurate data for population dynamics of all energy levels and is capable of semi-quantitatively describing both the short time behavior and the re-coherence around $t$=1800 a.u. The performance of NAF is robust when parameter $\gamma \in \left[ \left( \sqrt{F+1} - 1 \right) / F, 1/2 \right]$. It is encouraging that NAF semi-quantitatively describes the re-absorption and re-emission processes, while such processes are challenging for both Ehrenfest dynamics and SH methods.

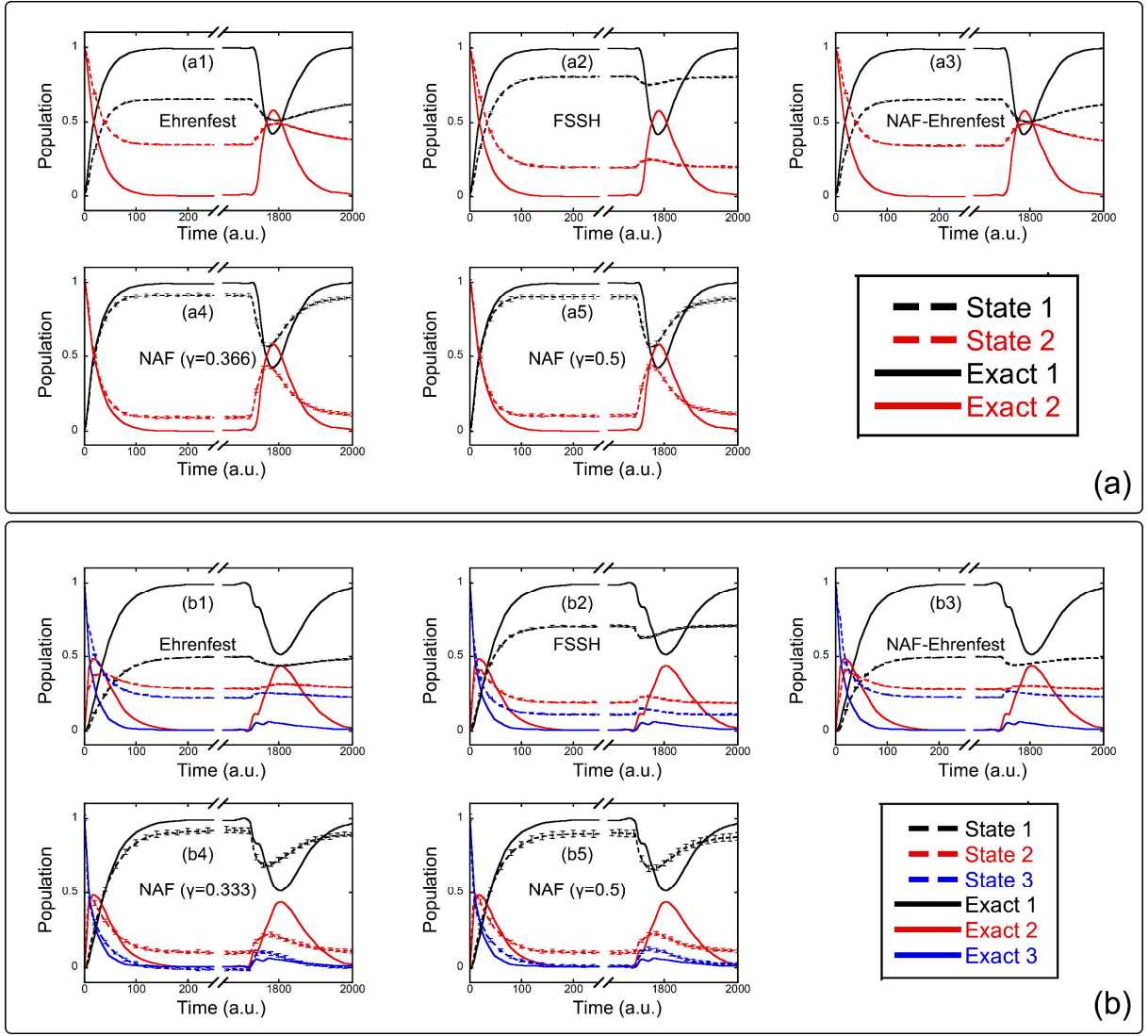

**Figure 3.** Panel (a): Population dynamics of 2-level atom-in-cavity model. Panels (a1)-(a5) denote the results of Ehrenfest dynamics, FSSH, NAF-Ehrenfest, NAF ($\gamma = 0.366$) and NAF ($\gamma = 0.5$), respectively, where the black and red dashed lines demonstrate the population of state 1 and 2, respectively, and the exact results (from refs [85, 86]) are demonstrated as solid lines with corresponding colors. Panel (b): Population dynamics of the 3-level atom-in-cavity model. Panels (b1)-(b5) denote the results of Ehrenfest dynamics, FSSH, NAF-Ehrenfest, NAF ($\gamma = 0.333$) and NAF ($\gamma = 0.5$), respectively, where the black, red and blue dashed lines demonstrate the population of state 1-3, respectively, and the exact results (from refs [85, 86]) are demonstrated as solid lines with corresponding colors. Four hundred standing-wave modes for the optical field are used to obtain converged data. Please see Section S1-C of the Supporting Information for more details.

**Singlet-Fission Model.** Singlet-fission (SF) depicts the conversion of a singlet exciton into two triplet excitons in molecular materials such as solar cells/organic semiconductors[24, 27, 89-91]. We use the three-state SF model of ref [90]. The system, which includes the singlet (S1) state, charge-transfer (CT) state, and double triplets (TT) state, is bilinearly coupled to a harmonic bath described by the Debye spectral density. The system-bath coupling of this SF model is relatively strong. It then offers a benchmark model to test non-adiabatic dynamics methods. The initial condition is that the bath modes are at the thermal equilibrium and the S1 state is excited. More details are available in Section S1-D of the Supporting Information.

Figure 4 shows that either of Ehrenfest dynamics and FSSH performs poorly for relatively long time. While NAF-Ehrenfest is similar to Ehrenfest dynamics for this case, NAF significantly outperforms either Ehrenfest dynamics or FSSH for the asymptotic behavior and leads to much more reasonable long-time results. The numerical results of NAF are nearly the same when parameter $\gamma \in \left[ \left( \sqrt{F+1} - 1 \right) / F, 1/2 \right]$.

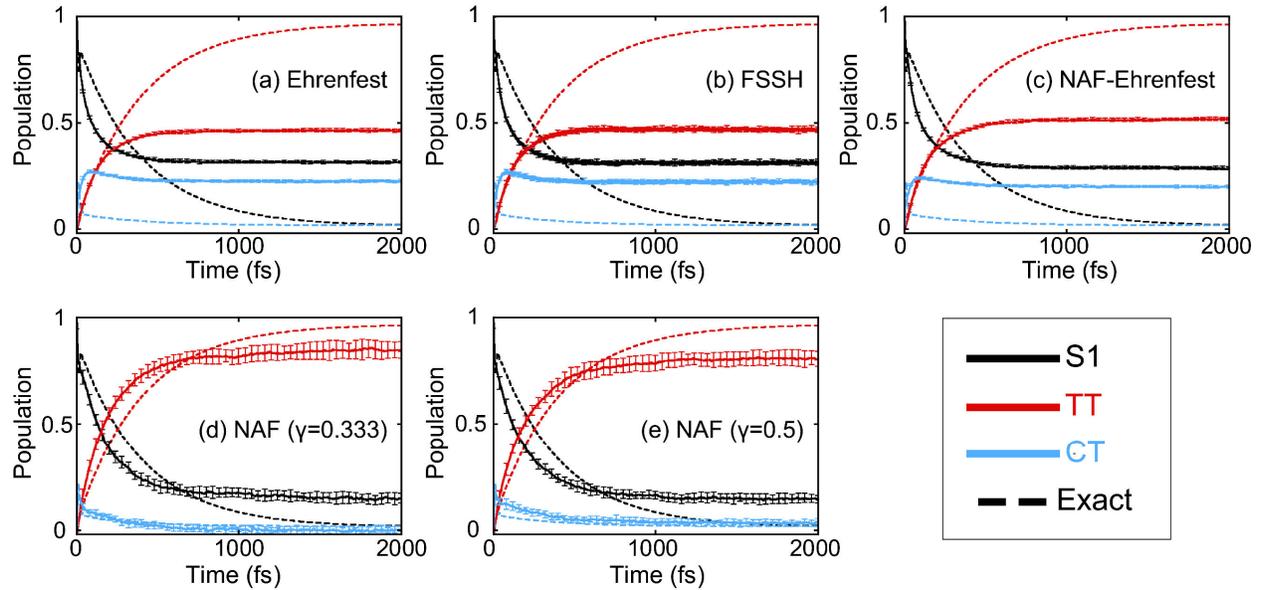

**Figure 4.** The population dynamics of the SF model at $T$=300 K. Panels (a)-(e) denote the results of Ehrenfest dynamics, FSSH, NAF-Ehrenfest, NAF ($\gamma = 0.333$) and NAF ($\gamma = 0.5$), respectively. In each panel, the black,

red and cyan solid lines represent the population of the S1, TT and CT states, respectively, and dashed lines with corresponding colors present the exact results produced by HEOM. More numerical details are presented in Section S1-D of the Supporting Information.

**Gas Phase Models with One Nuclear Degree of Freedom.** We further test NAF for gas phase models with asymptotic regions. We consider the coupled three-electronic-state photo-dissociation models of Miller and coworkers [92]. Each PES is described by a Morse oscillator and the coupling terms are depicted by Gaussian functions. The initial condition is a nuclear Gaussian wavepacket with electronic state 1 occupied. Numerically exact results for the models can obtained by the discrete variable representation (DVR) approach[93]. Please see Section S1-E of the Supporting Information for more numerical details.

It is shown in refs [92, 94] that, when the Ehrenfest trajectory is used in the linearized semi-classical initial value representation (LSC-IVR) of the Stock-Thoss interpretation[49] of the Meyer-Miller mapping Hamiltonian model[48], it fails to even qualitatively yield the nuclear momentum distribution or even electronic population dynamics; the more advanced forward-backward IVR or full SC-IVR[95-97] is necessary to describe the correct electronic and nuclear dynamics in the asymptotic region. It suggests that the interference between different Ehrenfest trajectories (e.g., in the SC-IVR framework) can lead to the correct nuclear dynamics behavior[94, 98]. While the LSC-IVR of the Stock-Thoss interpretation[49] of the Meyer-Miller mapping Hamiltonian model[48] is in principle equivalent to a trajectory-based approach on infinite (Wigner) coordinate-momentum phase space for both continuous nuclear DOFs and discrete electronic DOFs[17], NAF employs the generalized coordinate-momentum phase space. It is interesting to test the performance of NAF, which employs only independent trajectories, especially in the asymptotic region.

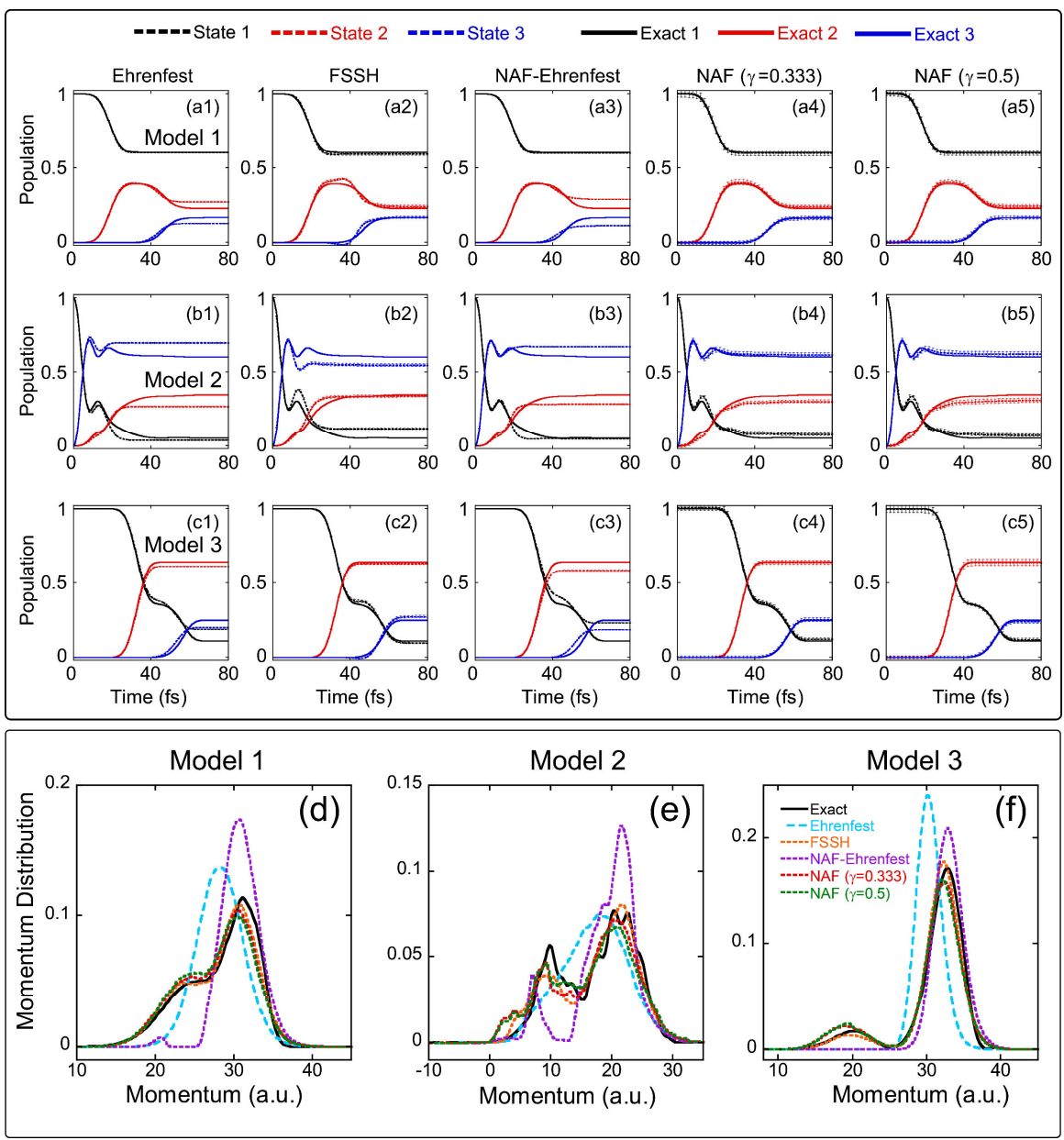

**Figure 5.** Results of the 3-state photo-dissociation models. The first three rows [Panels (a)-(c)] demonstrate the population dynamics of models 1-3, respectively, where the first to fifth columns denote the results of Ehrenfest dynamics, FSSH, NAF-Ehrenfest, NAF ($\gamma = 0.333$) and NAF ($\gamma = 0.5$), respectively. In Panels (a)-(c), the black, red and blue dashed lines demonstrate the population of states 1-3, respectively, and exact results are demonstrated as solid lines with corresponding colors. Panels (d)-(f) demonstrate the nuclear momentum distribution at 200 fs for models 1-3, respectively. Black solid lines: Exact results. Cyan long-dashed lines: Ehrenfest dynamics. Orange short-dashed lines: FSSH. Purple short-dashed lines: NAF-Ehrenfest. Red short-

dashed lines: NAF ( $\gamma = 0.333$ ). Green short-dashed lines: NAF ( $\gamma = 0.5$ ). More numerical details are presented in Section S1-E of the Supporting Information.

Figure 5 demonstrates the results for the three-state photo-dissociation models[92]. It is indicated that, while neither Ehrenfest dynamics nor NAF-Ehrenfest is capable of qualitatively producing the nuclear momentum distribution in the asymptotic region, both FSSH and NAF yield the correct correlation between electronic and nuclear dynamics. Figure 5 also shows that the numerical performance of NAF is insensitive to parameter $\gamma$ when $\gamma \in \left[ \left( \sqrt{F+1} - 1 \right) / F, 1/2 \right]$. Similarly, Figures S2 and S4 of the Supporting Information of the Perspective show the data for an asymmetric single avoided crossing (SAC) model. NAF is competent in producing the correct nuclear momentum distribution in the asymptotic region and is capable of performing well for describing electronic dynamics for transmission as well as reflection. Since it is often more expensive to include interference effects among trajectories, it is encouraging to see that NAF performs well even in the asymptotic region without involving the phase cancellation among trajectories.

**Linear Vibronic Coupling Models for Molecules Involving the Conical Intersection.** The linear vibronic coupling model (LVCM) is a simple but effective model that mimics molecular systems where the conical intersection (CI) region plays a critical role in such as light-driven phenomena. We test the two-electronic-state LVCM with three nuclear modes and that with 24 nuclear modes, which describe the S1/S2 conical intersection of the pyrazine molecule[99, 100]. The initial state is the cross-product of the vibronic ground state and the excited electronically diabatic state (S2). In addition, we study a typical three-electronic-state 2-nuclear-mode LVCM for the Cr(CO)$_5$ molecule[101]. The initial condition is the cross-product of the first excited electronically diabatic state and a Gaussian nuclear wave-packet. The center of the Gaussian wave-packet is located at the minimal point of the ground state of the $Cr(CO)_6$ molecule that will dissociate a carbonyl group and the width of each mode is from the corresponding vibrational frequencies [101]. The details are described in ref [101] and Section S1-F of the Supporting Information of this Perspective. While the diabatic representation is used for the initial condition and evaluation of dynamical properties (as done in MCTDH), the adiabatic representation is employed for real-time dynamics for the fair comparison among different non-adiabatic dynamics methods. Numerical results are plotted in Figures 6-7.

For population dynamics in all these LVCM cases, FSSH and NAF perform much better than Ehrenfest dynamics and NAF-Ehrenfest. NAF performs slightly better than FSSH for the 2-state 3-mode case of pyrazine. When the 2-state 24-mode case of pyrazine is studied, in comparison to FSSH, NAF yields slightly better nuclear dynamics results, but produces a little less accurate electronic population dynamics results. The test results for the 3-state 2-mode LVCM demonstrate that NAF is overall slightly better for population dynamics in three states. Figures 6-7 demonstrate that the overall performance of NAF is comparable to that of FSSH for the LVCM cases for realistic molecular systems. The NAF results are relatively independent of parameter $\gamma$ in region $\left[ \left( \sqrt{F+1} - 1 \right) / F, 1/2 \right]$.

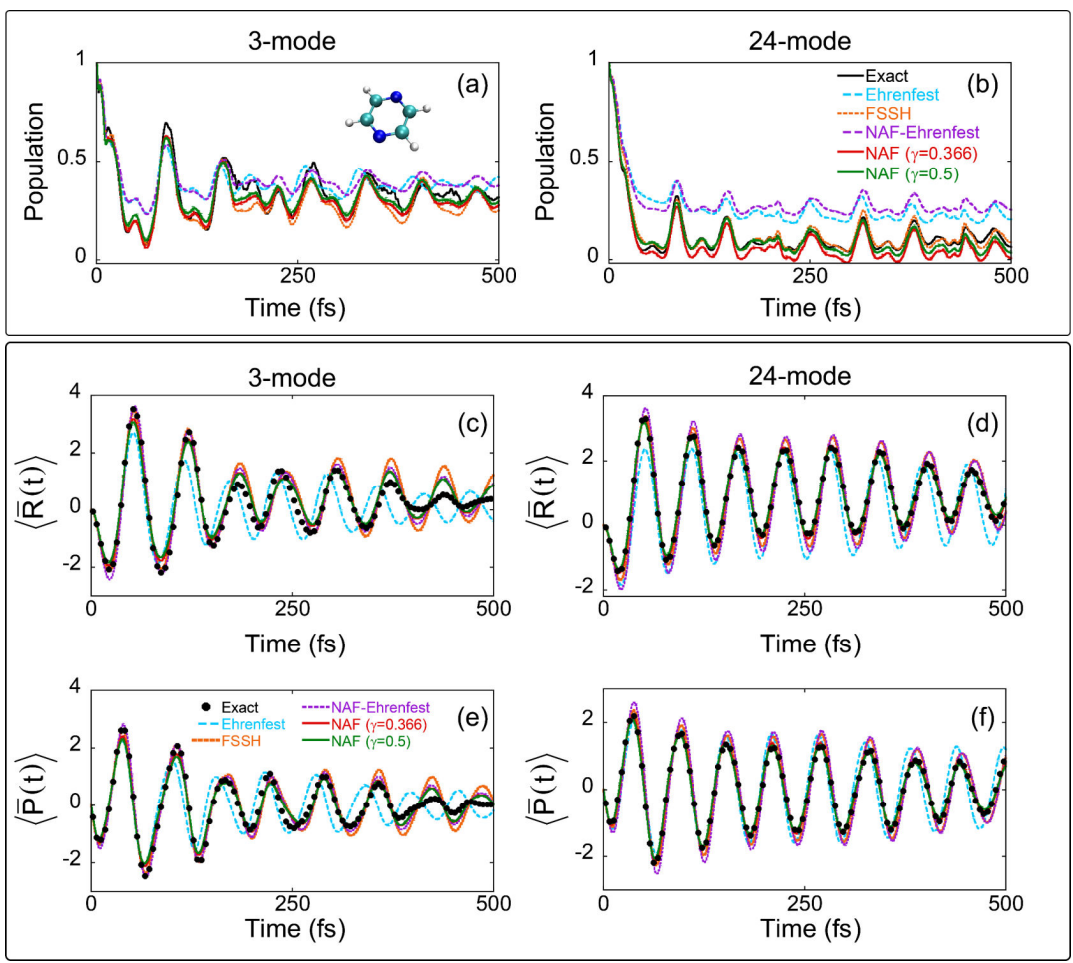

**Figure 6.** Panels (a)-(b) denote population dynamics of the second state of the 2-state LVCM with 3 modes for pyrazine[99] and that with 24 modes for the same molecule[100], respectively. Black solid lines: Exact results produced by MCTDH (ref [102]). Cyan long-dashed lines: Ehrenfest dynamics. Orange short-dashed lines: FSSH. Purple long-dashed lines: NAF-Ehrenfest. Red and green solid lines: NAF ($\gamma = 0.366$) and NAF ($\gamma = 0.5$), respectively. Panels (c)-(d) demonstrate the average coordinate $\langle \bar{R}(t) \rangle$ of the nuclear normal mode $\nu_{6a}$ of 3-mode and 24-mode LVCMs (for pyrazine), respectively. Black points: Exact results produced by MCTDH[102]. Cyan long-dashed lines: Ehrenfest. Orange short-dashed lines: FSSH. Purple short-dashed lines: NAF-Ehrenfest. Red and green solid lines: NAF ($\gamma = 0.366$) and NAF ($\gamma = 0.5$), respectively. Panels (e)-(f) are the same as Panels (c)-(d) but for the average momentum $\langle \bar{P}(t) \rangle$ of the nuclear normal mode $\nu_{6a}$. More numerical details are presented in Section S1-F of the Supporting Information.

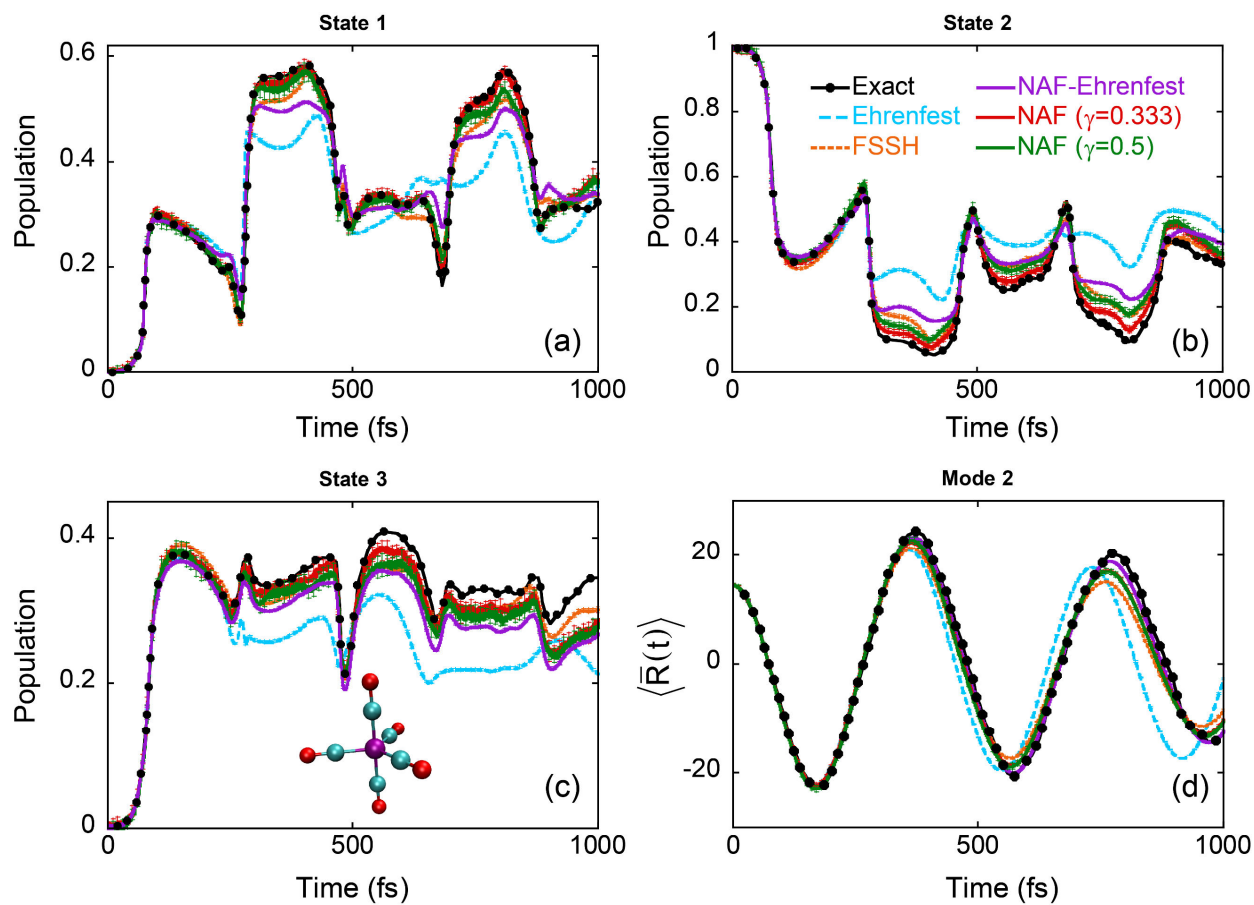

**Figure 7.** Panels (a)-(c) denote population dynamics of states 1-3 of the 2-nuclear-mode LVCM for the Cr(CO)$_5$ molecule[101], respectively. Panel (d) demonstrates the average coordinate $\langle \bar{R}(t) \rangle$ of the second nuclear normal mode. Black solid lines with black points: Exact results produced by MCTDH (obtained from ref [101]). Cyan long-dashed lines: Ehrenfest dynamics. Orange short-dashed lines: FSSH. Purple, red and green solid lines: NAF-Ehrenfest, NAF ($\gamma = 0.333$) and NAF ($\gamma = 0.5$), respectively. More numerical details are presented in Section S1-F of the Supporting Information.

In conclusion, the various benchmark model tests of typical gas phase and condensed phase systems (as shown in Figures 1-7) indicate that NAF with CPS is capable of capturing the correct pictures in the coupling region as well as the asymptotic region and shows overall better performance than FSSH as well as Ehrenfest dynamics. Other SH algorithms may lead to different results, but if one honestly treats nuclear DOFs on Wigner phase space to take care of nuclear quantum effects at the beginning, the condensed phase benchmark models (spin-boson, FMO, cQED, and singlet-fission models) are often challenging for SH methods. In comparison to NAF, the poor performance of NAF-Ehrenfest demonstrates that the integral expression for evaluation of the time-dependent physical property as well as the initial condition of the trajectory are two important factors for trajectory-based dynamics methods. NAF employs CPS for mapping the discrete electronic-state DOFs, which is rigorous for the initial condition on phase space and more consistent for evaluating dynamic properties. This is missing in the conventional mean field framework as well as in most SH methods. In addition, Section S7 of the Supporting Information demonstrates that fewest switches NAF (FS-NAF), which incorporates the non-adiabatic nuclear force term in the EOMs for nuclear variables of the FSSH algorithm, can systematically improve the numerical performance for condensed phase benchmark models. FS-NAF yields less accurate results than NAF though. It is expected that the NAF EOMs for nuclear variables are also capable of improving over various other SH methods where the non-adiabatic nuclear force term is never included.

In the Perspective, starting from the generalized exact coordinate-momentum phase space formulation of quantum mechanics, we propose NAF, a conceptually new trajectory-based approach, which does not employ the Ehrenfest trajectory or the Born-Oppenheimer trajectory as conventional trajectory-based approaches do. The EOMs of nuclear phase space variables of the NAF trajectory involve the non-adiabatic nuclear force term as well as the adiabatic nuclear force term of one single electronically adiabatic state. NAF is exact in the Born-Oppenheimer limit, frozen nuclei limit, and Landau-Zener limit, and consistently yields the correlation between electronic and nuclear dynamics in the asymptotic region as well as the state-coupling region. The evaluation of the nuclear force of the NAF trajectory requests no additional effort beyond that in Ehrenfest dynamics as well as FSSH even for *ab initio*-based simulations[103-106]. NAF involves only independent trajectories on quantum phase space and is practical for real complex/large molecular systems. Despite of its various pros, we expect that NAF of the current version is difficult, if not possible, to describe deep tunneling effects and quantum recurrence/coherence effects although such effects are often quenched in large molecular systems. E. g., it is expected that NAF is not adequate to accurately describe such as the one-dimensional scattering problem where deep tunneling and quantum resonances dominate[107]. The time-dependent multi-configuration approach with NAF trajectories will be a potential candidate to tackle these much more challenging quantum mechanical problems (with more computational effort). In addition to eq (10) or eq (15), other exact CPS expressions can also be used for NAF. (E.g., NAF with the CPS correlation function of ref [108] is presented in Section S6 of the Supporting Information.) More investigations in the future will shed light on the performance of NAF for more composite-systems in chemistry, biology, materials, quantum information and computation, and so forth.

# ■ ASSOCIATED CONTENT

**Supporting Information**.

Supporting Information is available free of charge via the Internet at the ACS website.

> Supporting Information includes eight sections: Simulation details for models in the main text; Additional results for models in the main text; Integrator of NAF for a finite time step; Comparisons of NAF and NAF(S) results; Comparisons of NAF and CMMcv results; Comparisons of GDTWA and NAF-GDTWA results; Comparisons of NAF, FS-NAF and FSSH results; Details of Ehrenfest dynamics. (PDF)

■ **AUTHOR INFORMATION**

**Corresponding Author**

*E-mail: jianliupku@pku.edu.cn

**BIOGRAPHIES**

**Baihua Wu** received his B.S. from the University of Science and Technology Beijing in 2019. He is currently a graduate student in theoretical chemistry at Peking University. His research interests are in phase space mapping approaches of nonadiabatic dynamics.

**Xin He** received his B.S. from Peking University in 2019. He is currently a graduate student in theoretical chemistry at Peking University. His research interests include phase space mapping theory of nonadiabatic dynamics.

**Jian Liu** is the Peking University Boya Distinguished Professor. He received his B.S. from the University of Science & Technology of China in 2000 and Ph.D. from the University of Illinois at Urbana−Champaign in 2005. He then did postdoctoral work at the University of California, Berkeley, and was a research associate at Stanford University before he joined Peking University in 2012. His research interests have been focused on the development of phase space formulations of quantum mechanics and trajectory-based methods for studying statistical mechanics and dynamics of complex (large) molecular systems.

**ORCID**

Baihua Wu: 0000-0002-1256-6859

Xin He: 0000-0002-5189-7204

Jian Liu: 0000-0002-2906-5858

**Notes**

The authors declare no competing financial interests.

## ◼ ACKNOWLEDGMENT

We thank Xiangsong Cheng and Youhao Shang for useful discussions. This work was supported by the National Science Fund for Distinguished Young Scholars Grant No. 22225304. We acknowledge the High-performance Computing Platform of Peking University, Beijing PARATERA Tech Co., Ltd., and Guangzhou supercomputer center for providing computational resources.

## ◼ References:

(1) Ehrenfest;, P.; Ehrenfest, T., Begriffliche Grundlagen Der Statistischen Auffassung in Der Mechanik. In *Encyklopädie Der Mathematischen Wissenschafte*, B. G. Teubner: Liepzig, Germany 1911; Vol. 4, pp 1-90.
(2) Nolte, D. D., The Tangled Tale of Phase Space. *Phys. Today* **2010**, *63*, 33-38. http://dx.doi.org/10.1063/1.3397041
(3) Goldstein, H.; Poole, C. P.; Safko, J. L., *Classical Mechanics (3rd Ed.)*. Addison-Wesley: San Francisco Munich, 2001.
(4) Arnol'd, V. I., *Mathematical Methods of Classical Mechanics*. Springer New York, NY: 2013.
(5) Weyl, H., Quantum Mechanics and Group Theory. *Z. Phys.* **1927**, *46*, 1-46. http://dx.doi.org/10.1007/bf02055756
(6) Wigner, E., On the Quantum Correction for Thermodynamic Equilibrium. *Phys. Rev.* **1932**, *40*, 749-759. http://dx.doi.org/10.1103/physrev.40.749
(7) Groenewold, H. J., On the Principles of Elementary Quantum Mechanics. *Physica* **1946**, *12*, 405-460. http://dx.doi.org/10.1016/s0031-8914(46)80059-4
(8) Moyal, J. E., Quantum Mechanics as a Statistical Theory. *Math. Proc. Cambridge Philos. Soc.* **1949**, *45*, 99-124. http://dx.doi.org/10.1017/s0305004100000487
(9) Lee, H. W., Theory and Application of the Quantum Phase-Space Distribution Functions. *Phys. Rep.* **1995**, *259*, 147-211. http://dx.doi.org/10.1016/0370-1573(95)00007-4
(10) Cohen, L., Generalized Phase-Space Distribution Functions. *J. Math. Phys.* **1966**, *7*, 781-786. http://dx.doi.org/10.1063/1.1931206
(11) Liu, J., A Unified Theoretical Framework for Mapping Models for the Multi-State Hamiltonian. *J. Chem. Phys.* **2016**, *145*, 204105. http://dx.doi.org/10.1063/1.4967815
(12) Liu, J., Isomorphism between the Multi-State Hamiltonian and the Second-Quantized Many-Electron Hamiltonian with Only 1-Electron Interactions. *J. Chem. Phys.* **2017**, *146*, 024110. http://dx.doi.org/10.1063/1.4973708
(13) He, X.; Liu, J., A New Perspective for Nonadiabatic Dynamics with Phase Space Mapping Models. *J. Chem. Phys.* **2019**, *151*, 024105. http://dx.doi.org/10.1063/1.5108736

(14) He, X.; Gong, Z.; Wu, B.; Liu, J., Negative Zero-Point-Energy Parameter in the Meyer-Miller Mapping Model for Nonadiabatic Dynamics. *J. Phys. Chem. Lett.* **2021**, *12*, 2496-2501. http://dx.doi.org/10.1021/acs.jpclett.1c00232
(15) He, X.; Wu, B.; Gong, Z.; Liu, J., Commutator Matrix in Phase Space Mapping Models for Nonadiabatic Quantum Dynamics. *J. Phys. Chem. A* **2021**, *125*, 6845-6863. http://dx.doi.org/10.1021/acs.jpca.1c04429
(16) Liu, J.; He, X.; Wu, B., Unified Formulation of Phase Space Mapping Approaches for Nonadiabatic Quantum Dynamics. *Acc. Chem. Res.* **2021**, *54*, 4215-4228. http://dx.doi.org/10.1021/acs.accounts.1c00511
(17) He, X.; Wu, B.; Shang, Y.; Li, B.; Cheng, X.; Liu, J., New Phase Space Formulations and Quantum Dynamics Approaches. *Wiley Interdiscip. Rev. Comput. Mol. Sci.* **2022**, *12*, e1619. http://dx.doi.org/10.1002/wcms.1619
(18) Wu, B.; He, X.; Liu, J., Phase Space Mapping Theory for Nonadiabatic Quantum Molecular Dynamics. In *Volume on Time-Dependent Density Functional Theory: Nonadiabatic Molecular Dynamics*, Zhu, C., Ed. Jenny Stanford Publishing: New York, 2022.
(19) Polli, D.; Altoe, P.; Weingart, O.; Spillane, K. M.; Manzoni, C.; Brida, D.; Tomasello, G.; Orlandi, G.; Kukura, P.; Mathies, R. A., et al., Conical Intersection Dynamics of the Primary Photoisomerization Event in Vision. *Nature* **2010**, *467*, 440-U88. http://dx.doi.org/10.1038/nature09346
(20) Martinez, T. J., Seaming Is Believing. *Nature* **2010**, *467*, 412-413. http://dx.doi.org/10.1038/467412a
(21) Domcke, W.; Yarkony, D. R.; Köppel, H., *Conical Intersections: Theory, Computation and Experiment*. World Scientific: Singapore, 2011.
(22) Scholes, G. D.; Fleming, G. R.; Olaya-Castro, A.; van Grondelle, R., Lessons from Nature About Solar Light Harvesting. *Nat. Chem.* **2011**, *3*, 763-774. http://dx.doi.org/10.1038/Nchem.1145
(23) Maiuri, M.; Ostroumov, E. E.; Saer, R. G.; Blankenship, R. E.; Scholes, G. D., Coherent Wavepackets in the Fenna-Matthews-Olson Complex Are Robust to Excitonic-Structure Perturbations Caused by Mutagenesis. *Nat. Chem.* **2018**, *10*, 177-183. http://dx.doi.org/10.1038/Nchem.2910
(24) Smith, M. B.; Michl, J., Singlet Fission. *Chem. Rev.* **2010**, *110*, 6891-6936. http://dx.doi.org/10.1021/cr1002613
(25) Scholes, G. D.; Rumbles, G., Excitons in Nanoscale Systems. *Nat. Mater.* **2006**, *5*, 683-696. http://dx.doi.org/10.1038/nmat1710
(26) Long, R.; Prezhdo, O. V.; Fang, W.-H., Nonadiabatic Charge Dynamics in Novel Solar Cell Materials. *Wiley Interdiscip. Rev. Comput. Mol. Sci.* **2017**, *7*, e1305. http://dx.doi.org/10.1002/wcms.1305
(27) Wang, Y.-C.; Ke, Y.; Zhao, Y., The Hierarchical and Perturbative Forms of Stochastic Schrodinger Equations and Their Applications to Carrier Dynamics in Organic Materials. *Wiley Interdiscip. Rev. Comput. Mol. Sci.* **2019**, *9*, e1375. http://dx.doi.org/10.1002/wcms.1375
(28) Garcia-Vidal, F. J.; Ciuti, C.; Ebbesen, T. W., Manipulating Matter by Strong Coupling to Vacuum Fields. *Science* **2021**, *373*, eabd0336. http://dx.doi.org/10.1126/science.abd0336
(29) Hammes-Schiffer, S., Theoretical Perspectives on Non-Born-Oppenheimer Effects in Chemistry. *Philos. Trans. Royal Soc. A* **2022**, *380*, 20200377. http://dx.doi.org/10.1098/rsta.2020.0377

(30) Ehrenfest, P., Bemerkung Über Die Angenäherte Gültigkeit Der Klassischen Mechanik Innerhalb Der Quantenmechanik. *Z. Phys.* **1927**, *45*, 455-457. http://dx.doi.org/10.1007/BF01329203
(31) Delos, J. B.; Thorson, W. R.; Knudson, S. K., Semiclassical Theory of Inelastic Collisions. I. Classical Picture and Semiclassical Formulation. *Phys. Rev. A* **1972**, *6*, 709-720. http://dx.doi.org/10.1103/PhysRevA.6.709
(32) Billing, G. D., On the Applicability of the Classical Trajectory Equations in Inelastic Scattering Theory. *Chem. Phys. Lett.* **1975**, *30*, 391-393. http://dx.doi.org/10.1016/0009-2614(75)80014-5
(33) Miller, W.; McCurdy, C., Classical Trajectory Model for Electronically Nonadiabatic Collision Phenomena. A Classical Analog for Electronic Degrees of Freedom. *J. Chem. Phys.* **1978**, *69*, 5163-5173. http://dx.doi.org/10.1063/1.436463
(34) Micha, D. A., A Self-Consistent Eikonal Treatment of Electronic Transitions in Molecular Collisions. *J. Chem. Phys.* **1983**, *78*, 7138-7145. http://dx.doi.org/10.1063/1.444753
(35) Zhu, C. Y.; Nobusada, K.; Nakamura, H., New Implementation of the Trajectory Surface Hopping Method with Use of the Zhu-Nakamura Theory. *J. Chem. Phys.* **2001**, *115*, 3031-3044. http://dx.doi.org/10.1063/1.1386811
(36) Wang, L.; Akimov, A.; Prezhdo, O. V., Recent Progress in Surface Hopping: 2011-2015. *J. Phys. Chem. Lett.* **2016**, *7*, 2100-2112. http://dx.doi.org/10.1021/acs.jpclett.6b00710
(37) Peng, J.; Xie, Y.; Hu, D.; Du, L.; Lan, Z., Treatment of Nonadiabatic Dynamics by on-the-Fly Trajectory Surface Hopping Dynamics. *Acta Phys.-Chim. Sin.* **2019**, *35*, 28-48. http://dx.doi.org/10.3866/PKU.WHXB201801042
(38) Barbatti, M., Nonadiabatic Dynamics with Trajectory Surface Hopping Method. *Wiley Interdiscip. Rev. Comput. Mol. Sci.* **2011**, *1*, 620-633. http://dx.doi.org/10.1002/wcms.64
(39) Cui, G.; Thiel, W., Generalized Trajectory Surface-Hopping Method for Internal Conversion and Intersystem Crossing. *J. Chem. Phys.* **2014**, *141*, 124101. http://dx.doi.org/10.1063/1.4894849
(40) Mai, S.; Marquetand, P.; Gonzlez, L., A General Method to Describe Intersystem Crossing Dynamics in Trajectory Surface Hopping. *Int. J. Quantum Chem.* **2015**, *115*, 1215-1231. http://dx.doi.org/10.1002/qua.24891
(41) Subotnik, J. E.; Jain, A.; Landry, B.; Petit, A.; Ouyang, W.; Bellonzi, N., Understanding the Surface Hopping View of Electronic Transitions and Decoherence. *Annu. Rev. Phys. Chem.* **2016**, *67*, 387-417. http://dx.doi.org/10.1146/annurev-physchem-040215-112245
(42) Tully, J. C.; Preston, R. K., Trajectory Surface Hopping Approach to Nonadiabatic Molecular Collisions: The Reaction of H+ with D2. *J. Chem. Phys.* **1971**, *55*, 562-572. http://dx.doi.org/10.1063/1.1675788
(43) Tully, J. C., Molecular Dynamics with Electronic Transitions. *J. Chem. Phys.* **1990**, *93*, 1061-1071. http://dx.doi.org/10.1063/1.459170
(44) Lang, H. Quantum Dynamics of Chemical Systems with Large Number of Degrees of Freedom: Linearized Phase Space Methods and Quantum Simulations. Ph.D. Dissertation, Ruprecht Karl University of Heidelberg, Heidelberg, Baden-Württemberg, Germany, 2022.
(45) Nakahara, M., *Geometry, Topology, and Physics*. 2 ed.; Institute of Physics Publishing: Bristol, 2003.
(46) Atiyah, M. F.; Todd, J. A., On Complex Stiefel Manifolds. *Math. Proc. Cambridge Philos. Soc.* **1960**, *56*, 342-353. http://dx.doi.org/10.1017/s0305004100034642
(47) Shang, Y.; Cheng, X.; Liu, J., **(to be submitted)**.

(48)     Meyer, H.-D.; Miller, W. H., A Classical Analog for Electronic Degrees of Freedom in Nonadiabatic Collision Processes. *J. Chem. Phys.* **1979**, *70*, 3214-3223. http://dx.doi.org/10.1063/1.437910
(49)     Stock, G.; Thoss, M., Semiclassical Description of Nonadiabatic Quantum Dynamics. *Phys. Rev. Lett.* **1997**, *78*, 578-581. http://dx.doi.org/10.1103/PhysRevLett.78.578
(50)     Das, A., *Field Theory*. World Scientific: Singapore, 2019.
(51)     Pacher, T.; Cederbaum, L. S.; Köppel, H., Adiabatic and Quasidiabatic States in a Gauge Theoretical Framework. *Adv. Chem. Phys.* **1993**, *84*, 293-391. http://dx.doi.org/10.1002/9780470141427.ch4
(52)     Cotton, S. J.; Liang, R.; Miller, W. H., On the Adiabatic Representation of Meyer-Miller Electronic-Nuclear Dynamics. *J. Chem. Phys.* **2017**, *147*, 064112. http://dx.doi.org/10.1063/1.4995301
(53)     Liu, X.; Liu, J., Path Integral Molecular Dynamics for Exact Quantum Statistics of Multi-Electronic-State Systems. *J. Chem. Phys.* **2018**, *148*, 102319. http://dx.doi.org/10.1063/1.5005059
(54)     Landau, L. D., On the Theory of Transfer of Energy at Collisions Ii. *Phys. Z. Sowjetunion* **1932**, *2*, 46-51.
(55)     Zener, C., Non-Adiabatic Crossing of Energy Levels. *Proc. R. Soc. London, Ser. A* **1932**, *137*, 696-702. http://dx.doi.org/10.1098/rspa.1932.0165
(56)     Stückelberg, E. C. G., Theory of Inelastic Collisions between Atoms. *Helv. Phys. Acta* **1932**, *5*, 369-423.
(57)     Kayanuma, Y., Nonadiabatic Transitions in Level Crossing with Energy Fluctuation. I. Analytical Investigations. *J. Phys. Soc. Jpn.* **1984**, *53*, 108-117. http://dx.doi.org/10.1143/JPSJ.53.108
(58)     Leggett, A. J.; Chakravarty, S.; Dorsey, A. T.; Fisher, M. P. A.; Garg, A.; Zwerger, W., Dynamics of the Dissipative Two-State System. *Rev. Mod. Phys.* **1987**, *59*, 1-85. http://dx.doi.org/10.1103/revmodphys.59.1
(59)     Makarov, D. E.; Makri, N., Path Integrals for Dissipative Systems by Tensor Multiplication. Condensed Phase Quantum Dynamics for Arbitrarily Long Time. *Chem. Phys. Lett.* **1994**, *221*, 482-491. http://dx.doi.org/10.1016/0009-2614(94)00275-4
(60)     Makri, N.; Makarov, D. E., Tensor Propagator for Iterative Quantum Time Evolution of Reduced Density Matrices. Ii. Numerical Methodology. *J. Chem. Phys.* **1995**, *102*, 4611-4618. http://dx.doi.org/10.1063/1.469509
(61)     Makri, N.; Makarov, D. E., Tensor Propagator for Iterative Quantum Time Evolution of Reduced Density Matrices. I. Theory. *J. Chem. Phys.* **1995**, *102*, 4600-4610. http://dx.doi.org/10.1063/1.469508
(62)     Makri, N., Small Matrix Path Integral with Extended Memory. *J. Chem. Theory. Comput.* **2021**, *17*, 1-6. http://dx.doi.org/10.1021/acs.jctc.0c00987
(63)     Makri, N., Small Matrix Disentanglement of the Path Integral: Overcoming the Exponential Tensor Scaling with Memory Length. *J. Chem. Phys.* **2020**, *152*, 041104. http://dx.doi.org/10.1063/1.5139473
(64)     Tanimura, Y.; Kubo, R., Time Evolution of a Quantum System in Contact with a Nearly Gaussian-Markoffian Noise Bath. *J. Phys. Soc. Jpn.* **1989**, *58*, 101-114. http://dx.doi.org/10.1143/JPSJ.58.101
(65)     Yan, Y.-A.; Yang, F.; Liu, Y.; Shao, J., Hierarchical Approach Based on Stochastic Decoupling to Dissipative Systems. *Chem. Phys. Lett.* **2004**, *395*, 216-221. http://dx.doi.org/10.1016/j.cplett.2004.07.036

(66) Xu, R.-X.; Cui, P.; Li, X.-Q.; Mo, Y.; Yan, Y., Exact Quantum Master Equation Via the Calculus on Path Integrals. *J. Chem. Phys.* **2005**, *122*, 041103. http://dx.doi.org/10.1063/1.1850899

(67) Shao, J., Stochastic Description of Quantum Open Systems: Formal Solution and Strong Dissipation Limit. *Chem. Phys.* **2006**, *322*, 187-192. http://dx.doi.org/10.1016/j.chemphys.2005.08.007

(68) Moix, J. M.; Cao, J., A Hybrid Stochastic Hierarchy Equations of Motion Approach to Treat the Low Temperature Dynamics of Non-Markovian Open Quantum Systems. *J. Chem. Phys.* **2013**, *139*, 134106. http://dx.doi.org/10.1063/1.4822043

(69) Meyer, H.-D.; Manthe, U.; Cederbaum, L. S., The Multi-Configurational Time-Dependent Hartree Approach. *Chem. Phys. Lett.* **1990**, *165*, 73-78. http://dx.doi.org/10.1016/0009-2614(90)87014-i

(70) Thoss, M.; Wang, H.; Miller, W. H., Self-Consistent Hybrid Approach for Complex Systems: Application to the Spin-Boson Model with Debye Spectral Density. *J. Chem. Phys.* **2001**, *115*, 2991-3005. http://dx.doi.org/10.1063/1.1385562

(71) Wang, H.; Thoss, M., Multilayer Formulation of the Multiconfiguration Time-Dependent Hartree Theory. *J. Chem. Phys.* **2003**, *119*, 1289-1299. http://dx.doi.org/10.1063/1.1580111

(72) Ren, J. J.; Li, W. T.; Jiang, T.; Wang, Y. H.; Shuai, Z. G., Time-Dependent Density Matrix Renormalization Group Method for Quantum Dynamics in Complex Systems. *Wiley Interdiscip. Rev. Comput. Mol. Sci.* **2022**, *12*, e1614. http://dx.doi.org/10.1002/wcms.1614

(73) Makri, N., The Linear Response Approximation and Its Lowest Order Corrections: An Influence Functional Approach. *J. Phys. Chem. B* **1999**, *103*, 2823-2829. http://dx.doi.org/10.1021/jp9847540

(74) Wang, H., Iterative Calculation of Energy Eigenstates Employing the Multilayer Multiconfiguration Time-Dependent Hartree Theory. *J. Phys. Chem. A* **2014**, *118*, 9253-9261. http://dx.doi.org/10.1021/jp503351t

(75) Fenna, R. E.; Matthews, B. W., Chlorophyll Arrangement in a Bacteriochlorophyll Protein from Chlorobium-Limicola. *Nature* **1975**, *258*, 573-577. http://dx.doi.org/10.1038/258573a0

(76) Engel, G. S.; Calhoun, T. R.; Read, E. L.; Ahn, T. K.; Mancal, T.; Cheng, Y. C.; Blankenship, R. E.; Fleming, G. R., Evidence for Wavelike Energy Transfer through Quantum Coherence in Photosynthetic Systems. *Nature* **2007**, *446*, 782-786. http://dx.doi.org/10.1038/nature05678

(77) Higgins, J. S.; Lloyd, L. T.; Sohail, S. H.; Allodi, M. A.; Otto, J. P.; Saer, R. G.; Wood, R. E.; Massey, S. C.; Ting, P. C.; Blankenship, R. E., et al., Photosynthesis Tunes Quantum-Mechanical Mixing of Electronic and Vibrational States to Steer Exciton Energy Transfer. *Proc. Natl. Acad. Sci.* **2021**, *118*, e2018240118. http://dx.doi.org/10.1073/pnas.2018240118

(78) Ishizaki, A.; Fleming, G. R., Theoretical Examination of Quantum Coherence in a Photosynthetic System at Physiological Temperature. *Proc. Natl. Acad. Sci.* **2009**, *106*, 17255-17260. http://dx.doi.org/10.1073/pnas.0908989106

(79) Tao, G.; Miller, W. H., Semiclassical Description of Electronic Excitation Population Transfer in a Model Photosynthetic System. *J. Phys. Chem. Lett.* **2010**, *1*, 891-894. http://dx.doi.org/10.1021/jz1000825

(80) Miller, W. H., Perspective: Quantum or Classical Coherence? *J. Chem. Phys.* **2012**, *136*, 210901. http://dx.doi.org/10.1063/1.4727849

(81) Cao, J. S.; Cogdell, R. J.; Coker, D. F.; Duan, H. G.; Hauer, J.; Kleinekathofer, U.; Jansen, T. L. C.; Mancal, T.; Miller, R. J. D.; Ogilvie, J. P., et al., Quantum Biology Revisited. *Sci. Adv.* **2020**, *6*, eaaz4888. http://dx.doi.org/10.1126/sciadv.aaz4888
(82) Haugland, T. S.; Ronca, E.; Kjønstad, E. F.; Rubio, A.; Koch, H., Coupled Cluster Theory for Molecular Polaritons: Changing Ground and Excited States. *Phys. Rev. X* **2020**, *10*, 041043. http://dx.doi.org/10.1103/PhysRevX.10.041043
(83) Toida, H.; Nakajima, T.; Komiyama, S., Vacuum Rabi Splitting in a Semiconductor Circuit Qed System. *Phys. Rev. Lett.* **2013**, *110*, 066802. http://dx.doi.org/10.1103/PhysRevLett.110.066802
(84) Guerin, W.; Santo, T.; Weiss, P.; Cipris, A.; Schachenmayer, J.; Kaiser, R.; Bachelard, R., Collective Multimode Vacuum Rabi Splitting. *Phys. Rev. Lett.* **2019**, *123*, 243401. http://dx.doi.org/10.1103/PhysRevLett.123.243401
(85) Hoffmann, N. M.; Schafer, C.; Rubio, A.; Kelly, A.; Appel, H., Capturing Vacuum Fluctuations and Photon Correlations in Cavity Quantum Electrodynamics with Multitrajectory Ehrenfest Dynamics. *Phys. Rev. A* **2019**, *99*, 063819. http://dx.doi.org/10.1103/PhysRevA.99.063819
(86) Hoffmann, N. M.; Schäfer, C.; Säkkinen, N.; Rubio, A.; Appel, H.; Kelly, A., Benchmarking Semiclassical and Perturbative Methods for Real-Time Simulations of Cavity-Bound Emission and Interference. *J. Chem. Phys.* **2019**, *151*, 244113. http://dx.doi.org/10.1063/1.5128076
(87) Li, T. E.; Chen, H. T.; Nitzan, A.; Subotnik, J. E., Quasiclassical Modeling of Cavity Quantum Electrodynamics. *Phys. Rev. A* **2020**, *101*, 033831. http://dx.doi.org/10.1103/PhysRevA.101.033831
(88) Saller, M. A. C.; Kelly, A.; Geva, E., Benchmarking Quasiclassical Mapping Hamiltonian Methods for Simulating Cavity-Modified Molecular Dynamics. *J. Phys. Chem. Lett.* **2021**, *12*, 3163-3170. http://dx.doi.org/10.1021/acs.jpclett.1c00158
(89) Singh, S.; Jones, W. J.; Siebrand, W.; Stoicheff, B. P.; Schneider, W. G., Laser Generation of Excitons and Fluorescence in Anthracene Crystals. *J. Chem. Phys.* **1965**, *42*, 330-342. http://dx.doi.org/10.1063/1.1695695
(90) Chan, W.-L.; Berkelbach, T. C.; Provorse, M. R.; Monahan, N. R.; Tritsch, J. R.; Hybertsen, M. S.; Reichman, D. R.; Gao, J.; Zhu, X.-Y., The Quantum Coherent Mechanism for Singlet Fission: Experiment and Theory. *Acc. Chem. Res.* **2013**, *46*, 1321-1329. http://dx.doi.org/10.1021/ar300286s
(91) Casanova, D., Theoretical Modeling of Singlet Fission. *Chem. Rev.* **2018**, *118*, 7164-7207. http://dx.doi.org/10.1021/acs.chemrev.7b00601
(92) Coronado, E. A.; Xing, J.; Miller, W. H., Ultrafast Non-Adiabatic Dynamics of Systems with Multiple Surface Crossings: A Test of the Meyer-Miller Hamiltonian with Semiclassical Initial Value Representation Methods. *Chem. Phys. Lett.* **2001**, *349*, 521-529. http://dx.doi.org/10.1016/s0009-2614(01)01242-8
(93) Colbert, D. T.; Miller, W. H., A Novel Discrete Variable Representation for Quantum Mechanical Reactive Scattering Via the S-Matrix Kohn Method. *J. Chem. Phys.* **1992**, *96*, 1982-1991. http://dx.doi.org/10.1063/1.462100
(94) Ananth, N.; Venkataraman, C.; Miller, W. H., Semiclassical Description of Electronically Nonadiabatic Dynamics Via the Initial Value Representation. *J. Chem. Phys.* **2007**, *127*, 084114. http://dx.doi.org/10.1063/1.2759932

(95) Miller, W. H., Spiers Memorial Lecture - Quantum and Semiclassical Theory of Chemical Reaction Rates. *Faraday Discuss.* **1998**, *110*, 1-21. http://dx.doi.org/10.1039/a805196h
(96) Sun, X.; Miller, W. H., Forward-Backward Initial Value Representation for Semiclassical Time Correlation Functions. *J. Chem. Phys.* **1999**, *110*, 6635-6644. http://dx.doi.org/10.1063/1.478571
(97) Miller, W. H., Quantum Dynamics of Complex Molecular Systems. *Proc. Natl. Acad. Sci.* **2005**, *102*, 6660-6664. http://dx.doi.org/10.1073/pnas.0408043102
(98) Miller, W. H., Electronically Nonadiabatic Dynamics Via Semiclassical Initial Value Methods. *J. Phys. Chem. A* **2009**, *113*, 1405-1415. http://dx.doi.org/10.1021/jp809907p
(99) Schneider, R.; Domcke, W., S1-S2 Conical Intersection and Ultrafast S2->S1 Internal Conversion in Pyrazine. *Chem. Phys. Lett.* **1988**, *150*, 235-242. http://dx.doi.org/10.1016/0009-2614(88)80034-4
(100) Krempl, S.; Winterstetter, M.; Plöhn, H.; Domcke, W., Path-Integral Treatment of Multi-Mode Vibronic Coupling. *J. Chem. Phys.* **1994**, *100*, 926-937. http://dx.doi.org/10.1063/1.467253
(101) Worth, G. A.; Welch, G.; Paterson, M. J., Wavepacket Dynamics Study of $Cr(Co)_5$ after Formation by Photodissociation: Relaxation through an $(E \oplus a) \otimes E$ Jahn–Teller Conical Intersection. *Mol. Phys.* **2006**, *104*, 1095-1105. http://dx.doi.org/10.1080/00268970500418182
(102) Worth, G. A.; Beck, M. H.; Jackle, A.; Meyer, H.-D.The MCTDH Package, Version 8.2, (2000). H.-D. Meyer, Version 8.3 (2002), Version 8.4 (2007). O. Vendrell and H.-D. Meyer Version 8.5 (2013). Version 8.5 contains the ML-MCTDH algorithm. See http://mctdh.uni-hd.de. (accessed on November 1st, 2023) Used version: 8.5.14.
(103) Li, X.; Tully, J. C.; Schlegel, H. B.; Frisch, M. J., Ab Initio Ehrenfest Dynamics. *J. Chem. Phys.* **2005**, *123*, 084106. http://dx.doi.org/10.1063/1.2008258
(104) Richter, M.; Marquetand, P.; Gonzalez-Vazquez, J.; Sola, I.; Gonzalez, L., Sharc: Ab Initio Molecular Dynamics with Surface Hopping in the Adiabatic Representation Including Arbitrary Couplings. *J. Chem. Theory. Comput.* **2011**, *7*, 1253-1258. http://dx.doi.org/10.1021/ct1007394
(105) Curchod, B. F. E.; Martínez, T. J., Ab Initio Nonadiabatic Quantum Molecular Dynamics. *Chem. Rev.* **2018**, *118*, 3305-3336. http://dx.doi.org/10.1021/acs.chemrev.7b00423
(106) Freixas, V. M.; White, A. J.; Nelson, T.; Song, H. J.; Makhov, D. V.; Shalashilin, D.; Fernandez-Alberti, S.; Tretiale, S., Nonadiabatic Excited-State Molecular Dynamics Methodologies: Comparison and Convergence. *J. Phys. Chem. Lett.* **2021**, *12*, 2970-2982. http://dx.doi.org/10.1021/acs.jpclett.1c00266
(107) He, X.; Wu, B. H.; Rivlin, T.; Liu, J.; Pollak, E., Transition Path Flight Times and Nonadiabatic Electronic Transitions. *J. Phys. Chem. Lett.* **2022**, *13*, 6966-6974. http://dx.doi.org/10.1021/acs.jpclett.2c01425
(108) Lang, H.; Vendrell, O.; Hauke, P., Generalized Discrete Truncated Wigner Approximation for Nonadiabatic Quantum-Classical Dynamics. *J. Chem. Phys.* **2021**, *155*, 024111. http://dx.doi.org/10.1063/5.0054696

■

# Supporting Information: Nonadiabatic Field on Quantum Phase Space: A Century after Ehrenfest

*Baihua Wu, Xin He, and Jian Liu\**

Beijing National Laboratory for Molecular Sciences, Institute of Theoretical and Computational Chemistry, College of Chemistry and Molecular Engineering,

Peking University, Beijing 100871, China

AUTHOR INFORMATION

**Corresponding Author**

\* Electronic mail: jianliupku@pku.edu.cn

## S1. Simulation Details for Models in the Main Text

### S1-A: Spin-Boson Model

The Hamiltonian for the spin-boson model reads

$$\hat{H} = \sum_{j=1}^{N_b} \frac{1}{2}\left(\hat{P}_j^2 + \omega_j^2 \hat{R}_j^2\right) + \left(\varepsilon + \sum_{j=1}^{N_b} c_j \hat{R}_j\right)\left(|1\rangle\langle 1| - |2\rangle\langle 2|\right) + \Delta\left(|1\rangle\langle 2| + |2\rangle\langle 1|\right), \quad \text{(S1)}$$

where $\varepsilon$ and $\Delta$ represent the energy bias and the tunneling between states $|1\rangle$ and $|2\rangle$, respectively. (We set $\hbar = 1$ throughout the Supporting Information.) The operators $\{\hat{P}_j, \hat{R}_j\}$ denote the mass-weighted momentum and coordinate of the $j$-th harmonic oscillator for the bath. The frequencies $\{\omega_j\}$ and the coupling coefficients $\{c_j\}$ are typically obtained by discretizing a given spectral density function. In this Perspective, we use the discretization scheme proposed in refs [1,2] for the Ohmic spectral density $J(\omega) = \frac{\pi}{2}\alpha\omega \exp(-\omega/\omega_c)$ with the Kondo parameter $\alpha$ and the cut-off frequency $\omega_c$, which results in the following expressions:

$$\begin{cases} \omega_j = -\omega_c \ln\left(1 - \dfrac{j}{1+N_b}\right) \\ c_j = \omega_j \sqrt{\dfrac{\alpha\omega_c}{N_b+1}} \end{cases}, \quad j = 1,\cdots,N_b. \quad \text{(S2)}$$

We employ four specific spin-boson models with $\varepsilon = \Delta = 1$ at low temperature ($\beta = 5$) of ref [3], which range from weak to strong system-bath coupling (small to large $\alpha$) and from low to high cut-off frequency $\omega_c$. Three hundred bath DOFs produced by eq (S2) are involved for guaranteeing full convergence. The initially occupied state is selected as the higher-level state $|1\rangle$, while the bath DOFs are sampled from the corresponding Wigner distribution

$$\rho_W(\mathbf{R}, \mathbf{P}) \propto \exp\left[-\sum_{j=1}^{N_b} \frac{\beta}{2Q(\omega_j)}\left(P_j^2 + \omega_j^2 R_j^2\right)\right] \tag{S3}$$

with $Q(\omega) = \frac{\beta\hbar\omega/2}{\tanh(\beta\hbar/2)}$ as the quantum corrector[4]. We investigate the dynamics of the electronic reduced density matrix, where the population difference $D(t) = P_{1\to 1}(t) - P_{1\to 2}(t)$ and the modulus of the off-diagonal term $|\rho_{12}(t)|$ are demonstrated. The exact results produced by extended HEOM (eHEOM)[5, 6] are taken from our previous work[3].

**S1-B: Seven-Site Model for the FMO Monomer**

The Fenna-Matthews-Olson (FMO) monomer is modeled as a 7-state site-exciton model. The total Hamiltonian of the site-exciton model is divided into three parts: the exciton part $\hat{H}_S$, the environment bath part $\hat{H}_B$ and the linear coupling term $\hat{H}_{S-B}$:

$$\hat{H} = \hat{H}_S + \hat{H}_B + \hat{H}_{S-B} . \tag{S4}$$

$$\hat{H}_S = \sum_{n,m=1}^{F} H_{S,nm} |n\rangle\langle m| . \tag{S5}$$

$$\hat{H}_B = \sum_{n=1}^{F}\sum_{j=1}^{N_b} \left(\hat{P}_{nj}^2 + \omega_j^2 \hat{R}_{nj}^2\right)/2 . \tag{S6}$$

$$\hat{H}_{S-B} = \sum_{n=1}^{F}\sum_{j=1}^{N_b} c_j \hat{R}_{nj} |n\rangle\langle n| . \tag{S7}$$

Similar to spin-boson models, the bath frequencies and system-bath coupling coefficients are determined by discretizing the spectral density. We employ the Debye spectral density $J(\omega) = 2\lambda\omega_c\omega/(\omega^2 + \omega_c^2)$ for each state, where $\lambda$ and $\omega_c$ denote the bath reorganization energy and the characteristic frequency, respectively. The corresponding discretization scheme from refs [7-9] is

$$\omega_j = \omega_c \tan\left(\frac{\pi}{2} - \frac{\pi j}{2(N_b+1)}\right), \quad j = 1, \cdots, N_b,  \quad (S8)$$

$$c_j = \omega_j \sqrt{\frac{2\lambda}{N_b+1}}, \quad j = 1, \cdots, N_b. \quad (S9)$$

The system Hamiltonian of the FMO model reads

$$\hat{H}_S = \begin{pmatrix} 12410 & -87.7 & 5.5 & -5.9 & 6.7 & -13.7 & -9.9 \\ -87.7 & 12530 & 30.8 & 8.2 & 0.7 & 11.8 & 4.3 \\ 5.5 & 30.8 & 12210 & -53.5 & -2.2 & -9.6 & 6.0 \\ -5.9 & 8.2 & -53.5 & 12320 & -70.7 & -17.0 & -63.3 \\ 6.7 & 0.7 & -2.2 & -70.7 & 12480 & 81.1 & -1.3 \\ -13.7 & 11.8 & -9.6 & -17.0 & 81.1 & 12630 & 39.7 \\ -9.9 & 4.3 & 6.0 & -63.3 & -1.3 & 39.7 & 12440 \end{pmatrix} \text{cm}^{-1}. \quad (S10)$$

The bath reorganization energy is $\lambda = 35 \text{ cm}^{-1}$ and the characteristic frequency is $\omega_c = 106.14 \text{ cm}^{-1}$. The number of bath DOFs for each site is chosen as 50. We investigate a challenging temperature $T = 77\text{K}$ as studied in our previous work[10]. The first site of the system is initially occupied, and the bath DOFs are sampled from the Wigner distributions of the corresponding harmonic oscillators. The dynamics of both population and coherence terms (i.e., the diagonal and off-diagonal elements of the electronic reduced density matrix) are demonstrated. Numerical exact results are produced by HEOM.

**S1-C: Atom-In-Cavity Models**

The total Hamiltonian for the atom-in-cavity models can be decomposed into three parts. The hydrogen atom is described by F atomic energy levels:

$$\hat{H}_a = \sum_{n=1}^{F} \varepsilon_n |n\rangle\langle n|, \tag{S11}$$

where $\varepsilon_n$ is the atomic energy level of the $n$-th atomic state. The optical field part reads

$$\hat{H}_p = \sum_{j=1}^{N} \frac{1}{2}\left(\hat{P}_j^2 + \omega_j^2 \hat{R}_j^2\right), \tag{S12}$$

where $\hat{R}_j, \hat{P}_j, \omega_j$ denote the canonical coordinate, canonical momentum, and frequency of the $j$-th optical field mode, respectively. The coupling term between the atom and the optical field can be expressed using the dipole approximation[11] as

$$\hat{H}_c = \sum_{j=1}^{N} \omega_j \lambda_j(r_0) \hat{R}_j \sum_{n \neq m} \mu_{nm} |n\rangle\langle m|. \tag{S13}$$

Here, $\mu_{nm}$ represents the transitional dipole moment between the $n$-th and $m$-th states, and the atom-optical field interaction reads

$$\lambda_j(r_0) = \sqrt{\frac{2}{\varepsilon_0 L}} \sin\left(\frac{j\pi r_0}{L}\right), \quad j = 1, \cdots, N, \tag{S14}$$

where $L$, $\varepsilon_0$ and $r_0$ denote the volume length of the cavity, the vacuum permittivity, and the location of the atom, respectively. We set $L = 236200$ a.u. and $r_0 = L/2$. Four hundred standing-wave modes are employed for the optical field, where the frequency of the $j$-th mode is $\omega_j = j\pi c/L$. (Here $c = 137.036$ a.u. is the light speed in vacuum). We employ a three-state model with the energy levels $\varepsilon_1 = -0.6738$, $\varepsilon_2 = -0.2798$, $\varepsilon_3 = -0.1547$, and the dipole moments $\mu_{12} = -1.034$, $\mu_{23} = -2.536$ (all in atomic units). A reduced two-state case that retains the two lowest atomic states is also investigated. At the beginning, the atom is in the highest atomic state, and each optical field mode is in the corresponding optical vacuum state, whose Wigner distribution reads

$$\rho_W(R_j, P_j) \propto \exp\left[-\left(P_j^2/\omega_j + \omega_j R_j^2\right)\right], \quad j=1,\cdots,N, \quad (S15)$$

The exact results produced by truncated configuration interaction calculations are available in refs [12, 13].

**S1-D: Singlet-Fission Model**

The singlet-fission (SF) model utilized in the Perspective is a three-state site-exciton model with the Debye spectral density. This model contains the high-energy singlet state (S1), the charge-transfer state (CT), and the multi-exciton state that will subsequently split into the double triplets (TT). The system Hamiltonian reads[14, 15],

$$\hat{H}_S = \begin{pmatrix} 0.2 & -0.05 & 0 \\ -0.05 & 0.3 & -0.05 \\ 0 & -0.05 & 0 \end{pmatrix} \text{eV} \quad \begin{matrix} \text{S1} \\ \text{CT} \\ \text{TT} \end{matrix} \quad . \quad (S16)$$

The bath reorganization energy is $\lambda = 0.1$ eV and the characteristic frequency is $\omega_c = 0.18$ eV. The number of bath modes is chosen as 200 for each state. The system is initially in the S1 state. The nuclear DOFs are sampled from the Wigner distributions of the corresponding harmonic oscillators at 300 K. We employ HEOM to obtain numerically exact results.

**S1-E: Gas Phase Models with One Nuclear Degree of Freedom**

We first test the three anharmonic 3-state photodissociation models of Miller and coworkers [16]. These models are composed of Morse potential energy surfaces and Gaussian coupling terms:

$$V_{ii}(R) = D_i\left[1-e^{-\beta_i(R-R_i)}\right]^2 + C_i, \quad i=1,2,3.$$
$$V_{ij}(R) = V_{ji}(R) = A_{ij}e^{-\alpha_{ij}(R-R_{ij})^2}, \quad i,j=1,2,3; \text{ and } i \neq j.$$
(S17)

The parameters taken from ref [16] are listed in Table S1:

**Table S1:** Parameters of 3-State Photodissociation Models[16]

| Parameters | Model 1 | Model 2 | Model 3 |
|---|---|---|---|
| $C_1, C_2, C_3$ | 0.00, 0.01, 0.006 | 0.00, 0.01, 0.02 | 0.02, 0.00, 0.02 |
| $D_1, D_2, D_3$ | 0.003, 0.004, 0.003 | 0.020, 0.010, 0.003 | 0.020, 0.020, 0.003 |
| $R_1, R_2, R_3$ | 5.0, 4.0, 6.0 | 4.5, 4.0, 4.4 | 4.0, 4.5, 6.0 |
| $\beta_1, \beta_2, \beta_3$ | 0.65, 0.60, 0.65 | 0.65, 0.40, 0.65 | 0.40, 0.65, 0.65 |
| $A_{12}, A_{23}, A_{31}$ | 0.002, 0.002, 0.0 | 0.005, 0.0, 0.005 | 0.005, 0.0, 0.005 |
| $R_{12}, R_{23}, R_{31}$ | 3.40, 4.80, 0.00 | 3.66, 0.00, 3.34 | 3.40, 0.00, 4.97 |
| $\alpha_{12}, \alpha_{23}, \alpha_{31}$ | 16.0, 16.0, 0.0 | 32.0, 0.0, 32.0 | 32.0, 0.0, 32.0 |
| $R_e$ | 2.9 | 3.3 | 2.1 |

The system is initially occupied in the first diabatic state, and the nuclear DOF is sampled from the Wigner distribution of the ground state:

$$\rho_W(R,P) \propto e^{-m\omega(R-R_e)^2 - P^2/m\omega}, \quad (S18)$$

where $m = 20000$ a.u. is the mass of the nuclear DOF, $\omega = 0.005$ a.u. is the vibrational frequency of the ground state, and the center $R_e$ is also listed in Table S1. Note that $R$ stands for the chemical bond length, implying that it should be positive-definite.

The three Tully models[17] in the diabatic representation are described as follows. The single avoided crossing (SAC) model reads

$$V_{11}(R) = A(1-e^{-B|R|})\operatorname{sgn}(R)$$
$$V_{22}(R) = -V_{11}(R) \quad (S19)$$
$$V_{12}(R) = V_{21}(R) = Ce^{-DR^2}$$

with $A = 0.01$, $B = 1.6$, $C = 0.005$ and $D = 1.0$. The dual avoided crossing (DAC) model reads

$$V_{11}(R) = 0$$
$$V_{22}(R) = -Ae^{-BR^2} + E_0 \quad (S20)$$
$$V_{12}(R) = V_{21}(R) = Ce^{-DR^2}$$

with $A = 0.1$, $B = 0.28$, $C = 0.015$, $D = 0.06$ and $E_0 = 0.05$. The extended coupling region (ECR) model reads

$$V_{11}(R) = +E_0$$
$$V_{22}(R) = -E_0 \quad (S21)$$
$$V_{12}(R) = V_{21}(R) = C\left[e^{BR}h(-R) + (2 - e^{-BR})h(R)\right]$$

with $B = 0.9$, $C = 0.1$, $E_0 = -0.0006$, and $h(R)$ denotes the Heaviside function. The system with mass $m = 2000$ a.u. is initially occupied in the electronic ground state in the adiabatic representation with the nuclear wavefunction

$$\psi(R) \propto e^{-\alpha(R-R_0)^2/2 + iP_0(R-R_0)}, \quad (S22)$$

whose corresponding Wigner distribution reads

$$\rho_W(R, P) \propto e^{-\alpha(R-R_0)^2 - (P-P_0)^2/\alpha}. \quad (S23)$$

The center of the wavefunction, $R_0$, for SAC, DAC, and ECR models is set to -3.8, -10, and -13, respectively. The width parameter $\alpha = 1$ while the initial momentum is $P_0$.

In addition to Tully models, an asymmetric SAC model is also considered in the Perspective. Such a model was tested in refs [18, 19], where the diabatic potential matrix elements read

$$V_{11}(R) = A_1(1+\tanh(BR))$$
$$V_{22}(R) = A_2(1-\tanh(BR)) \quad . \quad \quad (S24)$$
$$V_{12}(R) = V_{21}(R) = Ce^{-D(R+Q)^2}$$

Here we use $A_1 = 0.04$, $A_2 = 0.01$, $B = 1.0$, $C = 0.005$, $D = 1.0$ and $Q = 0.7$ as chosen in ref [19]. The nuclear mass is set to 1980 a.u. The initial conditions for the asymmetric SAC model are identical to those of Tully models (eq (S22)), except that $R_0 = -5$ and $\alpha = 0.25$.

The diabatic population dynamics of 3-state photodissociation models and the scattering probabilities of each channel in the adiabatic representation for the Tully models as well as the asymmetric SAC model are investigated. In addition, we investigate the nuclear momentum distribution. For trajectory-based dynamics methods, the nuclear momentum distribution can be described by the time correlation function

$$\begin{aligned}\rho(P,t) &= \text{Tr}\left[\hat{\rho}e^{i\hat{H}t}\delta(\hat{P}-P)e^{-i\hat{H}t}\right]\\ &= \frac{1}{2\pi}\int_{-\infty}^{\infty}dse^{-iPs}\text{Tr}\left[\hat{\rho}e^{i\hat{H}t}e^{i\hat{P}s}e^{-i\hat{H}t}\right]\\ &= \frac{1}{2\pi}\int_{0}^{\infty}dse^{-iPs}\text{Tr}\left[\hat{\rho}e^{i\hat{H}t}e^{i\hat{P}s}e^{-i\hat{H}t}\right] + \frac{1}{2\pi}\int_{-\infty}^{0}dse^{-iPs}\text{Tr}\left[\hat{\rho}e^{i\hat{H}t}e^{i\hat{P}s}e^{-i\hat{H}t}\right]\\ &= \frac{1}{2\pi}\int_{0}^{\infty}dse^{-iPs}\text{Tr}\left[\hat{\rho}e^{i\hat{H}t}e^{i\hat{P}s}e^{-i\hat{H}t}\right] + \frac{1}{2\pi}\int_{0}^{\infty}dse^{iPs}\text{Tr}\left[\hat{\rho}e^{i\hat{H}t}e^{-i\hat{P}s}e^{-i\hat{H}t}\right] \quad , \quad (S25)\\ &= \frac{1}{2\pi}\int_{0}^{\infty}dse^{-iPs}\text{Tr}\left[\hat{\rho}e^{i\hat{H}t}e^{i\hat{P}s}e^{-i\hat{H}t}\right] + \frac{1}{2\pi}\int_{0}^{\infty}ds\left(e^{-iPs}\text{Tr}\left[\hat{\rho}e^{i\hat{H}t}e^{i\hat{P}s}e^{-i\hat{H}t}\right]\right)^{*}\\ &= \frac{1}{\pi}\text{Re}\int_{0}^{\infty}dse^{-iPs}\text{Tr}\left[\hat{\rho}e^{i\hat{H}t}e^{i\hat{P}s}e^{-i\hat{H}t}\right]\end{aligned}$$

where $\hat{\rho}$ stands for the (Hermitian) initial density. Eq (S25) requires a Fourier transformation for the time correlation function $C(s,t) = \text{Tr}\left[\hat{\rho} e^{i\hat{H}t} e^{i\hat{P}s} e^{-i\hat{H}t}\right]$. To smooth the momentum distribution curves, we introduce a Gaussian damping term, $\exp(-as^2)$ with the damping factor $a$, for the Fourier transformation. We set $a = 0.05$ a.u. for the 3-state photodissociation models and $a = 0.01$ a.u. for the Tully models and the asymmetric SAC model.

All exact results for the gas phase models with one nuclear DOF are produced by Discrete Value Representation (DVR)[20]. The results of the 3-state photodissociation models are demonstrated in Figure 5 of the main text, while those of the Tully models and the asymmetric SAC model are presented in Figures S1-S4 of Section S2.

**S1-F: Linear Vibronic Coupling Models**

The Hamiltonian of the linear vibronic coupling model (LVCM) in the diabatic representation reads

$$\hat{H} = \sum_{k=1}^{N} \frac{\omega_k}{2}\left(\hat{\tilde{P}}_k^2 + \hat{\tilde{R}}_k^2\right) + \sum_{n=1}^{F}\left(E_n + \sum_{k=1}^{N} \kappa_k^{(n)} \hat{\tilde{R}}_k\right)|n\rangle\langle n| + \sum_{n \neq m}^{F}\left(\sum_{k=1}^{N} \lambda_k^{(nm)} \hat{\tilde{R}}_k\right)|n\rangle\langle m|, \text{(S26)}$$

where $\hat{\tilde{P}}_k$ and $\hat{\tilde{R}}_k$ $(k=1,...,N)$ are dimensionless weighted normal-mode momentum and coordinate of the $k$-th nuclear DOF, respectively, with the corresponding frequency $\omega_k$. $E_n$ $(n=1,...,F)$ represents the vertical excitation energy of the $n$-th state. $\kappa_k^{(n)}$ and $\lambda_k^{(nm)}$ are linear coupling terms of the $k$-th nuclear DOF for the corresponding diagonal and off-diagonal Hamiltonian elements, respectively.

We first employ two typical 2-state LVCMs for pyrazine—the 3-mode model of Schneiders and Domcke[21] and the 24-mode model of Krempl *et al*[22]. The non-zero parameters of these two models are listed in Tables S2 and S3.

**Table S2**. Parameters of 3-mode LVCM of Pyrazine (Unit: eV)[21]

| | |
|---|---|
| $E_1$, $E_2$ | 3.94, 4.84 |
| $\kappa_1^{(1)}$, $\kappa_2^{(1)}$, $\kappa_1^{(2)}$, $\kappa_2^{(2)}$ | 0.037, -0.105, -0.254, 0.149 |
| $\lambda_3^{(12)}$ | 0.262 |
| $\omega_1$, $\omega_2$, $\omega_3$ | 0.126, 0.074, 0.118 |

**Table S3**. Parameters of 24-mode LVCM of Pyrazine (Unit: eV)[22]

| | |
|---|---|
| $E_1$, $E_2$ | -0.4617, 0.4617 |
| $\lambda_1^{(12)}$ | 0.1825 |

| Mode | $\omega$ | $\kappa^{(1)}$ | $\kappa^{(2)}$ |
|---|---|---|---|
| 1 | 0.0936 | | |
| 2 | 0.074 | -0.0964 | 0.1194 |
| 3 | 0.1273 | 0.0470 | 0.2012 |
| 4 | 0.1568 | 0.1594 | 0.0484 |
| 5 | 0.1347 | 0.0308 | -0.0308 |
| 6 | 0.3431 | 0.0782 | -0.0782 |
| 7 | 0.1157 | 0.0261 | -0.0261 |
| 8 | 0.3242 | 0.0717 | -0.0717 |
| 9 | 0.3621 | 0.0780 | -0.0780 |
| 10 | 0.2673 | 0.0560 | -0.0560 |
| 11 | 0.3052 | 0.0625 | -0.0625 |
| 12 | 0.0968 | 0.0188 | -0.0188 |
| 13 | 0.0589 | 0.0112 | -0.0112 |
| 14 | 0.0400 | 0.0069 | -0.0069 |
| 15 | 0.1726 | 0.0265 | -0.0265 |
| 16 | 0.2863 | 0.0433 | -0.0433 |
| 17 | 0.2484 | 0.0361 | -0.0361 |
| 18 | 0.1536 | 0.0210 | -0.0210 |
| 19 | 0.2105 | 0.0281 | -0.0281 |
| 20 | 0.0778 | 0.0102 | -0.0102 |
| 21 | 0.2294 | 0.0284 | -0.0284 |
| 22 | 0.1915 | 0.0196 | -0.0196 |
| 23 | 0.4000 | 0.0306 | -0.0306 |
| 24 | 0.3810 | 0.0269 | -0.0269 |

We also test the LVCM proposed by Worth and coworkers[23] in the Perspective. The model includes 2 nuclear normal modes and 3 electronic states to describe the dynamics of $Cr(CO)_5$ through a Jahn–Teller conical intersection after photodissociation. The non-zero parameters for this model are listed in Table S4.

**Table S4.** Parameters of 2-mode LVCM of Cr(CO)$_5$ (Unit: eV)[23]

| | |
|---|---|
| $E_1$, $E_2$, $E_3$ | 0.0424, 0.0424, 0.4344 |
| $\kappa_2^{(1)}$, $\kappa_2^{(2)}$ | -0.0328, 0.0328 |
| $\lambda_1^{(12)}$, $\lambda_1^{(23)}$, $\lambda_2^{(13)}$ | 0.0328, -0.0978, -0.0978 |
| $\omega_1$, $\omega_2$ | 0.0129, 0.0129 |

When we study the models of the pyrazine molecule, the second diabatic (electronic) state is initially occupied, and the nuclear variables are sampled from the Wigner distribution of the vibrational ground state:

$$\rho_W\left(\overline{\mathbf{R}}, \overline{\mathbf{P}}\right) \propto \exp\left[-\sum_{k=1}^{N}\left(\overline{R}_k^2 + \overline{P}_k^2\right)\right] . \qquad (S27)$$

When we test the LVCM of Cr(CO)$_5$, the second diabatic (electronic) state is occupied at the beginning. The initial nuclear wavefunction leads to the corresponding Wigner distribution

$$\rho_W\left(\overline{\mathbf{R}}, \overline{\mathbf{P}}\right) \propto \exp\left[-\sum_{k=1}^{2}\left(\frac{\left(\overline{R}_k - r_k\right)^2}{2\alpha_k^2} + 2\alpha_k^2 \overline{P}_k^2\right)\right] . \qquad (S28)$$

where $r_1 = 0$, $r_2 = 14.3514$, $\alpha_1 = 0.4501$ and $\alpha_2 = 0.4586$. In simulations of LVCMs, trajectories are evolved in the adiabatic representation, where the kinematic nuclear momentum $\{P_k\}$ (or equivalently the mapping diabatic nuclear momentum) is used. The canonical (mass-weighted) nuclear coordinate and its corresponding canonical momentum $\{R_k, P_k\}$ in the diabatic representation can be obtained from the dimensionless nuclear variables $\{\overline{R}_k, \overline{P}_k\}$ by the relation

$$\bar{R}_k = \sqrt{\omega_k} R_k, \quad \bar{P}_k = P_k / \sqrt{\omega_k} . \tag{S29}$$

We study both the electronic and nuclear dynamics of the LVCMs. The former is shown by the time-dependent electronic population, and the latter is demonstrated by the mean value of the nuclear coordinate and that of the nuclear momentum as functions of time:

$$\langle \bar{R}_k(t) \rangle = \mathrm{Tr}\left[ \hat{\rho} e^{i\hat{H}t} \hat{\bar{R}}_k e^{-i\hat{H}t} \right] , \tag{S30}$$

$$\langle \bar{P}_k(t) \rangle = \mathrm{Tr}\left[ \hat{\rho} e^{i\hat{H}t} \hat{\bar{P}}_k e^{-i\hat{H}t} \right] . \tag{S31}$$

Here $\hat{\rho}$ denotes the initial density operator for both nuclear and electronic DOFs. For simplicity, we only demonstrate one nuclear DOF for each model. The results of the normal mode $v_{6a}$ of pyrazine models (that is, the second normal mode in Tables S2-S3 with the corresponding frequency $\omega_2 = 0.074$ eV) and those of the second normal mode of the LVCM of Cr(CO)$_5$ are presented. We perform MCTDH calculations for two pyrazine models by using the Heidelberg MCTDH package (V8.5)[24], while the MCTDH results of the Cr(CO)$_5$ model are taken from ref [23].

**S1-G: Additional Details in the Simulations**

In the tests of all trajectory-based methods, independent trajectories evolve in the adiabatic representation. When we study the spin-boson models, FMO model, atom-in-cavity models, SF model, three-state photodissociation models, and LVCMs, we have to perform the adiabatic-to-diabatic transformation to yield results for the time correlation functions in the diabatic representation, because numerically exact results are available only in the diabatic representation. We find that a significantly smaller time step size is required for several models (e.g., FMO and SF models). A more efficient approach is to evolve the electronic DOFs in the diabatic representation while concurrently evolving the nuclear DOFs in the adiabatic representation. (See details of the integrator of NAF in Section S3.) The time step size and the number of trajectories for each model are listed in Table S5.

Table S5. The time step and the number of trajectories for each model.

| Model | Time step size | Number of trajectories |
|---|---|---|
| Spin-boson models | 0.01 a.u. | $10^5$ |
| FMO model | 0.1 fs | $10^5$ |
| Atom-in-cavity models | 0.1 a.u. | $10^5$ |
| SF model | 0.001 fs | 24000 |
| 3-state photodissociation models | 0.01 fs | $10^5$ |
| Tully models | 0.01 fs | $10^5$ |
| Asymmetric SAC model | 0.01 fs | $10^5$ |
| LVCMs | 0.01 fs | $10^5$ |

## S2. Additional Results for Models in the Main Text

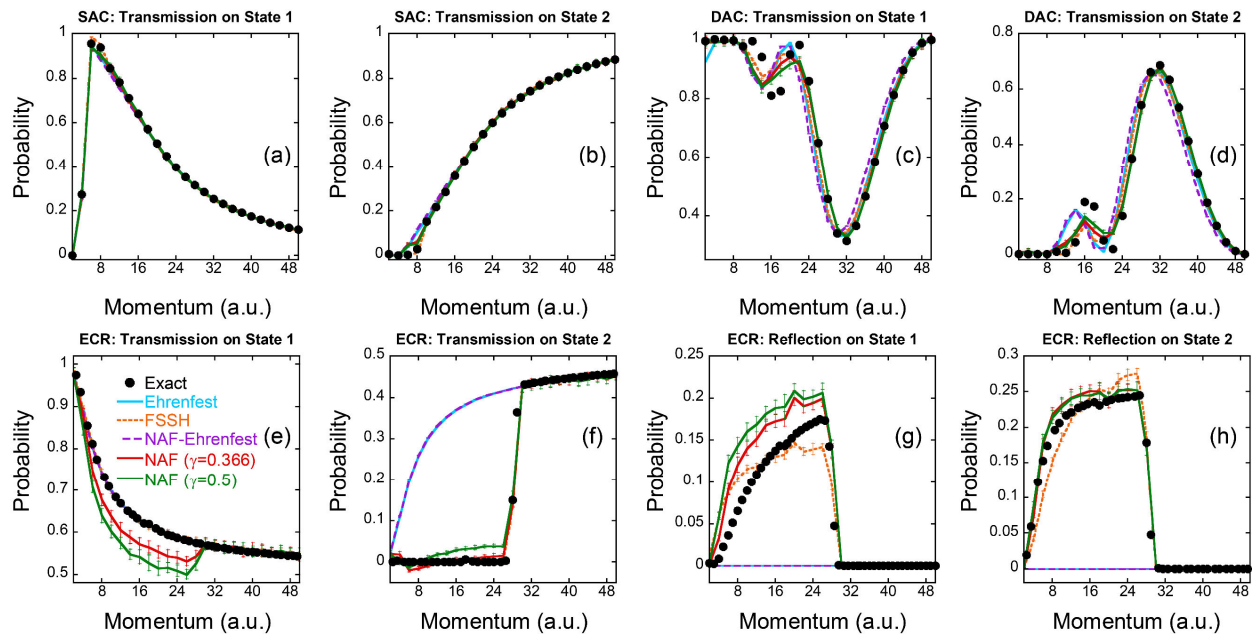

**Figure S1**. Results of the scattering probability of the three Tully models as a function of the initial momentum for each channel. Panels (a), (c), and (e) denote the transmission probabilities on the adiabatic ground state of the SAC, DAC, and ECR models, respectively. Panels (b), (d), and (f) are the same as Panels (a), (c), and (e), respectively, but for the transmission probabilities on adiabatic excited states. Panels (g)-(h) are the same as Panels (e)-(f), respectively, but for the reflection probabilities of the ECR model. Black points: Exact results by DVR. Cyan solid lines: Ehrenfest dynamics. Orange short-dashed lines: FSSH. Purple long-dashed lines: NAF-Ehrenfest. Red and green solid lines: NAF ($\gamma = 0.366$) and NAF ($\gamma = 0.5$), respectively.

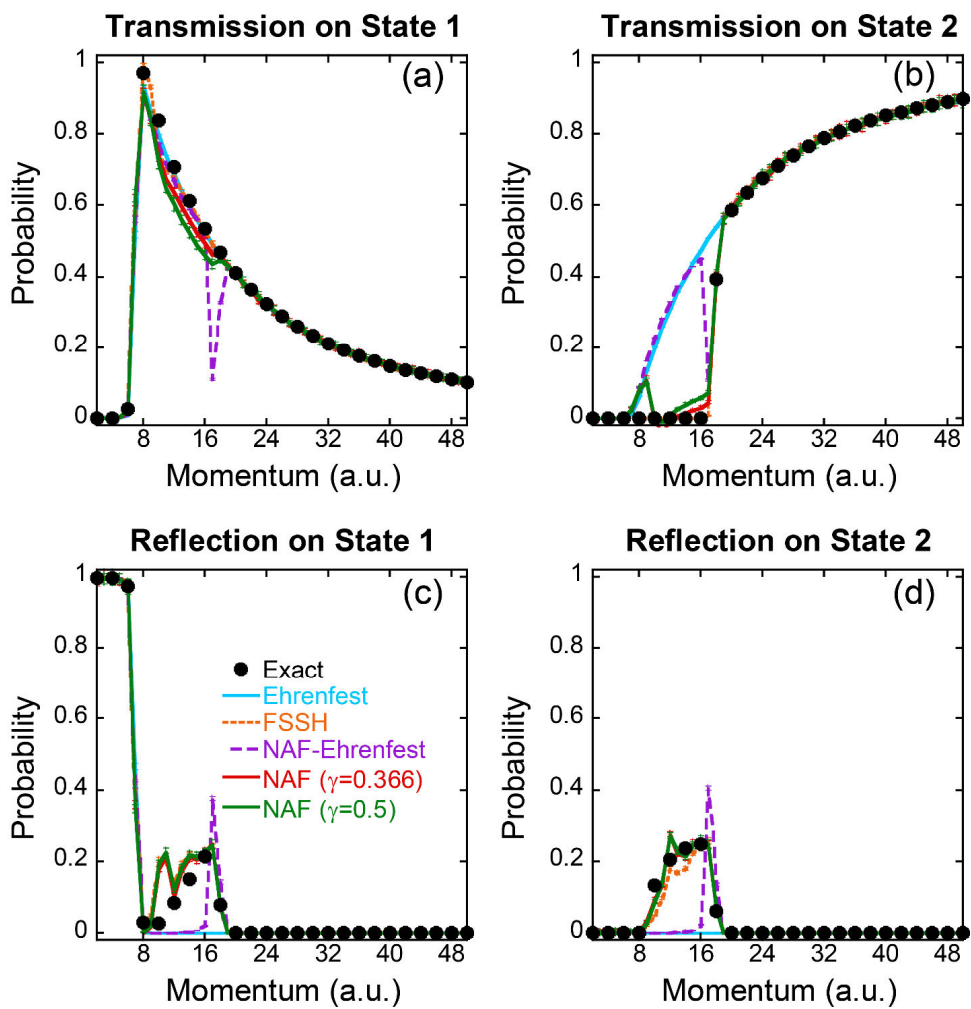

**Figure S2**. Results of the scattering probability of the asymmetric SAC model as a function of the initial momentum for each channel. Panels (a)-(b) denote the transmission probabilities on the adiabatic ground and excited states, respectively. Panels (c)-(d) are the same as Panels (a)-(b), respectively, but for the reflection probabilities. Black points: Exact results by DVR. Cyan solid lines: Ehrenfest dynamics. Orange short-dashed lines: FSSH. Purple long-dashed lines: NAF-Ehrenfest. Red and green solid lines: NAF with $\gamma = 0.366$ and NAF with $\gamma = 0.5$, respectively.

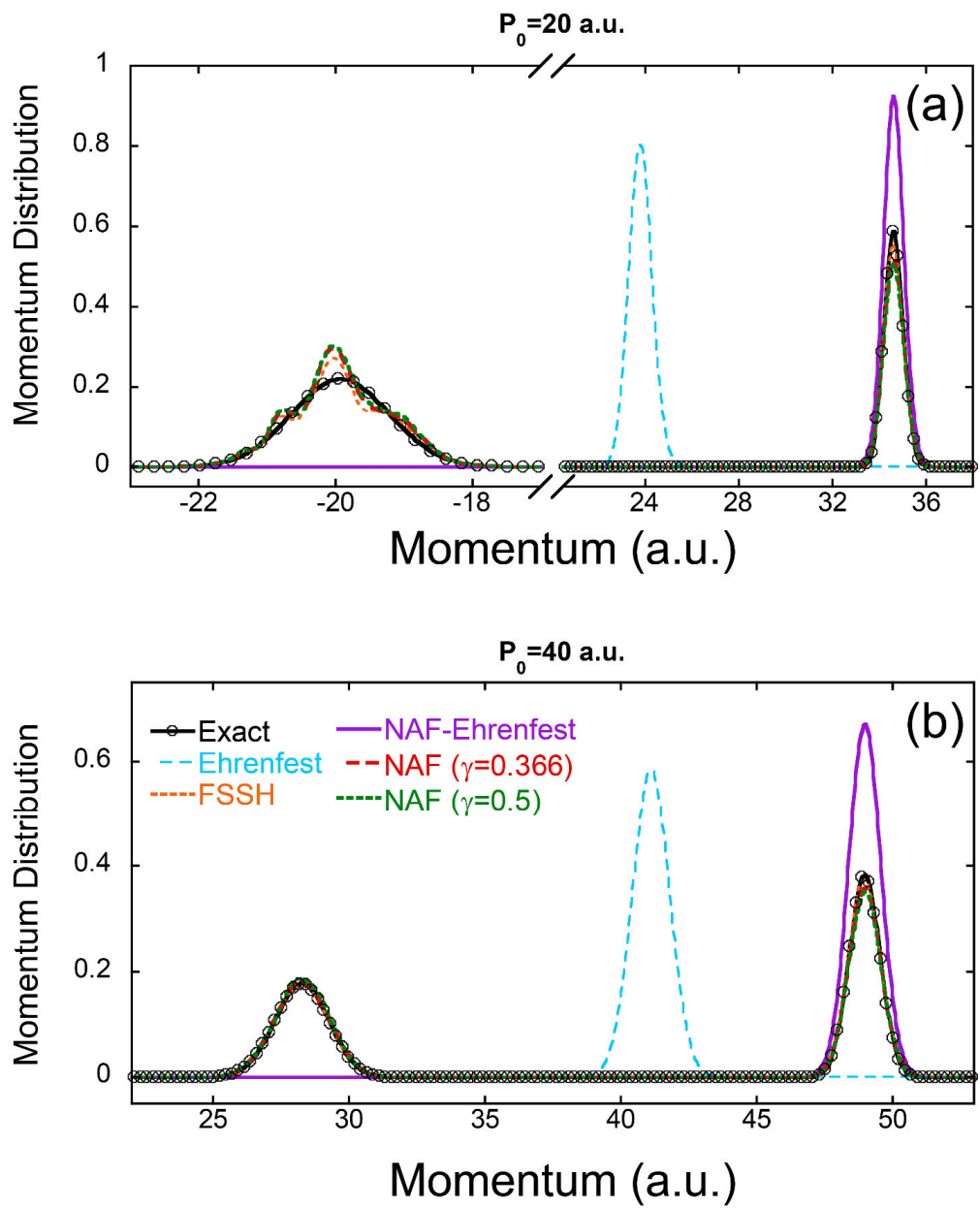

**Figure S3**. Nuclear momentum distribution of the ECR model after scattering. Panels (a)-(b) denote the results of the initial momentum $P_0 = 20$ a.u. and $P_0 = 40$ a.u., respectively. Black solid lines with black circles: Exact results by DVR. Cyan long-dashed lines: Ehrenfest dynamics. Orange short-dashed lines: FSSH. Purple solid lines: NAF-Ehrenfest. Red long-dashed lines: NAF ($\gamma = 0.366$). Green short-dashed lines: NAF ($\gamma = 0.5$).

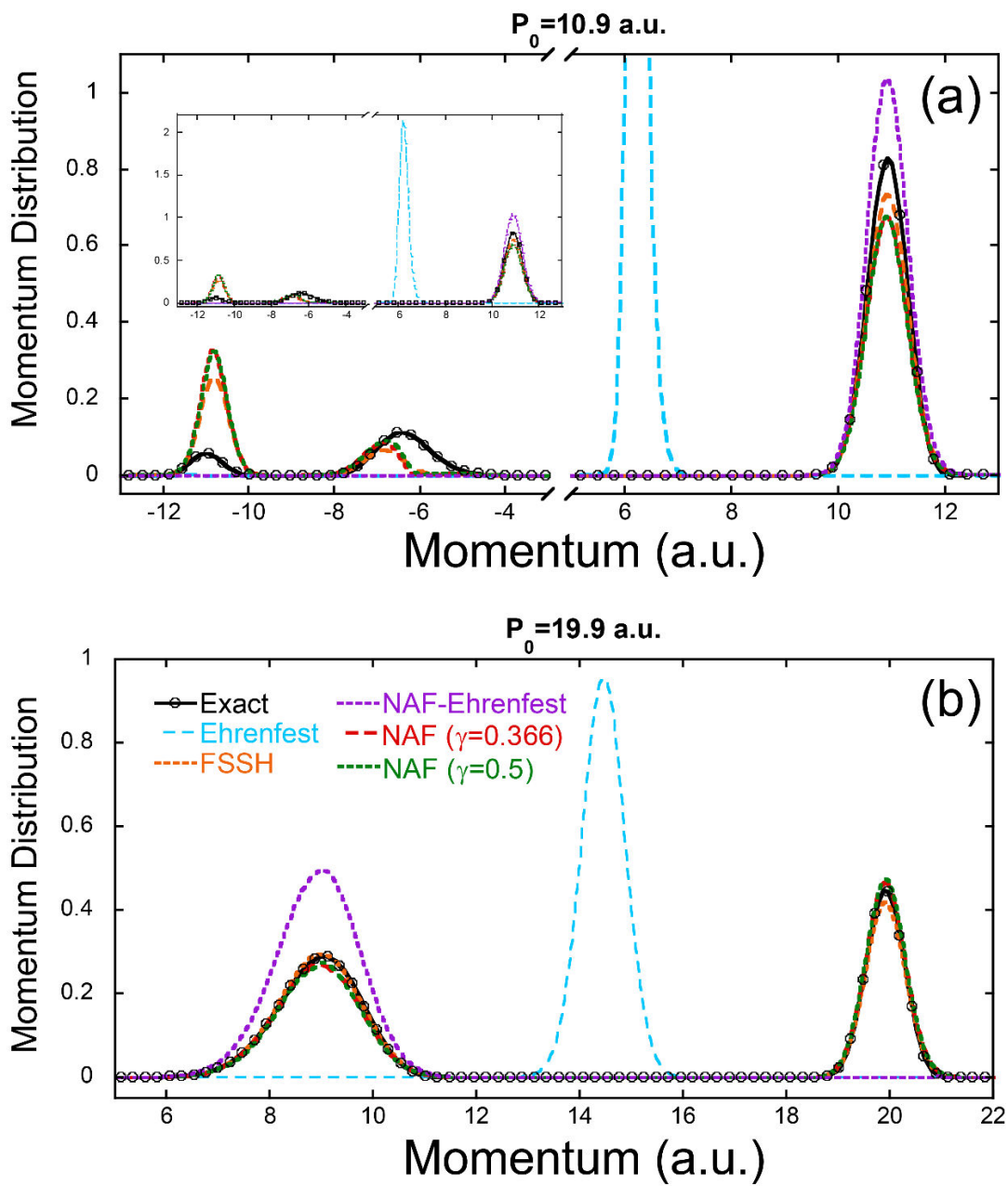

**Figure S4**. Nuclear momentum distribution of the asymmetric SAC model after scattering. Panels (a)-(b) denote the results of the initial momentum $P_0 = 10.9$ a.u. and $P_0 = 19.9$ a.u., respectively. Black solid lines with black circles: Exact results by DVR. Cyan long-dashed lines: Ehrenfest dynamics. Orange short-dashed lines: FSSH. Purple short-dashed lines: NAF-Ehrenfest. Red long-dashed lines: NAF ($\gamma = 0.366$). Green short-dashed lines: NAF ($\gamma = 0.5$).

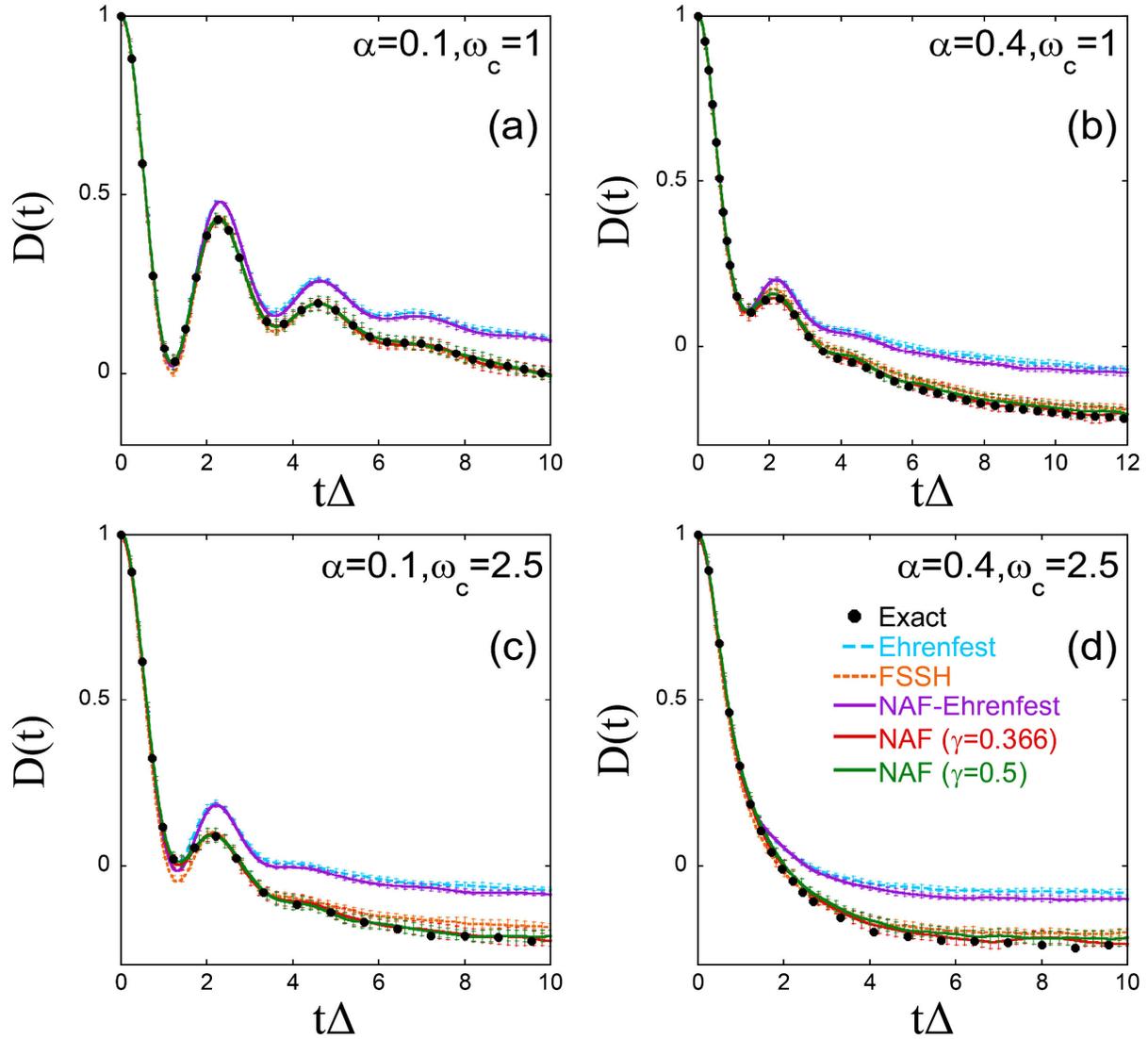

**Figure S5**. Results for spin-boson models which are identical to Panels (a)-(d) of Figure 1 in the main text but with a higher temperature $\beta = 0.25$.

## S3. Integrator of NAF for a Finite Time Step

The integrator of the EOMs of NAF is described as follows.

1. Update the nuclear kinematic momentum (equivalently, the diabatic nuclear momentum) within a half time step $\Delta t/2$

$$\mathbf{P}_{t+\Delta t/2} \leftarrow \mathbf{P}_t - \left( \nabla_{\mathbf{R}} E_{j_{old}}(\mathbf{R}_t) + \sum_{k \neq l} \left[ (E_k(\mathbf{R}_t) - E_l(\mathbf{R}_t)) \mathbf{d}_{lk}(\mathbf{R}_t) \right] \tilde{\rho}_{kl}(t) \right) \frac{\Delta t}{2} . \quad (S32)$$

2. Update the nuclear coordinate within a full-time step $\Delta t$

$$\mathbf{R}_{t+\Delta t} \leftarrow \mathbf{R}_t + \mathbf{M}^{-1} \mathbf{P}_{t+\Delta t/2} \Delta t . \quad (S33)$$

3. Update phase variables of electronic DOFs within a full-time step $\Delta t$ according to

$$\tilde{\mathbf{g}}_{t+\Delta t} \leftarrow \tilde{\mathbf{U}}(\mathbf{R}_{t+\Delta t}, \mathbf{P}_{t+\Delta t/2}; \Delta t) \tilde{\mathbf{g}}_t . \quad (S34)$$

$$\tilde{\mathbf{\Gamma}}_{t+\Delta t} \leftarrow \tilde{\mathbf{U}}(\mathbf{R}_{t+\Delta t}, \mathbf{P}_{t+\Delta t/2}; \Delta t) \tilde{\mathbf{\Gamma}}_t \tilde{\mathbf{U}}^{\dagger}(\mathbf{R}_{t+\Delta t}, \mathbf{P}_{t+\Delta t/2}; \Delta t) . \quad (S35)$$

4. Determine a new occupied state $j_{new}$ based on the statements in the main text and rescale $\mathbf{P}$ if $j_{new} \neq j_{old}$,

$$\mathbf{P}_{t+\Delta t/2} \leftarrow \mathbf{P}_{t+\Delta t/2} \sqrt{\left( H_{NAF}(\mathbf{R}_{t+\Delta t}, \mathbf{P}_{t+\Delta t/2}, \tilde{\mathbf{x}}_{t+\Delta t}, \tilde{\mathbf{p}}_{t+\Delta t}) - E_{j_{new}}(\mathbf{R}_{t+\Delta t}) \right) / \left( \mathbf{P}^T_{t+\Delta t/2} \mathbf{M}^{-1} \mathbf{P}_{t+\Delta t/2} / 2 \right)} . \quad (S36)$$

If $H_{NAF}(\mathbf{R}_{t+\Delta t}, \mathbf{P}_{t+\Delta t/2}, \tilde{\mathbf{x}}_{t+\Delta t}, \tilde{\mathbf{p}}_{t+\Delta t}) < E_{j_{new}}(\mathbf{R}_{t+\Delta t})$, the switching of the adiabatic nuclear force component is frustrated. In such a case we keep $j_{new} = j_{old}$ and the rescaling step eq (S36) is skipped.

5. Update the nuclear kinematic momentum within the other half time step $\Delta t/2$

$$\mathbf{P}_{t+\Delta t} \leftarrow \mathbf{P}_{t+\Delta t/2} - \left( \nabla_{\mathbf{R}} E_{j_{new}} (\mathbf{R}_{t+\Delta t}) + \sum_{k \neq l} \left[ (E_k(\mathbf{R}_{t+\Delta t}) - E_l(\mathbf{R}_{t+\Delta t})) \mathbf{d}_{lk}(\mathbf{R}_{t+\Delta t}) \right] \tilde{\rho}_{kl}(t+\Delta t) \right) \frac{\Delta t}{2} . \quad (S37)$$

6. Rescale the nuclear kinematic momentum **P** again to satisfy the mapping energy conservation

$$\mathbf{P}_{t+\Delta t} \leftarrow \mathbf{P}_{t+\Delta t} \sqrt{\left( H_{NAF}(\mathbf{R}_0, \mathbf{P}_0, \tilde{\mathbf{x}}_0, \tilde{\mathbf{p}}_0) - E_{j_{new}}(\mathbf{R}_{t+\Delta t}) \right) / \left( \mathbf{P}_{t+\Delta t}^T \mathbf{M}^{-1} \mathbf{P}_{t+\Delta t} / 2 \right)} . \quad (S38)$$

If $H_{NAF}(\mathbf{R}_0, \mathbf{P}_0, \tilde{\mathbf{x}}_0, \tilde{\mathbf{p}}_0) < E_{j_{new}}(\mathbf{R}_{t+\Delta t})$, it indicates that the time step size $\Delta t$ is relatively large for the integrator from time $t$ to time $t + \Delta t$. In such a case, one should then choose a smaller time step size $\Delta t$ and repeat Steps 1-6 for the update of $(\mathbf{R}_{t+\Delta t}, \mathbf{P}_{t+\Delta t}, \tilde{\mathbf{x}}_{t+\Delta t}, \tilde{\mathbf{p}}_{t+\Delta t})$ from $(\mathbf{R}_t, \mathbf{P}_t, \tilde{\mathbf{x}}_t, \tilde{\mathbf{p}}_t)$. The time step size $\Delta t$ should be adjusted in the region where the sum of adiabatic and nonadiabatic nuclear force terms is large.

As described in Section S1-G, it is sometimes more efficient to evolve the electronic mapping variables in the diabatic representation, where Step 3 of the integrator above is replaced by

$$\mathbf{g}_t \leftarrow \mathbf{T}(\mathbf{R}_t) \tilde{\mathbf{g}}_t . \quad (S39)$$

$$\mathbf{\Gamma}_t \leftarrow \mathbf{T}(\mathbf{R}_t) \tilde{\mathbf{\Gamma}}_t \mathbf{T}^\dagger(\mathbf{R}_t) . \quad (S40)$$

$$\mathbf{g}_{t+\Delta t} \leftarrow \exp\left[-i\Delta t \mathbf{V}(\mathbf{R}_{t+\Delta t})\right] \mathbf{g}_t . \quad (S41)$$

$$\mathbf{\Gamma}_{t+\Delta t} \leftarrow \exp\left[-i\Delta t \mathbf{V}(\mathbf{R}_{t+\Delta t})\right] \mathbf{\Gamma}_t \exp\left[i\Delta t \mathbf{V}(\mathbf{R}_{t+\Delta t})\right] . \quad (S42)$$

$$\tilde{\mathbf{g}}_{t+\Delta t} \leftarrow \mathbf{T}^\dagger(\mathbf{R}_{t+\Delta t}) \mathbf{g}_{t+\Delta t} . \quad (S43)$$

$$\tilde{\mathbf{\Gamma}}_{t+\Delta t} \leftarrow \mathbf{T}^\dagger(\mathbf{R}_{t+\Delta t}) \mathbf{\Gamma}_{t+\Delta t} \mathbf{T}(\mathbf{R}_{t+\Delta t}) . \quad (S44)$$

Here $\mathbf{T}(\mathbf{R})$ is the diabatic-to-adiabatic transformation matrix with elements $T_{nm}(\mathbf{R}) = \langle n | \phi_m(\mathbf{R}) \rangle$. Similarly, such a strategy can be applied to Ehrenfest dynamics and FSSH. In the diabatic-to-

adiabatic transformation, especially in the coupling region, we carefully trace both the sign and order of the adiabatic basis $|\phi_n(\mathbf{R}_{t+\Delta t})\rangle$ based on the values from the previous time step. This guarantees that the value of $\langle \phi_n(\mathbf{R}_{t+\Delta t}) | \phi_m(\mathbf{R}_t) \rangle - \delta_{nm}$ remains small enough to make the adiabatic basis change smoothly.

## S4. Comparisons of NAF and NAF(S) Results

We demonstrate the results of NAF with a stochastically selected single-state adiabatic nuclear force component, denoted as NAF(S). In NAF(S), the probability of choosing the single-state adiabatic nuclear force contributed by the $j$-th (adiabatic) state at time $t$ is $|\tilde{\rho}_{jj}(t)| / \sum_{k=1}^{F} |\tilde{\rho}_{kk}(t)|$. Figures S6-S8 illustrate comparisons of NAF and NAF(S) results for the spin-boson models, FMO model, and SF model, respectively. The overall performance of NAF(S) for these models is similar to that of NAF.

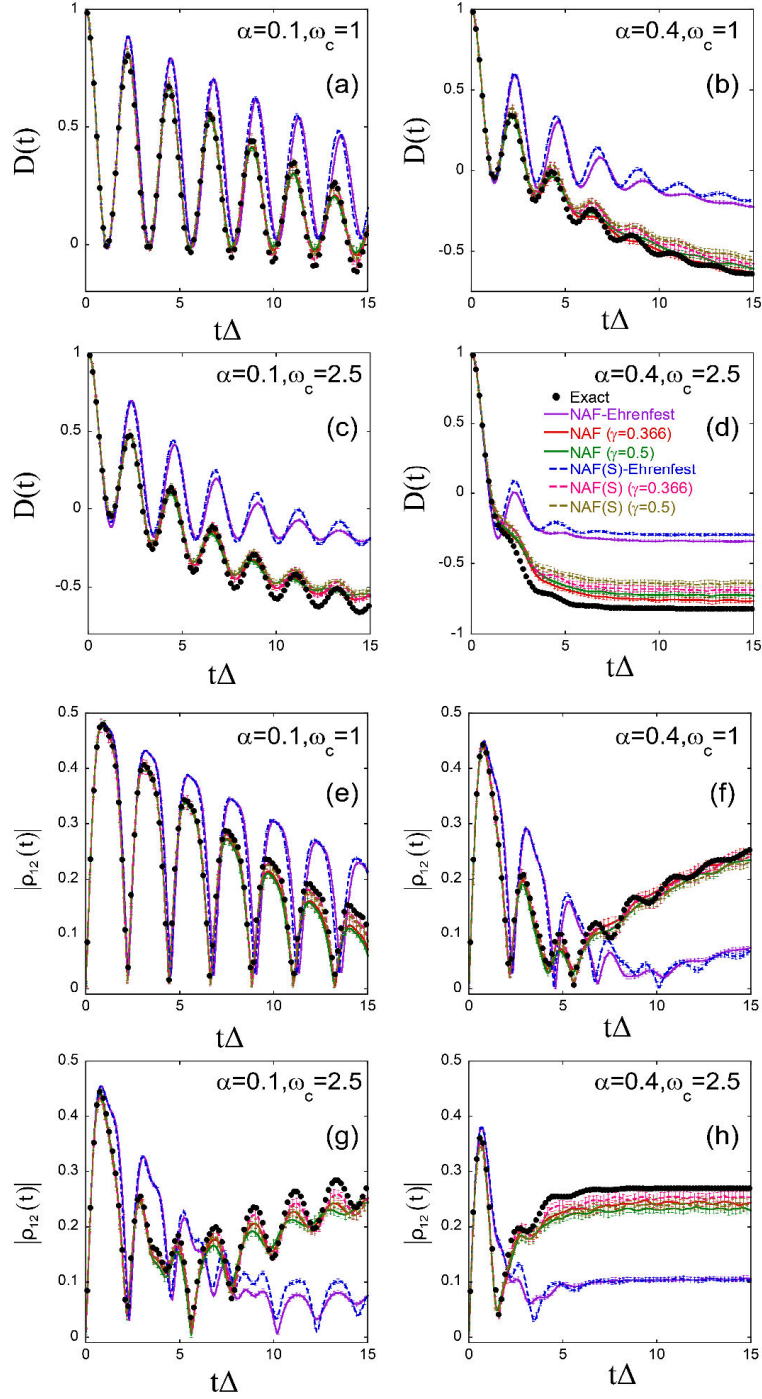

**Figure S6**. Each panel is the same as that in Figure 1 of the main text, but with the comparison between NAF and NAF(S). Black points: Exact results produced by eHEOM. Purple, red, and green solid lines: NAF-Ehrenfest, NAF ($\gamma = 0.366$), and NAF ($\gamma = 0.5$), respectively. Blue, magenta, and brown dashed lines: NAF(S)-Ehrenfest, NAF(S) ($\gamma = 0.366$) and NAF(S) ($\gamma = 0.5$), respectively.

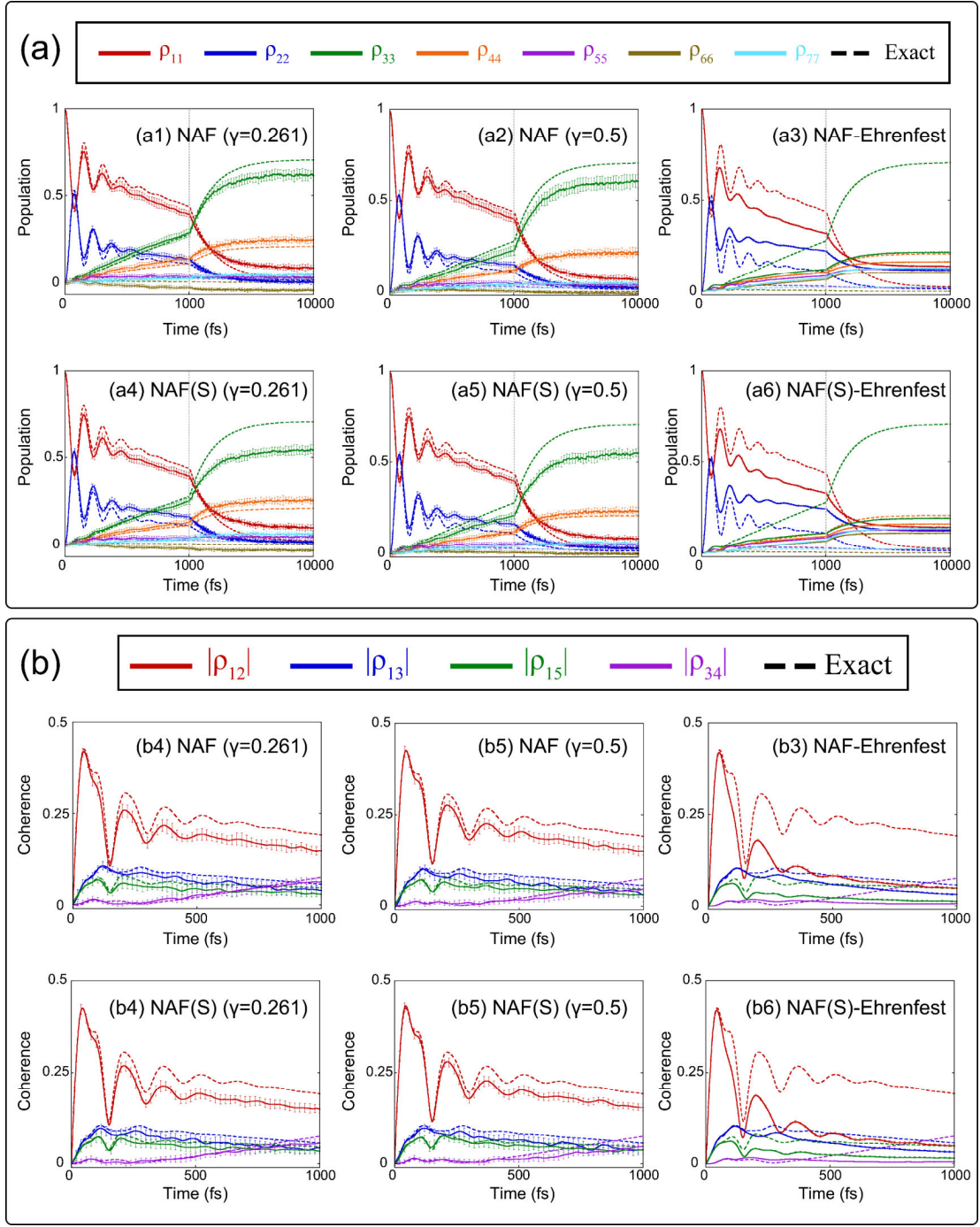

**Figure S7**. Similar to Figure 2 in the main text, but Panels (a1)-(a6) represent the results of NAF ($\gamma = 0.261$), NAF ($\gamma = 0.5$), NAF-Ehrenfest, NAF(S) ($\gamma = 0.261$), NAF(S) ($\gamma = 0.5$) and NAF(S)-Ehrenfest, respectively. Panel (b) is the same as Panel (a) but for the coherence terms.

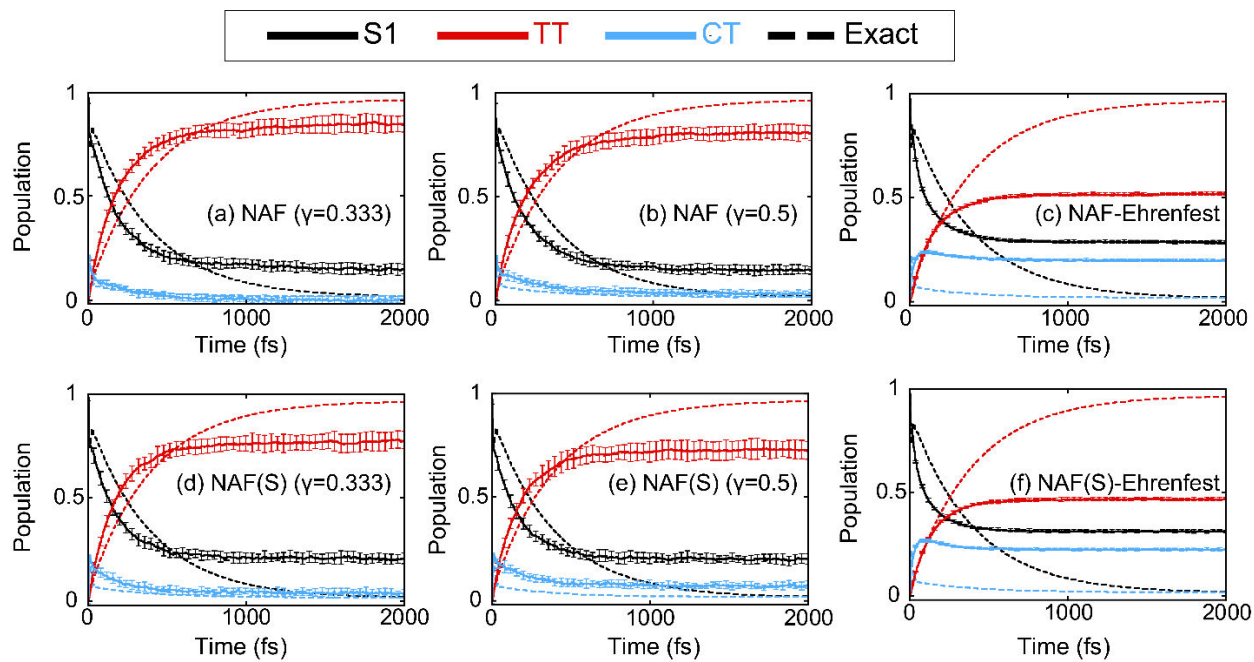

**Figure S8**. Similar to Figure 4 of the main text, but Panels (a)-(f) represent the results of NAF ($\gamma = 0.333$), NAF ($\gamma = 0.5$), NAF-Ehrenfest, NAF(S) ($\gamma = 0.333$), NAF(S) ($\gamma = 0.5$) and NAF(S)-Ehrenfest, respectively.

## S5. Comparisons of NAF and CMMcv Results

Figure S9 compares NAF to CMMcv[10] for the 3-state photodissociation models. In the CMMcv simulation, a hard wall potential $U(R) = \begin{cases} 0, & R > 0 \\ \infty, & R \leq 0 \end{cases}$ is added to the original Morse potential energy surfaces. That is, when $R \leq 0$ and $P \leq 0$, we let $P \leftarrow -P$ for each trajectory. The hard wall potential is to prevent the bond length $R$ from being negative, which is unphysical. The strategy was employed in our previous CMM/CMMcv investigation of the same models[10]. In the NAF simulation, it is not necessary to add such a hard wall potential.

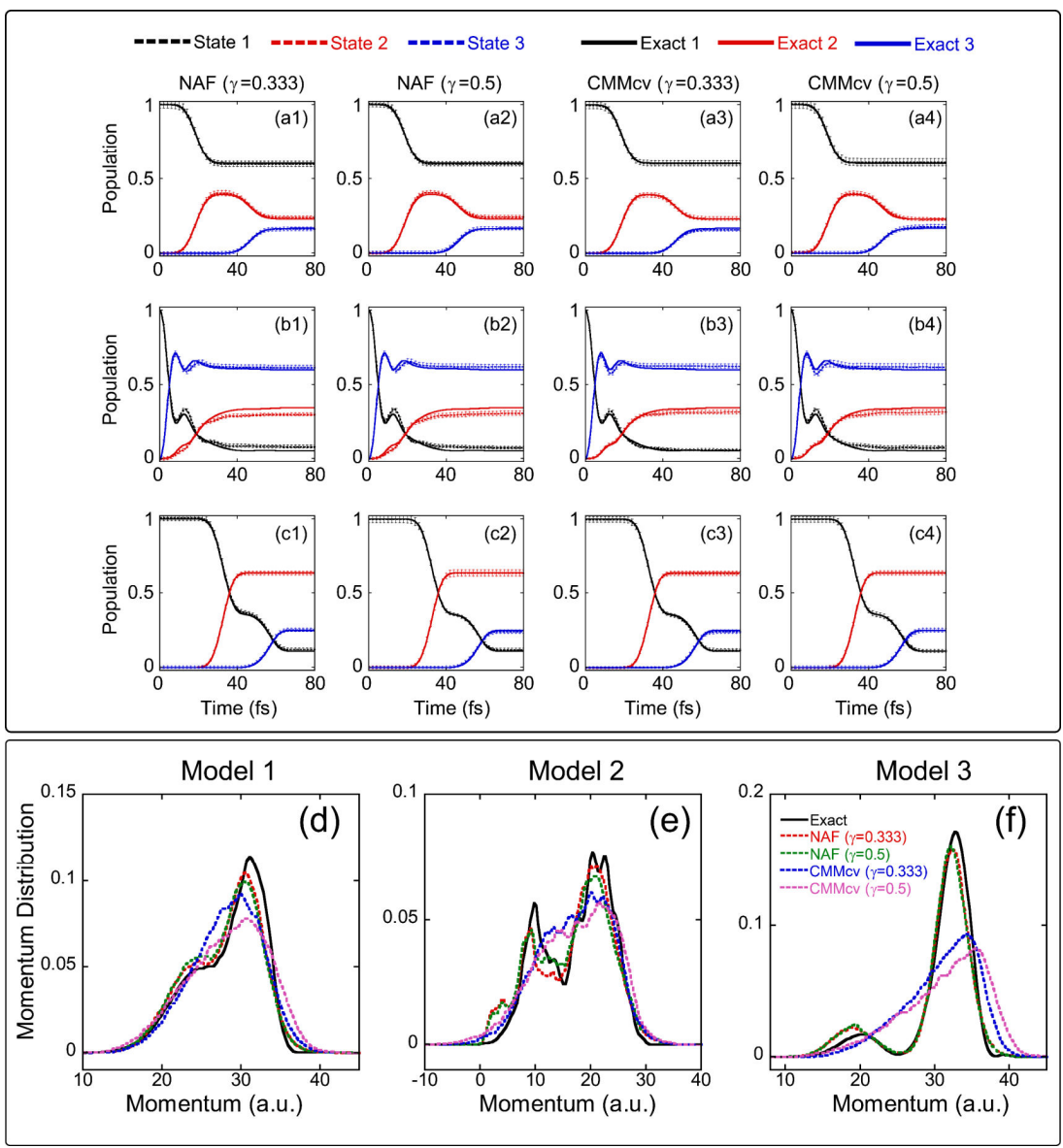

**Figure S9**. Similar to Figure 5 in the main text, but the first to fourth columns in Panels (a)-(c) represent the results of NAF ($\gamma = 0.333$), NAF ($\gamma = 0.5$), CMMcv ($\gamma = 0.333$), and CMMcv ($\gamma = 0.5$), respectively. In Panels (d)-(f), the blue and pink dashed lines denote the results of CMMcv ($\gamma = 0.333$) and CMMcv ($\gamma = 0.5$), respectively. A hard wall potential is applied in the CMMcv simulation to avoid the negative bond length for the three models.

## S6. Comparisons of GDTWA and NAF-GDTWA Results

As mentioned in the main text, in ref [25] we show that the mathematical structure of the mapping constraint (coordinate-momentum) phase space (CPS) of our recent work[3, 10, 26-29] is related to the quotient space $\mathrm{U}(F)/\mathrm{U}(F-r)$, namely the complex Stiefel manifold[25, 30, 31]. Here, $1 \leq r < F$. It is straightforward to show the phase space of the generalized discrete truncated Wigner approximation (GDTWA)[32] developed by Lang *et al.* can be a discrete subset of the manifold $\mathrm{U}(F)/\mathrm{U}(F-2)$. The electronic mapping kernel of GDTWA follows the form given in eq (7) of the main text and is identical to its inverse mapping kernel. The initial condition of the electronic mapping kernel reads

$$\hat{K}_{ele} = \begin{bmatrix} 0 & \cdots & 0 & \frac{1}{\sqrt{2}}e^{-i\theta_1} & 0 & \cdots & 0 \\ \vdots & \ddots & \vdots & \vdots & \vdots & \ddots & \vdots \\ 0 & \cdots & 0 & \frac{1}{\sqrt{2}}e^{-i\theta_{j_{occ}-1}} & 0 & \cdots & 0 \\ \frac{1}{\sqrt{2}}e^{i\theta_1} & \cdots & \frac{1}{\sqrt{2}}e^{i\theta_{j_{occ}-1}} & 1 & \frac{1}{\sqrt{2}}e^{i\theta_{j_{occ}+1}} & \cdots & \frac{1}{\sqrt{2}}e^{i\theta_F} \\ 0 & \cdots & 0 & \frac{1}{\sqrt{2}}e^{-i\theta_{j_{occ}+1}} & 0 & \cdots & 0 \\ \vdots & \ddots & \vdots & \vdots & \vdots & \ddots & \vdots \\ 0 & \cdots & 0 & \frac{1}{\sqrt{2}}e^{-i\theta_F} & 0 & \cdots & 0 \end{bmatrix}, \quad (S45)$$

where $j_{occ}$ denotes the index of the initially occupied state, and each $\theta_n\,(n=1,\cdots,F)$ variable is uniformly sampled from $\{\pi/4,\ 3\pi/4,\ 5\pi/4,\ 7\pi/4\}$. When we use the EOMs of NAF with the initial condition and expression for the evaluation of time-dependent (electronic) properties of GDTWA, we obtain the NAF-GDTWA method. Figures S10-S12 present comparisons of GDTWA to NAF-GDTWA for the 3-state photodissociation models, LVCM of Cr(CO)$_5$ and FMO model, respectively.

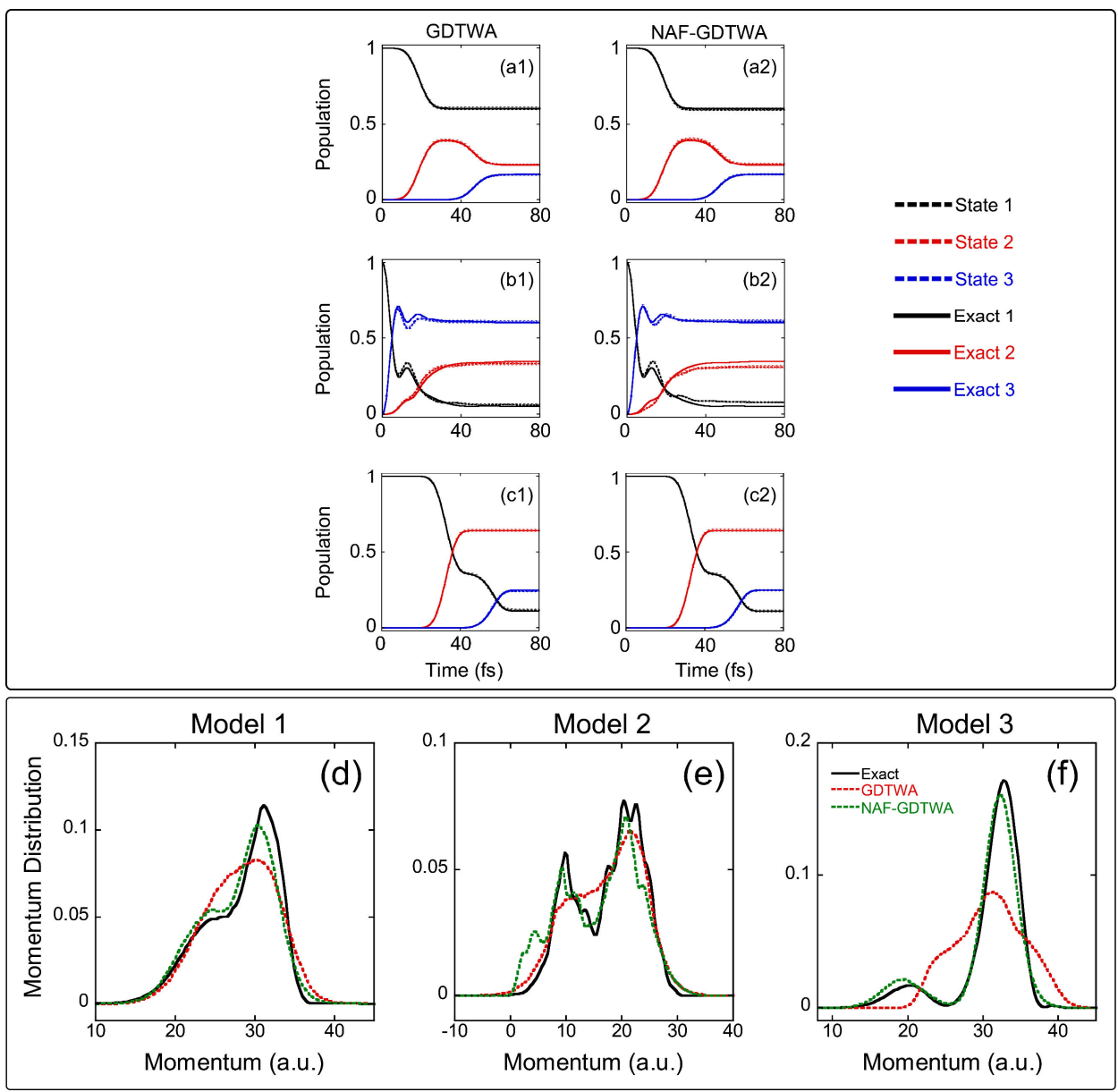

**Figure S10**. Similar to Figure 5 in the main text, but the left and right columns in Panels (a)-(c) represent the results of GDTWA and NAF-GDTWA, respectively. In Panels (d)-(f), the red and green dashed lines represent the GDTWA and NAF-GDTWA results, respectively. A hard wall potential is applied in the GDTWA simulation to avoid the negative bond length for the three models. Such a strategy is not necessary in the NAF-GDTWA simulation.

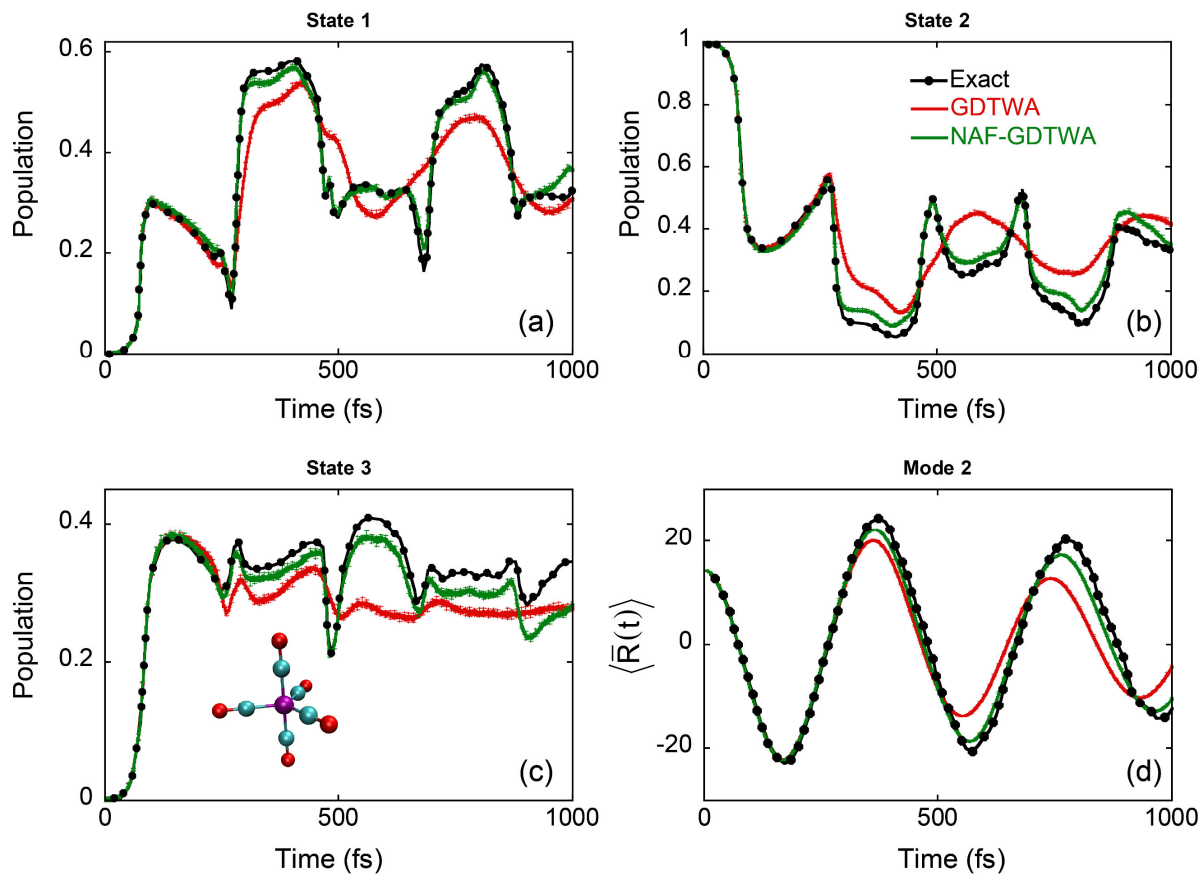

**Figure S11**. Similar to Figure 7 in the main text, but the red and green solid lines represent the GDTWA and NAF-GDTWA results, respectively.

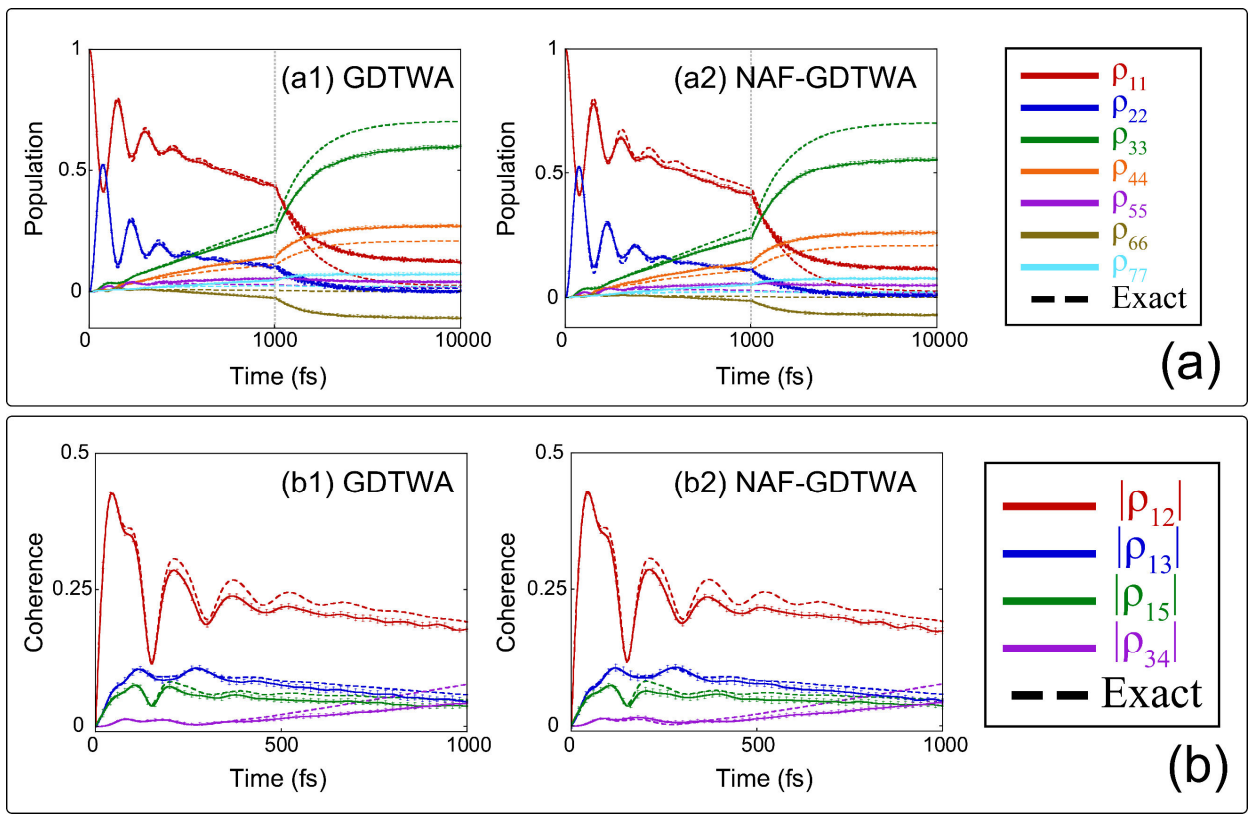

**Figure S12**. Similar to Figure 2 in the main text, but Panels (a1)-(a2) represent the GDTWA and NAF-GDTWA results, respectively. Panel (b) is the same as Panel (a) but for the coherence terms.

## S7. Comparisons of NAF, FS-NAF, and FSSH Results

In this section, we introduce the Fewest Switches NAF (FS-NAF) approach, where we use the EOMs of NAF with the initial condition and expression for the evaluation of time-dependent properties of the FSSH method. It incorporates the nonadiabatic nuclear force term in the EOMs for nuclear variables of the FSSH algorithm[17, 33]. The strategy of nuclear momentum rescaling is also employed to ensure energy conservation. The algorithm of FS-NAF reads:

1. The initial values of nuclear variables $\{\mathbf{R}_0, \mathbf{P}_0\}$ are sampled from the corresponding Wigner distribution. When the $j_{occ}$-th adiabatic state of the system $|\phi_{j_{occ}}(\mathbf{R}_0)\rangle$ is initially occupied, the initial electronic amplitude vector $\tilde{\mathbf{c}}$ is $\{\tilde{c}_n = e^{i\theta}\delta_{j_{occ}n}\}$, where $\theta$ is uniformly sampled in $[0, 2\pi)$. The index of the adiabatic state that offers the single-state adiabatic nuclear force component for FS-NAF is set as $j = j_{occ}$.

    If the $j_{occ}$-th adiabatic state of the system $|j_{occ}\rangle$ is initially occupied instead, the initial electronic amplitude vector $\tilde{\mathbf{c}}$ in the adiabatic representation is obtained by the diabatic-to-adiabatic transformation $\tilde{\mathbf{c}} = \mathbf{T}^\dagger(\mathbf{R}_0)\mathbf{c}$ with $\{c_n = e^{i\theta}\delta_{j_{occ}n}\}$. The index of the adiabatic state that provides the single-state adiabatic nuclear force component, $j$, is randomly sampled from $\{k = 1, \cdots, F\}$ according to the probability $|T_{j_{occ}k}(\mathbf{R}_0)|^2$ as suggested in ref [34].

2. Calculate the initial energy $H_0 = \mathbf{P}_0^T \mathbf{M}^{-1} \mathbf{P}_0 / 2 + E_j(\mathbf{R}_0)$. Set time $t = 0$.

3. The reduced electronic density matrix in the adiabatic representation is

$$\tilde{\rho}_{nm}(t) = \begin{cases} \delta_{nj}, & n = m \\ \tilde{c}_n(t)\tilde{c}_m^*(t), & n \neq m \end{cases}, \quad (S46)$$

while the corresponding electronic density matrix in the diabatic representation reads

$$\boldsymbol{\rho}(t) = \mathbf{T}(\mathbf{R}_t)\tilde{\boldsymbol{\rho}}(t)\mathbf{T}^\dagger(\mathbf{R}_t) \quad . \tag{S47}$$

4. Update the nuclear kinematic momentum (equivalently, the diabatic nuclear momentum) within a half time step $\Delta t/2$

$$\mathbf{P}_{t+\Delta t/2} \leftarrow \mathbf{P}_t - \left(\nabla_\mathbf{R} E_j(\mathbf{R}_t) + \sum_{n \neq m}^F \left[(E_n(\mathbf{R}_t) - E_m(\mathbf{R}_t))\mathbf{d}_{mn}(\mathbf{R}_t)\right]\tilde{\rho}_{nm}(t)\right)\frac{\Delta t}{2} \quad . \tag{S48}$$

5. Update the nuclear coordinate within a full-time step $\Delta t$

$$\mathbf{R}_{t+\Delta t} \leftarrow \mathbf{R}_t + \mathbf{M}^{-1}\mathbf{P}_{t+\Delta t/2}\Delta t \quad . \tag{S49}$$

6. Update the electronic amplitude within a full-time step $\Delta t$ according to

$$\tilde{\mathbf{c}}(t + \Delta t) \leftarrow \tilde{\mathbf{U}}(\mathbf{R}_{t+\Delta t}, \mathbf{P}_{t+\Delta t/2}; \Delta t)\tilde{\mathbf{c}}(t) \quad . \tag{S50}$$

7. Evaluate the switching probability as the hopping probability of the FSSH algorithm:

$$\omega_{j \to k} = \begin{cases} 0, & j = k \\ \max\left(\dfrac{2\,\text{Im}\left[i\tilde{c}_k(t+\Delta t)\tilde{c}_j^*(t+\Delta t)\mathbf{M}^{-1}\mathbf{P}_{t+\Delta t/2} \cdot \mathbf{d}_{jk}(\mathbf{R}_{t+\Delta t})\right]}{|\tilde{c}_j(t+\Delta t)|^2}\Delta t, 0\right), & j \neq k \end{cases} \quad . \tag{S51}$$

If the switching probability $\omega_{j \to k}$ is greater than 1, we set $\omega_{j \to k} = 1$. Generate a uniform random number $\xi$ in $[0,1]$. If $\xi$ falls in the region, $\left[\sum_{n=1}^{k-1}\omega_{j \to n}, \sum_{n=1}^{k}\omega_{j \to n}\right]$, then we try the switching $j \to k$, i.e., the electronic state that contributes to the adiabatic nuclear force component is switched to State $k$ from State $j$. The nuclear kinematic momentum (equivalently, the diabatic nuclear momentum) is adjusted along the direction of the nonadiabatic coupling vector as done in FSSH,

$$\frac{1}{2}\mathbf{P}_{t+\Delta t/2}^T\mathbf{M}^{-1}\mathbf{P}_{t+\Delta t/2} + E_j(\mathbf{R}_{t+\Delta t})$$
$$= \frac{1}{2}\left(\mathbf{P}_{t+\Delta t/2} + \lambda\mathbf{d}_{jk}(\mathbf{R}_{t+\Delta t})\right)^T\mathbf{M}^{-1}\left(\mathbf{P}_{t+\Delta t/2} + \lambda\mathbf{d}_{jk}(\mathbf{R}_{t+\Delta t})\right) + E_k(\mathbf{R}_{t+\Delta t}) \quad . \tag{S52}$$

If $\lambda$ of eq (S52) has no real solution, the switching of the adiabatic nuclear force component is frustrated and such a switching event is abandoned. Otherwise, we set $j = k$ and adjust the nuclear kinematic momentum $\mathbf{P}_{t+\Delta t/2} \leftarrow \mathbf{P}_{t+\Delta t/2} + \lambda_{\min}\mathbf{d}_{jk}(\mathbf{R})$, where $\lambda_{\min}$ is the root of eq (S52) with the smaller absolute value.

8. Update the nuclear kinematic momentum within the other half time step $\Delta t/2$

$$\mathbf{P}_{t+\Delta t} \leftarrow \mathbf{P}_{t+\Delta t/2} - \left(\nabla_{\mathbf{R}}E_j(\mathbf{R}_{t+\Delta t}) + \sum_{n \neq m}^{F}\left[\left(E_n(\mathbf{R}_{t+\Delta t}) - E_m(\mathbf{R}_{t+\Delta t})\right)\mathbf{d}_{mn}(\mathbf{R}_{t+\Delta t})\right]\tilde{c}_n(t+\Delta t)\tilde{c}_m^*(t+\Delta t)\right)\frac{\Delta t}{2}. \quad (S53)$$

9. Rescale the nuclear kinematic momentum to satisfy the energy conservation

$$\mathbf{P}_{t+\Delta t} \leftarrow \mathbf{P}_{t+\Delta t}\sqrt{(H_0 - E_j(\mathbf{R}_{t+\Delta t}))/(\mathbf{P}_{t+\Delta t}^T\mathbf{M}^{-1}\mathbf{P}_{t+\Delta t}/2)}. \quad (S54)$$

If $H_0 < E_j(\mathbf{R}_{t+\Delta t})$, it suggests that the time step size $\Delta t$ is relatively large for the integrator from time $\Delta t$ to time $t + \Delta t$. In such a case, one should then choose a smaller time step size $\Delta t$ and repeat Steps 4-8 for the update of $(\mathbf{R}_{t+\Delta t}, \mathbf{P}_{t+\Delta t}, \tilde{\mathbf{c}}(t+\Delta t))$ from $(\mathbf{R}_t, \mathbf{P}_t, \tilde{\mathbf{c}}(t))$. The time step size $\Delta t$ should be adjusted in the region where the sum of adiabatic and nonadiabatic nuclear force terms is large.

10. Update the time variable, $t \leftarrow t + \Delta t$. Repeat Steps 3-9 until the evolution of the trajectory ends.

If one removes the nonadiabatic nuclear force term in the RHS of eq (S48) and that of eq (S53), and skips Step 9, then the algorithm above becomes the conventional FSSH algorithm.

Figures S13-S15 illustrate comparisons of FS-NAF, FSSH, and NAF results for spin-boson models, FMO model, and SF model, respectively. It is evident that FS-NAF systematically improves over FSSH.

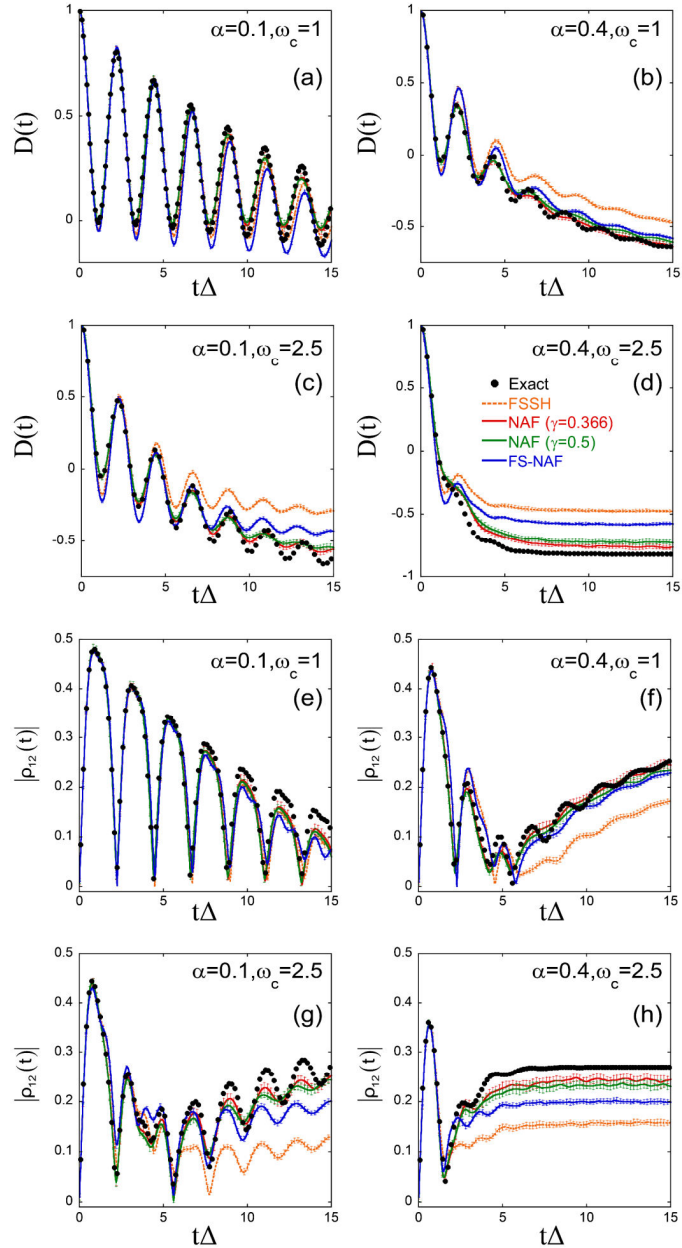

**Figure S13**. Each panel is identical to that in Figure 1 of the main text, but for comparison among FSSH, NAF, and FS-NAF. Black points: Exact results produced by eHEOM. Orange short-dashed lines: FSSH. Red and green solid lines: NAF ($\gamma = 0.366$) and NAF ($\gamma = 0.5$), respectively. Blue solid lines: FS-NAF.

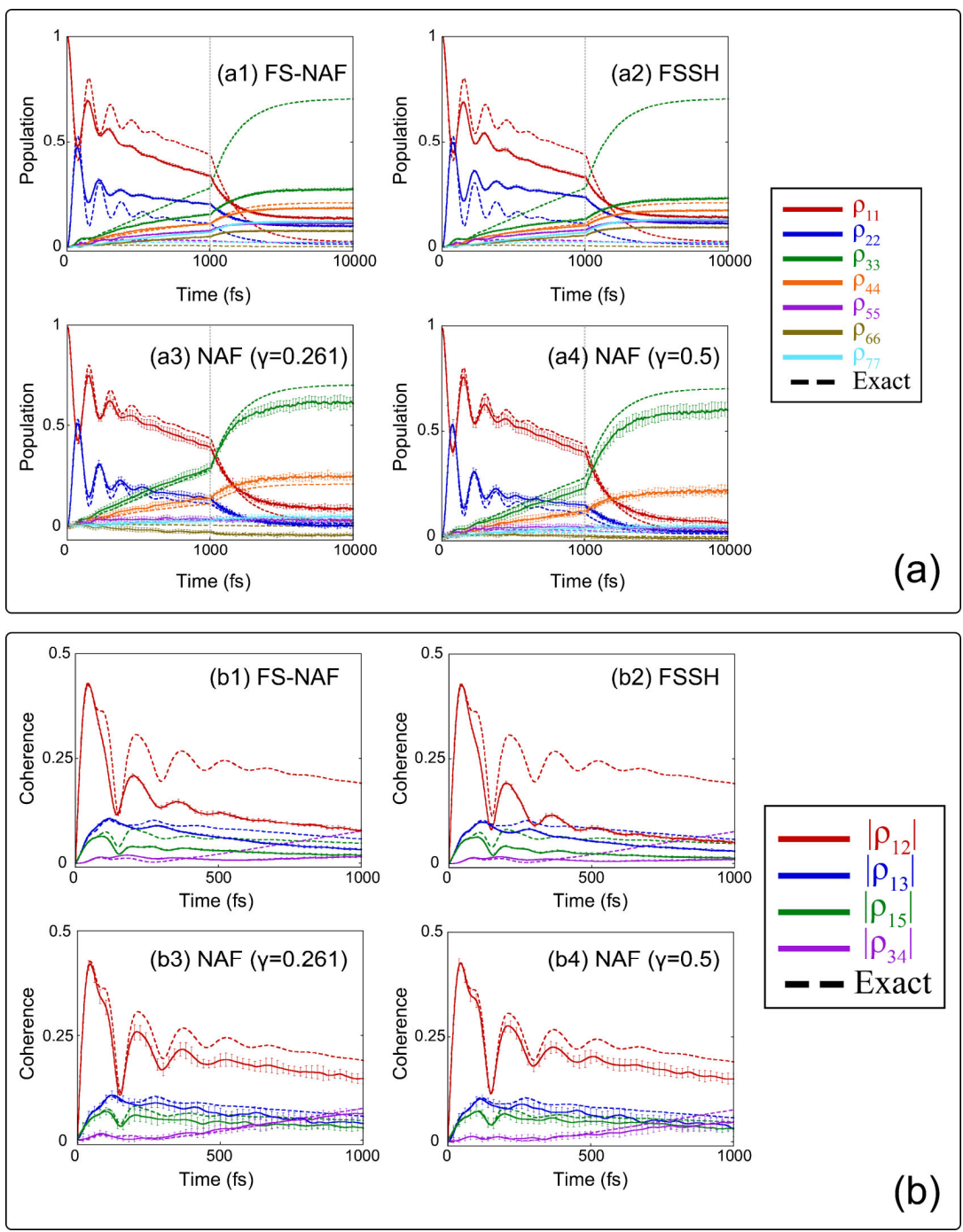

**Figure S14**. Similar to Figure 2 of the main text, but Panels (a1)-(a4) represent the results of FS-NAF, FSSH, NAF ($\gamma = 0.261$), and NAF ($\gamma = 0.5$), respectively. Panel (b) is the same as Panel (a) but for the coherence terms.

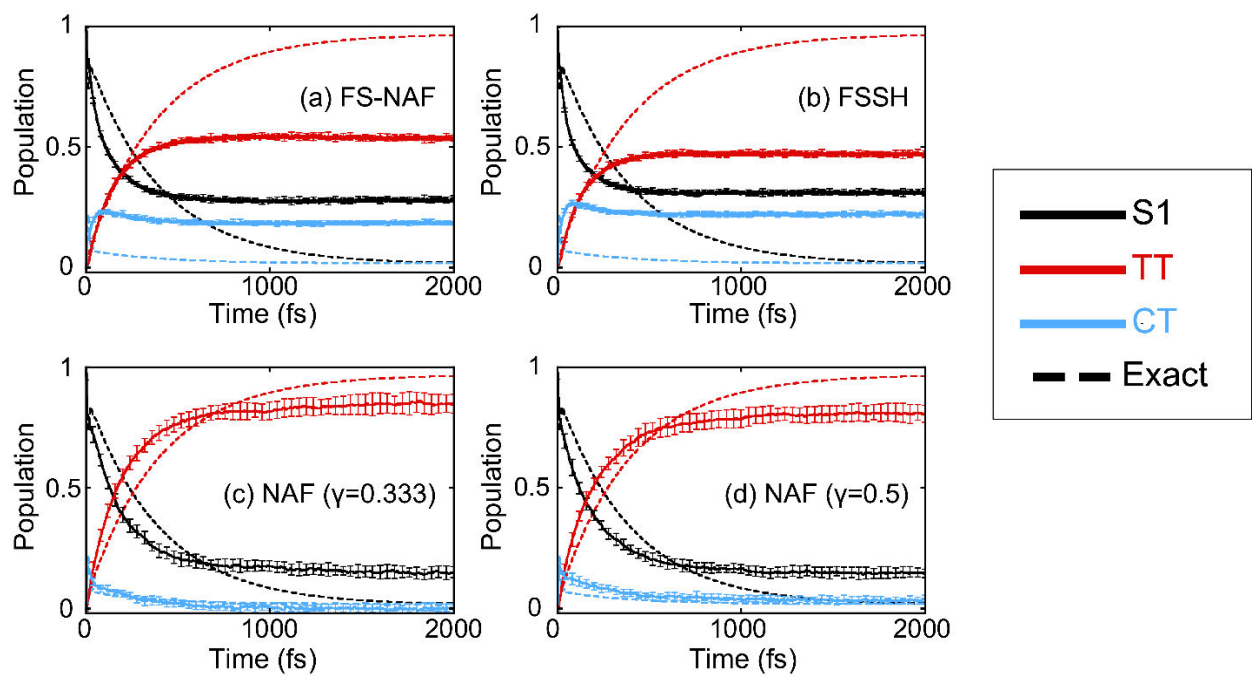

**Figure S15**. Similar to Figure 4 of the main text, but Panels (a)-(d) represent the results of FS-NAF, FSSH, NAF ($\gamma = 0.333$), and NAF ($\gamma = 0.5$), respectively.

## S8. Details of Ehrenfest Dynamics

We describe the details of conventional Ehrenfest dynamics for nonadiabatic transitions. We use $\mathbf{c}$ to denote the electronic amplitude vector in the diabatic representation, whose initial condition is $\{c_n = e^{i\theta}\delta_{j_{occ}n}\}$, where $j_{occ}$ is the index of the initially occupied diabatic state and $\theta$ can be randomly sampled from $[0, 2\pi)$. The electronic amplitude in the adiabatic representation is $\tilde{\mathbf{c}} = \mathbf{T}^{\dagger}(\mathbf{R})\mathbf{c}$. The nuclear DOFs are sampled from the Wigner distribution of the corresponding initial nuclear density operator. The EOMs of Ehrenfest dynamics in the diabatic representation read

$$\dot{\mathbf{c}} = -i\mathbf{V}(\mathbf{R})\mathbf{c} . \tag{S55}$$

$$\dot{\mathbf{R}} = \mathbf{M}^{-1}\mathbf{P} . \tag{S56}$$

$$\dot{\mathbf{P}} = -\sum_{n,m=1}^{F} \nabla_{\mathbf{R}} V_{nm}(\mathbf{R}) c_n c_m^* . \tag{S57}$$

while in the adiabatic representation read

$$\dot{\tilde{\mathbf{c}}} = -i\mathbf{V}^{(\text{eff})}(\mathbf{R}, \mathbf{P})\tilde{\mathbf{c}} . \tag{S58}$$

$$\dot{\mathbf{R}} = \mathbf{M}^{-1}\mathbf{P} . \tag{S59}$$

$$\dot{\mathbf{P}} = -\sum_{n=1}^{F} \nabla_{\mathbf{R}} E_n(\mathbf{R})|\tilde{c}_n|^2 - \sum_{n\neq m}^{F}\left[(E_n(\mathbf{R}) - E_m(\mathbf{R}))\mathbf{d}_{mn}(\mathbf{R})\right]\tilde{c}_n\tilde{c}_m^* . \tag{S60}$$

The electronic density matrix is $\mathbf{cc}^{\dagger}$ in the diabatic representation and $\tilde{\mathbf{c}}\tilde{\mathbf{c}}^{\dagger}$ in the adiabatic representation for each trajectory, respectively. Provided that the electronic diabatic basis sets are well-defined, it is trivial to show that the force in the RHS of eq (S57) is equivalent to that of the RHS of eq (S60) by applying the covariance relation under the diabatic-to-adiabatic transformation (as explicitly discussed in Section 4.1 of ref [27]). Nuclear dynamics is then independent of the representation of the electronic basis sets. In eq (S60) **P** is the diabatic nuclear momentum, or equivalently the nuclear kinematic momentum in the adiabatic representation. One can follow Section 4.1 of ref [27] and Appendix 2 of the Supporting Information of ref [27] to show the relation. An earlier discussion is available in ref [35].

The nuclear force in the RHS of eq (S57) or eq (S60) is the expectation value weighted by the electronic density matrix. It is in the spirit of the Ehrenfest theorem[36], $\frac{d}{dt}\langle \hat{\mathbf{P}} \rangle = -\langle \nabla_{\mathbf{R}} \hat{V} \rangle$ in quantum mechanics. Early examples of the application of the Ehrenfest theorem to chemical dynamics include refs [37-40]. It is not clear to us who first employed the Ehrenfest theorem for electronically nonadiabatic transition processes though.

■ **AUTHOR INFORMATION**

**Corresponding Author**

*E-mail: jianliupku@pku.edu.cn

**ORCID**

Baihua Wu: 0000-0002-1256-6859

Xin He: 0000-0002-5189-7204

Jian Liu: 0000-0002-2906-5858

**Notes**

The authors declare no competing financial interest.

■ **ACKNOWLEDGMENT**

We thank Xiangsong Cheng and Youhao Shang for useful discussions. This work was supported by the National Science Fund for Distinguished Young Scholars Grant No. 22225304. We acknowledge the High-performance Computing Platform of Peking University, Beijing PARATERA Tech Co., Ltd., and Guangzhou Supercomputer Center for providing computational resources.

■ **References:**

(1) Makri, N., The Linear Response Approximation and Its Lowest Order Corrections: An Influence Functional Approach. *J. Phys. Chem. B* **1999**, *103*, 2823-2829. http://dx.doi.org/10.1021/jp9847540
(2) Wang, H., Iterative Calculation of Energy Eigenstates Employing the Multilayer Multiconfiguration Time-Dependent Hartree Theory. *J. Phys. Chem. A* **2014**, *118*, 9253-9261. http://dx.doi.org/10.1021/jp503351t
(3) He, X.; Gong, Z.; Wu, B.; Liu, J., Negative Zero-Point-Energy Parameter in the Meyer-Miller Mapping Model for Nonadiabatic Dynamics. *J. Phys. Chem. Lett.* **2021**, *12*, 2496-2501. http://dx.doi.org/10.1021/acs.jpclett.1c00232
(4) Liu, J.; Miller, W. H., A Simple Model for the Treatment of Imaginary Frequencies in Chemical Reaction Rates and Molecular Liquids. *J. Chem. Phys.* **2009**, *131*, 074113. http://dx.doi.org/10.1063/1.3202438
(5) Duan, C. R.; Tang, Z. F.; Cao, J. S.; Wu, J. L., Zero-Temperature Localization in a Sub-Ohmic Spin-Boson Model Investigated by an Extended Hierarchy Equation of Motion. *Phys. Rev. B* **2017**, *95*, 214308. http://dx.doi.org/10.1103/PhysRevB.95.214308
(6) Tang, Z.; Ouyang, X.; Gong, Z.; Wang, H.; Wu, J., Extended Hierarchy Equation of Motion for the Spin-Boson Model. *J. Chem. Phys.* **2015**, *143*, 224112. http://dx.doi.org/10.1063/1.4936924
(7) Wang, H.; Song, X.; Chandler, D.; Miller, W. H., Semiclassical Study of Electronically Nonadiabatic Dynamics in the Condensed-Phase: Spin-Boson Problem with Debye Spectral Density. *J. Chem. Phys.* **1999**, *110*, 4828-4840. http://dx.doi.org/10.1063/1.478388
(8) Thoss, M.; Wang, H.; Miller, W. H., Self-Consistent Hybrid Approach for Complex Systems: Application to the Spin-Boson Model with Debye Spectral Density. *J. Chem. Phys.* **2001**, *115*, 2991-3005. http://dx.doi.org/10.1063/1.1385562
(9) Craig, I. R.; Thoss, M.; Wang, H., Proton Transfer Reactions in Model Condensed-Phase Environments: Accurate Quantum Dynamics Using the Multilayer Multiconfiguration Time-Dependent Hartree Approach. *J. Chem. Phys.* **2007**, *127*, 144503. http://dx.doi.org/10.1063/1.2772265
(10) He, X.; Wu, B.; Gong, Z.; Liu, J., Commutator Matrix in Phase Space Mapping Models for Nonadiabatic Quantum Dynamics. *J. Phys. Chem. A* **2021**, *125*, 6845-6863. http://dx.doi.org/10.1021/acs.jpca.1c04429
(11) Scully, M. O.; Zubairy, M. S., *Quantum Optics*. 5th ed.; Cambridge University Press: Cambridge, England, 2006.
(12) Hoffmann, N. M.; Schäfer, C.; Säkkinen, N.; Rubio, A.; Appel, H.; Kelly, A., Benchmarking Semiclassical and Perturbative Methods for Real-Time Simulations of Cavity-Bound Emission and Interference. *J. Chem. Phys.* **2019**, *151*, 244113. http://dx.doi.org/10.1063/1.5128076

(13) Hoffmann, N. M.; Schafer, C.; Rubio, A.; Kelly, A.; Appel, H., Capturing Vacuum Fluctuations and Photon Correlations in Cavity Quantum Electrodynamics with Multitrajectory Ehrenfest Dynamics. *Phys. Rev. A* **2019**, *99*, 063819. http://dx.doi.org/10.1103/PhysRevA.99.063819

(14) Chan, W.-L.; Berkelbach, T. C.; Provorse, M. R.; Monahan, N. R.; Tritsch, J. R.; Hybertsen, M. S.; Reichman, D. R.; Gao, J.; Zhu, X.-Y., The Quantum Coherent Mechanism for Singlet Fission: Experiment and Theory. *Acc. Chem. Res.* **2013**, *46*, 1321-1329. http://dx.doi.org/10.1021/ar300286s

(15) Tao, G. H., Electronically Nonadiabatic Dynamics in Singlet Fission: A Quasi-Classical Trajectory Simulation. *J. Phys. Chem. C* **2014**, *118*, 17299-17305. http://dx.doi.org/10.1021/jp5038602

(16) Coronado, E. A.; Xing, J.; Miller, W. H., Ultrafast Non-Adiabatic Dynamics of Systems with Multiple Surface Crossings: A Test of the Meyer-Miller Hamiltonian with Semiclassical Initial Value Representation Methods. *Chem. Phys. Lett.* **2001**, *349*, 521-529. http://dx.doi.org/10.1016/s0009-2614(01)01242-8

(17) Tully, J. C., Molecular Dynamics with Electronic Transitions. *J. Chem. Phys.* **1990**, *93*, 1061-1071. http://dx.doi.org/10.1063/1.459170

(18) Ananth, N.; Venkataraman, C.; Miller, W. H., Semiclassical Description of Electronically Nonadiabatic Dynamics Via the Initial Value Representation. *J. Chem. Phys.* **2007**, *127*, 084114. http://dx.doi.org/10.1063/1.2759932

(19) Church, M. S.; Hele, T. J. H.; Ezra, G. S.; Ananth, N., Nonadiabatic Semiclassical Dynamics in the Mixed Quantum-Classical Initial Value Representation. *J. Chem. Phys.* **2018**, *148*, 102326. http://dx.doi.org/10.1063/1.5005557

(20) Colbert, D. T.; Miller, W. H., A Novel Discrete Variable Representation for Quantum Mechanical Reactive Scattering Via the S-Matrix Kohn Method. *J. Chem. Phys.* **1992**, *96*, 1982-1991. http://dx.doi.org/10.1063/1.462100

(21) Schneider, R.; Domcke, W., S1-S2 Conical Intersection and Ultrafast S2->S1 Internal Conversion in Pyrazine. *Chem. Phys. Lett.* **1988**, *150*, 235-242. http://dx.doi.org/10.1016/0009-2614(88)80034-4

(22) Krempl, S.; Winterstetter, M.; Plöhn, H.; Domcke, W., Path-Integral Treatment of Multi-Mode Vibronic Coupling. *J. Chem. Phys.* **1994**, *100*, 926-937. http://dx.doi.org/10.1063/1.467253

(23) Worth, G. A.; Welch, G.; Paterson, M. J., Wavepacket Dynamics Study of Cr(CO)$_5$ after Formation by Photodissociation: Relaxation through an (E $\oplus$ A) $\otimes$ e Jahn–Teller Conical Intersection. *Mol. Phys.* **2006**, *104*, 1095-1105. http://dx.doi.org/10.1080/00268970500418182

(24) Worth, G. A.; Beck, M. H.; Jackle, A.; Meyer, H.-D.The MCTDH Package, Version 8.2, (2000). H.-D. Meyer, Version 8.3 (2002), Version 8.4 (2007). O. Vendrell and H.-D. Meyer Version 8.5 (2013). Version 8.5 contains the ML-MCTDH algorithm. See http://mctdh.uni-hd.de. (accessed on November 1st, 2023) Used version: 8.5.14.

(25) Shang, Y.; Cheng, X.; Liu, J., **(to be submitted)**.

(26) Wu, B.; He, X.; Liu, J., Phase Space Mapping Theory for Nonadiabatic Quantum Molecular Dynamics. In *Volume on Time-Dependent Density Functional Theory: Nonadiabatic Molecular Dynamics*, Zhu, C., Ed. Jenny Stanford Publishing: New York, 2022.

(27) He, X.; Wu, B.; Shang, Y.; Li, B.; Cheng, X.; Liu, J., New Phase Space Formulations and Quantum Dynamics Approaches. *Wiley Interdiscip. Rev. Comput. Mol. Sci.* **2022**, *12*, e1619. http://dx.doi.org/10.1002/wcms.1619

(28) Liu, J.; He, X.; Wu, B., Unified Formulation of Phase Space Mapping Approaches for Nonadiabatic Quantum Dynamics. *Acc. Chem. Res.* **2021**, *54*, 4215-4228. http://dx.doi.org/10.1021/acs.accounts.1c00511
(29) He, X.; Liu, J., A New Perspective for Nonadiabatic Dynamics with Phase Space Mapping Models. *J. Chem. Phys.* **2019**, *151*, 024105. http://dx.doi.org/10.1063/1.5108736
(30) Nakahara, M., *Geometry, Topology, and Physics*. 2 ed.; Institute of Physics Publishing: Bristol, 2003.
(31) Atiyah, M. F.; Todd, J. A., On Complex Stiefel Manifolds. *Math. Proc. Cambridge Philos. Soc.* **1960**, *56*, 342-353. http://dx.doi.org/10.1017/s0305004100034642
(32) Lang, H.; Vendrell, O.; Hauke, P., Generalized Discrete Truncated Wigner Approximation for Nonadiabatic Quantum-Classical Dynamics. *J. Chem. Phys.* **2021**, *155*, 024111. http://dx.doi.org/10.1063/5.0054696
(33) Peng, J.; Xie, Y.; Hu, D.; Du, L.; Lan, Z., Treatment of Nonadiabatic Dynamics by on-the-Fly Trajectory Surface Hopping Dynamics. *Acta Phys.-Chim. Sin.* **2019**, *35*, 28-48. http://dx.doi.org/10.3866/PKU.WHXB201801042
(34) Landry, B. R.; Falk, M. J.; Subotnik, J. E., Communication: The Correct Interpretation of Surface Hopping Trajectories: How to Calculate Electronic Properties. *J. Chem. Phys.* **2013**, *139*, 211101. http://dx.doi.org/10.1063/1.4837795
(35) Cotton, S. J.; Liang, R.; Miller, W. H., On the Adiabatic Representation of Meyer-Miller Electronic-Nuclear Dynamics. *J. Chem. Phys.* **2017**, *147*, 064112. http://dx.doi.org/10.1063/1.4995301
(36) Ehrenfest, P., Bemerkung Über Die Angenäherte Gültigkeit Der Klassischen Mechanik Innerhalb Der Quantenmechanik. *Z. Phys.* **1927**, *45*, 455-457. http://dx.doi.org/10.1007/BF01329203
(37) Delos, J. B.; Thorson, W. R.; Knudson, S. K., Semiclassical Theory of Inelastic Collisions. I. Classical Picture and Semiclassical Formulation. *Phys. Rev. A* **1972**, *6*, 709-720. http://dx.doi.org/10.1103/PhysRevA.6.709
(38) Billing, G. D., On the Applicability of the Classical Trajectory Equations in Inelastic Scattering Theory. *Chem. Phys. Lett.* **1975**, *30*, 391-393. http://dx.doi.org/10.1016/0009-2614(75)80014-5
(39) Miller, W.; McCurdy, C., Classical Trajectory Model for Electronically Nonadiabatic Collision Phenomena. A Classical Analog for Electronic Degrees of Freedom. *J. Chem. Phys.* **1978**, *69*, 5163-5173. http://dx.doi.org/10.1063/1.436463
(40) Micha, D. A., A Self-Consistent Eikonal Treatment of Electronic Transitions in Molecular Collisions. *J. Chem. Phys.* **1983**, *78*, 7138-7145. http://dx.doi.org/10.1063/1.444753